\documentclass[onecolumn, numberedappendix]{aastex702}
\usepackage{graphicx}
\usepackage{color}

\newcommand{\av}{$A_V$}

\newcommand{\eg}{{\it e.g.}}
\newcommand{\etal}{et~al.}
\newcommand{\ic}{$I_{\rm C}$}

\newcommand{\ks}{$K_{\rm s}$}
\newcommand{\vmkz}{$(V-K_{\rm s})_0$}
\newcommand{\vmk}{$(V-K_{\rm s})$}

\newcommand{\vsini}{$v \sin i$}
\newcommand{\mum}{$\mu$m}

\begin{document}

\title{A Catalog of Young Stellar Objects in the Lagoon Nebula}

\reportnum{Version from \today}

\author[0000-0001-6381-515X]{L.~M.~Rebull}
\affiliation{Infrared Science Archive (IRSA) and 
NASA/IPAC Teacher Archive Research Project (NITARP), IPAC, MS 100-22, 1200 E.\
California Blvd., California Institute of Technology, Pasadena, CA
91125, USA}
\email{rebull@ipac.caltech.edu}

\author[0000-0002-4115-0318]{L.~Venuti}
\affiliation{SETI Institute, 339 Bernardo Ave., Suite 200, Mountain View, CA 94043, USA}
\affiliation{Visiting Fellow, School of Physics, UNSW Science, Kensington, NSW 2052, Australia}
\email{lvenuti@seti.org}

\author[0000-0002-3656-6706]{A.~M.~Cody}
\affiliation{SETI Institute, 339 Bernardo Ave., Suite 200, Mountain View, CA 94043, USA}
\email{acody@seti.org}

\begin{abstract}
M8, the Lagoon Nebula, has been known for more than 100 years as a 
region forming stars. It is (in projection) very close to if not
in the Galactic Bulge. As a result, the source surface density is very
high and source confusion is a substantial challenge. Because of the 
location, there have been many surveys that serendipitously observed
M8, in addition to studies specifically targeting it. But to this point,
there has been no effort to resolve source confusion across all the
existing catalogs, spanning many different wavelengths, sensitivities,
and spatial resolutions. We have 
done all of this source reconciliation, and present the merged catalog
here.  The net result is 3019 YSOs with largely photometric data
and/or YSO selection from $\sim$50 different origins, primarily
ranging from $g$-band through 25 \mum, but with a significant number
of X-ray sources, and some data of limited sensitivity out
through 70-450 \mum. The catalog includes several flags that enable 
users to select, e.g., the YSOs that have been identified as young
by the most studies, or the YSOs with the most multi-wavelength 
counterparts, or the ones with the clearest IR excesses. We use
this catalog in a companion paper; it will also be of use in, e.g., 
future SPHEREx and Roman work.

\end{abstract}

\section{Introduction}
\label{sec:intro}

\begin{figure}[htb!]
\epsscale{1.0}
\plotone{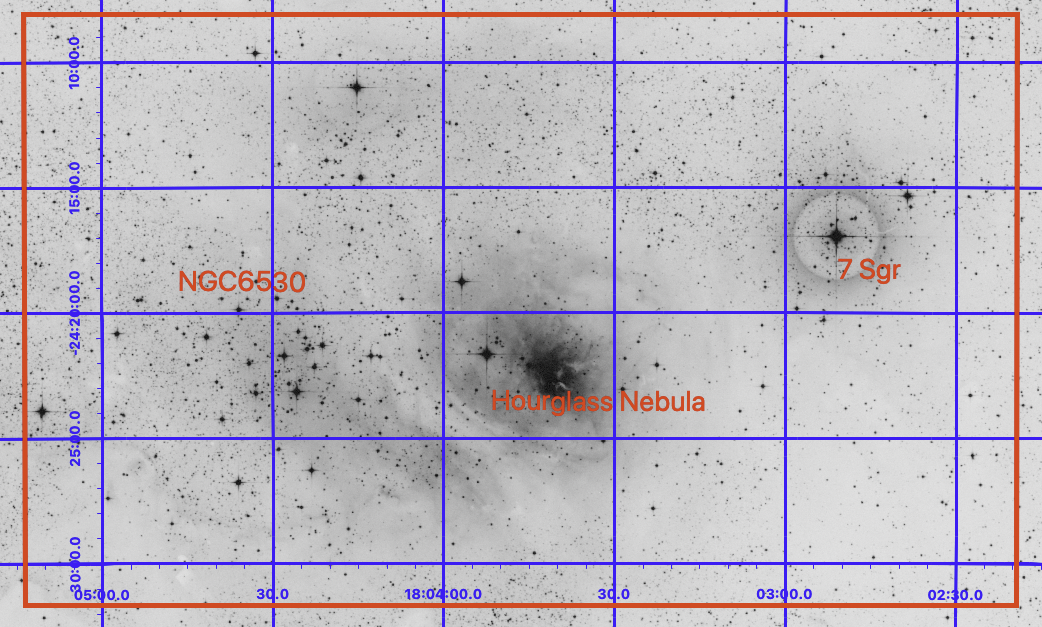}
\caption{The Lagoon Nebula star-forming region. The background image
in the Palomar Observatory Sky Survey (POSS) Digitized Sky Survey 
(DSS) IR image in reverse greyscale; north is up and east is
left. The large red square is  the region studied here, 
$\sim0.66\arcdeg \times \sim0.38\arcdeg$. NGC~6530 is an open cluster
just in front of the cloud forming stars; 7 Sgr is the bright star
in the upper right.  \citet{tothill2008} describe more of  the
features of this region.}
\label{fig:where}
\end{figure}

The Lagoon Nebula, also known as M8, has been studied for more than
100 years as we have sought to understand the star formation going on
within it.  In one of the first studies, \citet{Herbig1957} commented
on the many ``T~Tauri-like'' stars in M8, citing stars that are bright
in H$\alpha$ emission and/or variable.   Despite the well-more than
600 articles on M8, and more than 400 on NGC 6530, the open cluster
slightly in front of the nebulosity that is also part of M8 (see
Fig.~\ref{fig:where}), this region is difficult to study because it is
more or less (in projection) in the Galactic Bulge. With Galactic coordinates of
$l=005.957\arcdeg, b=-01.166\arcdeg$, the source surface density is
very high, especially when moving to fainter magnitude limits or into
the near-infrared (NIR) where the dust is more transparent than in the
optical. The stars in this region are known to be young
\citep[$\sim$1-2 Myr;][]{2019AA...623A.159P}, although some distance
away at $\sim$1250 pc \citep{2019AA...623A.159P}. 

We are carrying on the tradition begun by \citet{Herbig1957} of
exploring the young stars in M8, particularly their variability properties. 
In \citet{venuti2021,venuti2024}, we have discussed results from multicolor
ground- and space-based monitoring of the Lagoon Nebula, and explored
the influence of environment and mass on circumstellar disk accretion. 
In the process of those analyses, we assembled a catalog of
known young stellar objects (YSOs) and candidate YSOs from the literature. 
This paper consolidates all of that catalog assembly process into one
place so as to make it easier for future investigators, particularly
as upcoming surveys like SPHEREx and Roman will need well-vetted catalogs
in order to disentangle and interpret the observations.

Since the late 1990s, there have been $\sim$30 papers published in
refereed major astronomy journals studying significant numbers of
young stars in M8 and its immediate surroundings. There are an
additional $\sim$20 large catalogs over roughly that same timescale
that overlapped this region as part of an all-sky survey, a Galactic
Plane survey, and/or a Galactic Bulge Survey. The net result is
largely photometric data and/or YSO selection from $\sim$50 different
origins, primarily ranging from $g$-band through 25 \mum, but with a
significant number of X-ray sources, and some data of limited
sensitivity out to 70-450 \mum. There are many very bright sources
here at all of the wavelengths considered, as well as unresolved
multiple sources at all bands (some obvious multiples, some not).
Different catalogs had varying degrees of success when it came to
removing artifacts from bright and/or unresolved and/or high surface
density sources.  Different articles had different approaches for
selecting members or young stars, including position and proper motion, 
H$\alpha$ emission, infrared (IR) excess, X-ray emission, and variability. 

We wished to aggregate all of the photometry and YSO identifications 
from the literature into a master catalog so that we could move 
forward with additional analyses. 
We optimized our work here for the region over which we had
space-based K2 \citep{k2howell} light curves \citep{venuti2021}, that
is, the box from (18:02:19.33, $-$24:08:05.9) to (18:05:13.48,
$-$24:31:40.1); see Figure~\ref{fig:where}.  This is a region
$\sim0.66\arcdeg \times \sim0.38\arcdeg$.  There are legitimate
members of M8 and/or NGC 6530 beyond these boundaries, but
consideration of these objects is beyond the scope of the present
paper.

In this paper, we primarily discuss how we went about merging the many
different catalogs and checking those merges (Sec.~\ref{sec:datasets}). 
Identifying the stars with IR excesses (disks) is covered in Sec.~\ref{sec:disks}.  
The general properties of the final catalog are explored in 
Sec.~\ref{sec:properties}. Sec.~\ref{sec:reddening} explores issues
related to reddening in this region, because it is patchy 
and occasionally substantial. We close with some ideas on 
reliability and completeness (Sec.~\ref{sec:reliability}), and 
wind up with some final thoughts (sec.~\ref{sec:theend}).

\begin{longrotatetable}
\begin{deluxetable}{p{5.5cm}p{5.5cm}ccccccc}
\tabletypesize{\scriptsize}
\tablecaption{Literature Included In The Catalog, Part 1\tablenotemark{a} \label{tab:literature} }
\tablewidth{0pt}
\tablehead{\colhead{Reference} &  \colhead{{Paper or catalog focus \& notes}}& 
 \colhead{X-rays} & \colhead{optical}
& \colhead{NIR}& \colhead{MIR}& \colhead{FIR, submm}}
\startdata
IRAS/PSC \citep{iraspaper1984,irasdata2019}& all-sky catalog & \ldots & \ldots  & \ldots & IRAS 12& IRAS 25,60,100 \\
\citet{1997AAS..123...63V}&  PPM to find members, SpTy& \ldots  & $UBVR_cI_c$& \ldots & \ldots & \ldots \\
\citet{2000AJ....120..333S}&  members via $R-H\alpha$ but did not tabulate $H\alpha$ separately& \ldots  & $UBVI_c$& \ldots & \ldots & \ldots \\
\citet{2002AA...395..499R}& X-rays &  XMM& \ldots & \ldots & \ldots & \ldots \\
MSX \citep{Eganmsx,msxdata}& Galactic Plane catalog&  \ldots  & \ldots & \ldots & MSXAB1B2CDE&  \ldots \\
\citet{2004ApJ...608..781D}&  X-rays plus archival data from prior studies&   CXO&  \ldots & \ldots & \ldots & \ldots  \\
\citet{2005AA...430..941P}&  finding members; new optical data plus archival data from prior studies& \ldots  & $BVI_c$&\ldots  & \ldots & \ldots \\
DENIS \citep{denispaper,denisdata} & big catalog&  \ldots   & $i$& $JK_s$& \ldots &  \ldots \\
2MASS/PSC \citep{2masspaper,2masspsc}& all-sky catalog&  \ldots   & \ldots & $JH$\ks& \ldots &  \ldots \\
\citet{2006ChJAA...6..435Z} & PPM to find members & \ldots & photog & \ldots & \ldots & \ldots \\
\citet{2006MNRAS.366..739A}& tiny field of view&\ldots &  \ldots  &  $JH$\ks& \ldots & \ldots \\
\citet{2007MNRAS.374.1253A} &  follow-up on a few members  & \ldots & \ldots & \ldots & \ldots & \ldots \\
AKARI IRC \citep{akaripaper,akariirc}& all-sky catalog &  \ldots  & \ldots & \ldots & Akari 9,18&  \ldots \\
AKARI FIS \citep{akaripaper,akarifis}& all-sky catalog & \ldots  & \ldots & \ldots & \ldots &  Akari 65,90,140,160 \\
\citet{2007AJ....134.1368C}&  PPM to find members&  \ldots & $B$& \ldots & \ldots &  \ldots \\
\citet{2007AA...462..123P}&  spectroscopic follow up (\vsini, Li) &  \ldots & $BVI_c$& \ldots & \ldots & \ldots   \\
UKIDSS \citep{ukidsspaper,ukidssdata} & big catalog&  \ldots & \ldots & $JHK$& \ldots &  \ldots \\
Spitzer/GLIMPSE \citep{glimpseprelaunch,glimpsepaper,glimpse2,glimpse3d} & Galactic Plane catalog&\ldots &  \ldots  & \ldots & IRAC1234& \ldots  \\
\citet{2010MNRAS.407.1170K} & members selected from IR excess &  \ldots & \ldots & \ldots & IRAC1234&  \ldots \\
WISE/AllWISE \citep{wisepaper,allwisedata} & all-sky catalog & \ldots & \ldots & \ldots & WISE1234& \ldots  \\
\citet{2012ApJ...747...51H} &  rotation periods   & \ldots& $VI_c$& \ldots & \ldots &  \ldots \\
VVV \citep{2012AA...537A.107S,2017yCat.2348....0M} & big catalog & \ldots  & \ldots & $ZYJHK_s$& \ldots &  \ldots \\
MYStIX: \citet{2013ApJS..209...31P} &  Chandra+IRAC&  CXO& \ldots & $JH$\ks& IRAC1234& \ldots  \\
MYStIX: \citet{2013ApJS..209...32B} &  Chandra+IRAC; larger than prior paper; smaller area in general& CXO& \ldots & $JH$\ks& IRAC1234&  \ldots \\
Spitzer/SEIP \citep{seipcatalog} &  catalog over large portion of sky & \ldots & \ldots & \ldots & IRAC1234, M1& \ldots  \\
\citet{2013MNRAS.434..806B}&  catalog with everything and with member subset& \ldots & $Ugri$& $Z$& \ldots &  \ldots \\
VPHAS+ \citep{2014MNRAS.440.2036D,2016yCat.2341....0D} & big catalog& \ldots  & $ugriH\alpha$& \ldots & \ldots &  \ldots \\
WISE/unWISE \citep{2014AJ....147..108L,2017AJ....153...38M,2017AJ....154..161M,2019ApJS..240...30S} & all-sky catalog   &  & & & WISE12&   \\
\citet{2013PhDT........36C}&  Chandra, archival data from prior work& CXO& \ldots & \ldots & \ldots & \ldots  \\
\citet{2015MNRAS.453.1026K}&  members from H$\alpha$, UV excess, derived mass accretion&  \ldots & \ldots & \ldots & \ldots & \ldots  \\
MYStIX: \citet{2016ApJ...833..193R}& Chandra+IRAC+Herschel; youngest YSOs & CXO & \ldots & $JH$\ks & IRAC1234 & SPIRE 500 \\
PanSTARRS \citep{panstarrs,2017yCat.2349....0C} & big catalog& \ldots  & $gri$& $zy$& \ldots & \ldots  \\
Herschel/HPDP PSC 70 \citep{martonexplansupp,pacspsc70}&  catalog over large portion of sky&  \ldots & \ldots & \ldots  & \ldots & PACS 70 \\
Herschel/HPDP PSC 160 \citep{martonexplansupp,pacspsc160}&  catalog over large portion of sky&  \ldots & \ldots & \ldots  & \ldots & PACS 160 \\
SCUBA \citep{2008ApJS..175..277D,2009yCat..21750277D}&   catalog& \ldots  & \ldots & \ldots & \ldots & SCUBA 450,850 \\
\citet{2019AA...623A.159P} &  spectroscopic follow up (\vsini, Li), Gaia-ESO survey& \ldots  & $VI_c$& \ldots & \ldots & \ldots  \\
\citet{2019AA...623A..25D}&  PPM members (Gaia DR2) & CXO& \ldots & \ldots & \ldots & \ldots  \\
\citet{2019MNRAS.486.2477W} & PPM members (Gaia DR2), Gaia-ESO survey & \ldots & Gaia, spec & \ldots & \ldots & \ldots \\
MYStIX: \citet{2020AA...633A.155R}  & spectroscopic follow up & \ldots &   spec &  \ldots & \ldots & \ldots\\
SPICY: \citet{2021ApJS..254...33K,2021ipac.data.I556K} & YSOs in Gal.~Plane; IRAC+Gaia+ZTF variab. & \ldots & \ldots & \ldots& IRAC1234 & \ldots\\
WISE/CatWISE \citep{2021ApJS..253....8M,2020ApJS..247...69E,catwise} & all-sky catalog & \ldots & \ldots& \ldots & WISE12&  \ldots \\
\citet{2022MNRAS.509.1664J} & PPM (Gaia EDR3), Gaia-ESO survey & \ldots  & Gaia, spec & \ldots & \ldots & \ldots\\
Gaia DR3 \citep{gaiadr3,gaiadr3data} & all-sky catalog&  & Gaia $BGR$  & \ldots & \ldots & \ldots  \\
Gaia DR3/distances \citep{2021AJ....161..147B,2021yCat.1352....0B} & all-sky catalog; distances from Bayesian statistics& \ldots & Gaia & \ldots & \ldots  & \ldots \\
\citet{2022MNRAS.517.6191O} & \ks\ monitoring, periods & \ldots & \ldots& \ks\ & \ldots & \ldots \\
\citet{venuti2021} & K2 and ground-based monitoring& \ldots & $ugriH\alpha$ & \ldots & \ldots& \ldots\\
\citet{venuti2024} & disk accretion& \ldots & $ugriH\alpha$ & \ldots & \ldots& \ldots\\
\enddata
\tablenotetext{a}{The bands in the last 5 columns are the values we retained, as opposed to
a more comprehensive list of the values used by the authors. In many cases, the authors
used values from other references in this list -- for example, \citet{2021AJ....161..147B}
used the data from \citet{gaiadr3}, but we obtained the Gaia photometry from
its original source \citep{gaiadr3}, so the photometry (``Gaia $BGR$") is not listed in the 
in the row corresponding to \citet{2021AJ....161..147B}. }
\end{deluxetable}
\end{longrotatetable}
\clearpage

\begin{longrotatetable}
\begin{deluxetable}{p{5.5cm}ccccp{4cm}}
\tabletypesize{\scriptsize}
\tablecaption{Literature Included In The Catalog, Part 2 \label{tab:literature2} }
\tablewidth{0pt}
\tablehead{\colhead{Reference} &  \colhead{Area covered}& 
\colhead{\# srcs/\# ysos\tablenotemark{a}} & \colhead{YSO flag\tablenotemark{b}} & 
\colhead{Sp.\ Res.\tablenotemark{c}} & \colhead{Depth\tablenotemark{d}}
}
\startdata
IRAS/PSC \citep{iraspaper1984,irasdata2019}& whole area & 19/5&  & $\sim1-3\arcmin$ & to $\gtrsim$1.2 Jy at 12 $\mu$m \\
\citet{1997AAS..123...63V}& most of area & 96/96& x& $\sim16\arcsec$ & \ks$\sim$11 mag \\
\citet{2000AJ....120..333S}& smaller area & 900/446& x& $\sim2\arcsec$ & \ks$\sim$12 mag   \\
\citet{2002AA...395..499R}& smaller area & 105/105& (x) & $\sim6\arcsec$ & \ks$\sim$12 mag  \\
MSX \citep{Eganmsx,msxdata}&  whole area & 292/35&  & $\sim20\arcsec$ & to $\gtrsim$0.07 Jy at 4.29 $\mu$m   \\
\citet{2004ApJ...608..781D}& smaller area & 884/884& (x) & $\sim1\arcsec$ & \ks$\sim$13.2 mag   \\
\citet{2005AA...430..941P}& smaller area & 1,329/926& x& $\sim0.8\arcsec$ & \ks$\sim$13.2 mag \\
DENIS \citep{denispaper,denisdata} &  most of area & 24,550/1,550&  & $\sim2\arcsec$ & $K_D\sim13$ mag \\
2MASS/PSC \citep{2masspaper,2masspsc}& whole area& 29,501/2,069&  & $\sim1.5\arcsec$ & \ks$\sim$13.2 mag  \\
\citet{2006ChJAA...6..435Z} & most of area & 112/112 & x & $\sim1\arcsec$? & \ks$\sim$10.5 mag\\
\citet{2006MNRAS.366..739A}&  small area & 744/127&  & $<1\arcsec$ & \ks$\sim$15.7 mag\\
\citet{2007MNRAS.374.1253A} &  small area & 44/44& x& $\sim1\arcsec$ & \ks$\sim$13 mag\\
AKARI IRC \citep{akaripaper,akariirc}&  whole area& 79/27&  & $\sim5\arcsec$ &  to $\gtrsim$ 0.1 Jy at 9 $\mu$m \\
AKARI FIS \citep{akaripaper,akarifis}&  whole area& 61/1&  & $\gtrsim2\arcsec$ & to $\gtrsim$ 0.75 Jy at 65 $\mu$m  \\
\citet{2007AJ....134.1368C}&  most of area& 147/147& x& $\sim1.5\arcsec$? & \ks$\sim$10.7 mag\\
\citet{2007AA...462..123P}&  smaller area& 332/332& x& \ldots & \ks$\sim$12.4 mag \\
UKIDSS \citep{ukidsspaper,ukidssdata} & most of area& 157,995/2,580&  & $\sim0.8\arcsec$ & \ks$\sim$16 mag   \\
Spitzer/GLIMPSE \citep{glimpseprelaunch,glimpsepaper,glimpse2,glimpse3d} & whole area& 69,080/2,210&  & 
$\sim1.6\arcsec$ & [3.6]$\sim$14.0 mag \\
\citet{2010MNRAS.407.1170K} &  most of area& 221/221& x& $\sim6\arcsec$ & \ks$\sim$11.6 mag \\
WISE/AllWISE \citep{wisepaper,allwisedata} &  whole area& 6,006/284&  & $\sim6\arcsec$ & [3.4]$\sim$11 mag \\
\citet{2012ApJ...747...51H} &  smaller area& 290/290& (x) & $\sim1.5\arcsec$ & \ks$\sim$13.2 mag  \\
VVV \citep{2012AA...537A.107S,2017yCat.2348....0M} & whole area& 168,338/2,665&  & $\sim1\arcsec$ & \ks$\sim$16.5 mag \\
MYStIX: \citet{2013ApJS..209...31P} &  smaller area & 1,327/1,235& x& $\sim1\arcsec$ & \ks$\sim$13 mag  \\
MYStIX: \citet{2013ApJS..209...32B} &  smaller area & 1,914/1,914& x& $\sim1\arcsec$ & \ks$\sim$13 mag  \\
Spitzer/SEIP \citep{seipcatalog} &   whole area & 15,778/993&  & $\sim1.6\arcsec$ & [3.6]$\sim$13.0 mag \\
\citet{2013MNRAS.434..806B}&  smaller area & 8,567/936& x& $\sim1.5\arcsec$ & \ks$\sim$15.0 mag   \\
VPHAS+ \citep{2014MNRAS.440.2036D,2016yCat.2341....0D} & whole area& 59,473/2,280 & & $\sim1\arcsec$ & $i\sim$21 mag  \\
WISE/unWISE \citep{2014AJ....147..108L,2017AJ....153...38M,2017AJ....154..161M,2019ApJS..240...30S} & whole area & 24,759/628 &  & $\sim6\arcsec$ & [3.4]$\sim$12.5 mag \\
\citet{2013PhDT........36C}&  smaller area& 1,350/1,303& (x) & $\sim1\arcsec$  & \ks$\sim$13.5 mag \\
\citet{2015MNRAS.453.1026K}&  most of area& 195/195& x&  $\sim1\arcsec$ & \ks$\sim$12.7 mag \\
MYStIX: \citet{2016ApJ...833..193R}& smaller area & 48/48 & x & $\sim1\arcsec$ & \ks$\sim$13 mag \\
PanSTARRS \citep{panstarrs,2017yCat.2349....0C} & whole area& 84,118/2,494&  & $\sim1\arcsec$ & $y\sim$19 mag   \\
Herschel/HPDP PSC 70 \citep{martonexplansupp,pacspsc70}&  basically off-edge& 1/0& & $\sim6\arcsec$ & \ldots \\
Herschel/HPDP PSC 160 \citep{martonexplansupp,pacspsc160}&  basically off-edge& 2/0&  & $\sim11\arcsec$ & \ldots \\
SCUBA \citep{2008ApJS..175..277D,2009yCat..21750277D}&   smaller area& 33/0& & $\sim9-14\arcsec$ & \ldots \\
\citet{2019AA...623A.159P} & smaller area& 616/616& x& $\sim0.5\arcsec$ (Gaia) & \ks$\sim$13.2  mag \\
\citet{2019AA...623A..25D}&  most of area & 2,901/2,031& x& $\sim1\arcsec$& \ks$\sim$13.6 mag \\
\citet{2019MNRAS.486.2477W} & most of area & 776/776 & x & $\sim0.5\arcsec$ (Gaia) & \ks$\sim$13.6 mag\\
MYStIX: \citet{2020AA...633A.155R}  & smaller area & 86/86 & x & $\sim1\arcsec$ &  \ks$\sim$13 mag\\
SPICY: \citet{2021ApJS..254...33K,2021ipac.data.I556K} & whole area & 466/466 & x &$\sim1\arcsec$  & \ks$\sim$13.0 mag\\
WISE/CatWISE \citep{2021ApJS..253....8M,2020ApJS..247...69E,catwise} & whole area & 16,926/447 &   & $\sim6\arcsec$ & [3.4]$\sim$12.5 mag \\
\citet{2022MNRAS.509.1664J} & whole area & 1458/735 & x & $\sim0.5\arcsec$ (Gaia)  & \ks$\sim$13.2 mag \\
Gaia DR3 \citep{gaiadr3,gaiadr3data} & whole area & 36,939/2,374&  & $\sim0.5\arcsec$ & $G\sim$20.7 mag\\
Gaia DR3/distances \citep{2021AJ....161..147B,2021yCat.1352....0B} & whole area & 27,693/2,202& & $\sim0.5\arcsec$ & $G\sim$20.5 mag  \\
\citet{2022MNRAS.517.6191O} & whole area & 361/361 & (x) & & \ks$\sim$13.2 mag \\
\citet{venuti2021} & whole area & 275/271 & x & $\sim1\arcsec$& \ks$\sim$13.2 mag \\
\citet{venuti2024} & whole area & 943/928 & x & $\sim1\arcsec$& \ks$\sim$13.2 mag \\
\enddata
\tablenotetext{a}{Total number of sources within the specified RA/Dec region, and 
total number of literature YSOs or candidate YSOs in final, merged catalog, separated by a `/.'}
\tablenotetext{b}{We retained from this paper or catalog the indication of whether or 
not a given source was a cluster member. Some papers or catalogs publish only the members, 
so their catalogs are entirely YSOs; other catalogs have far more entries than YSOs. Notation 
of `(x)' indicates an implicit YSO identification, e.g., that all X-ray detections or all 
variables are taken as YSOs. }
\tablenotetext{c}{Approximate spatial resolution, in arcseconds; see the text.}
\tablenotetext{d}{Approximate depth of survey; see the text.}
\end{deluxetable}
\end{longrotatetable}
\clearpage

\section{Specific Data Sets and the Merging Thereof}
\label{sec:datasets}

\subsection{Introduction}

Because the surface density of point sources in and around M8 is very
high, exactly how we merge catalogs (and in what order) matters; it
can substantially affect which sources remain at the end of the
merging process, and what cross-IDs are assigned to which objects.  We
have developed considerable skill in this kind of merging across
heterogeneous catalogs \citep[see, e.g.,][and the discussions 
therein]{1333catalogpaper,ic417paper}; this is particularly difficult
when merging between the output of large, well-tested pipelines with
high-quality astrometry (e.g., those from large sky surveys with many
point sources) and those from specific surveys of small regions (which
may not have the same level of coordinate accuracy and precision), and
across wavelengths with a variety of spatial resolutions and depths 
(where, among other things, there may not be many point sources to
anchor the astrometry). Tools available at NASA's Infrared Science
Archive (IRSA) are critical for this, because they allow for
interactive exploration  of images and catalogs, including overlaying
multiple catalogs on the same images.  We took special care in this
very dense region, repeatedly merging, testing the fidelity of the 
merge (see discussion below), and re-merging (and re-testing). 
Sources were only consolidated (or removed) if we were certain that 
they should be consolidated (or removed). 

In this section, we describe the datasets merged and the order in
which they  were combined (unless noted otherwise). Where possible, we
ingested entire catalogs; otherwise, we incorporated YSO-only catalogs
as provided. Across all datasets,  we tracked cross-identifications
and known or candidate YSOs.  Because our primary goal is to compile a
reliable list of YSOs, we  focused our manual cross-match quality and
detailed source assessment -- including  assembling spectral energy
distributions (SEDs) and image inspection --  exclusively on the YSO
candidates.

Specifically because there are so many sources of data, 
Tables~\ref{tab:literature} and \ref{tab:literature2} consolidate all
of the most important information to compare all the constituent
catalogs, in chronological order (as opposed to the order in which
they  were combined). Note that the constituent catalogs sometimes
cover only portions of our field (see Table~\ref{tab:literature2} and
text below), and these portions are not necessarily the center of the
field; sometimes additional constraints are imposed by the
nebulosity.  Plots of each catalog's spatial coverage are included in
the subsections below. In each case, we have plotted the individual
catalogs (black) on top of the  entire catalog (light grey) as an
indication of sky coverage. If the catalog has contributed both a
source catalog and a catalog of members or YSO candidates, those are
highlighted in red.  Artifacts of both astrophysical (nebulosity) and
instrumental (bright source masking; tile boundaries) origins are
apparent.

The final catalog is described in Sec.~\ref{sec:properties} and 
appears online, delivered to IRSA and 
described in Appendix~\ref{app:datatable} and Table~\ref{tab:bigdata}.

\subsection{Infrared - the Anchors}

\subsubsection{2MASS}

The Two Micron All-Sky Survey \citep[2MASS;][]{2masspaper} surveyed
the entire sky in $JH$\ks.  It provides a highly reliable large-scale
coordinate system across the sky, and can provide a reliable position
reference for both optical and IR surveys. 2MASS is used to establish
coordinate systems by many recent surveys.  We started with the 2MASS
catalog \citep{2masspsc}, retaining all of the sources from that
catalog. The spatial resolution of the 2MASS survey is usually cited
as $\sim$1.5$\arcsec$ \citep{2masspaper}. 

\begin{figure}[htb!]
\epsscale{0.5}
\plotone{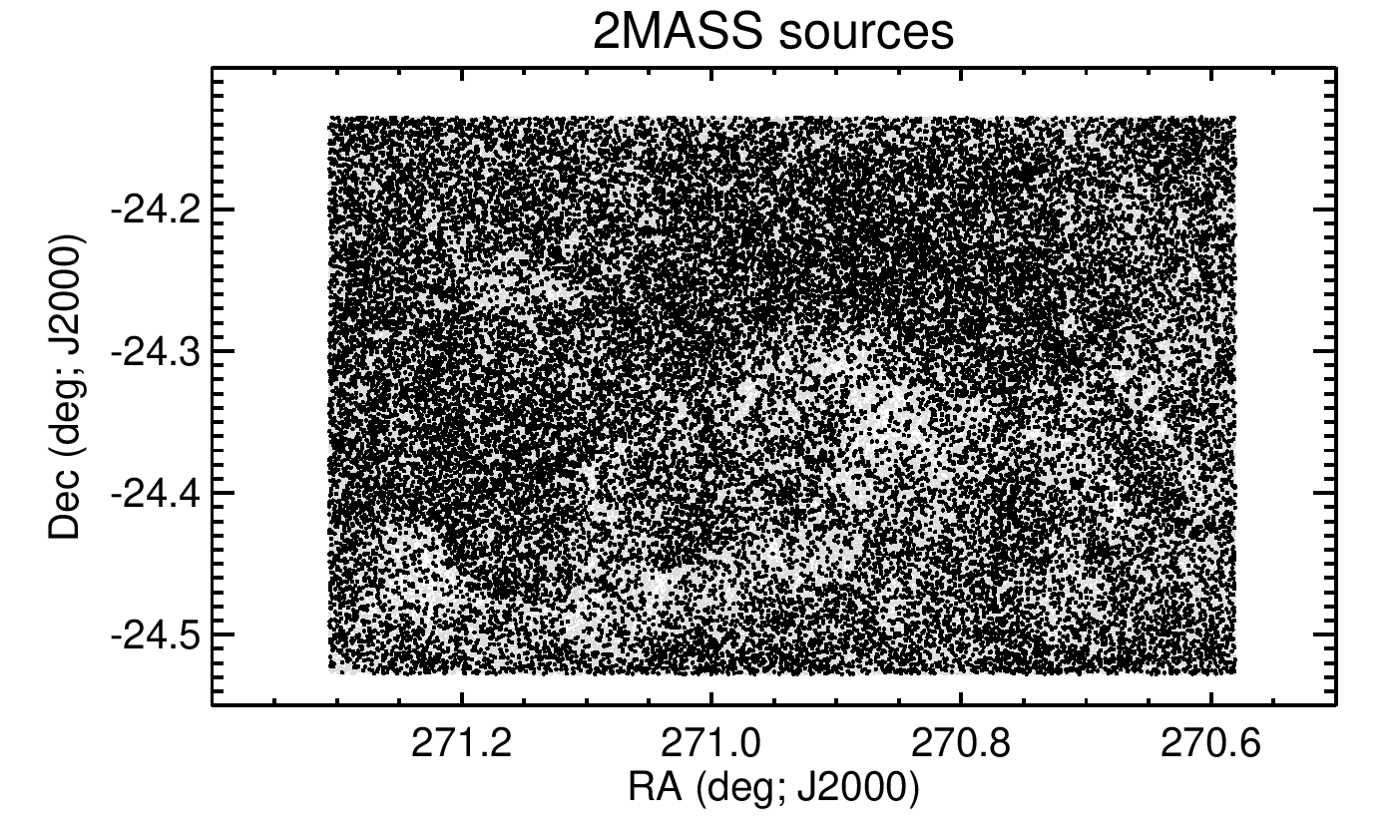}
\caption{2MASS/PSC \citep{2masspaper,2masspsc} sources (black) on the 
entire catalog (light grey). Because there are 2MASS data densely distributed
over the field, it is very hard to see the light grey. Places where there 
are bright nebulosity and extinction (compare to optical image in 
Fig.~\ref{fig:where}) are evident as lower source densities.}
\label{fig:wheretwomass}
\end{figure}

\textbf{ Coverage and numbers.} 2MASS covers our entire survey region,
with 29,501 sources. They are plotted in Fig.~\ref{fig:wheretwomass}. 
The depth at \ks\ is usually cited as $\sim$15.3 mag
\citep{2masspaper}, but here in this incredibly complicated region, a
histogram of \ks\ peaks at $\sim$13.2 mag.  While no YSOs are
identified as part of the 2MASS catalog itself, 2,069 2MASS sources in
this region are identified in the literature as YSOs or candidates,
vetted through the analysis below. All of the sources from this
catalog are retained in the master catalog. 

\subsubsection{Spitzer/GLIMPSE}

The Spitzer Space Telescope \citep{mainspitzer} pipelines produced
positions that are anchored to those from 2MASS, so we next worked
with Spitzer data. The Spitzer Legacy program called Galactic Legacy
Infrared Mid-Plane Survey Extraordinare
\citep[GLIMPSE;][]{glimpsepaper} included the Galactic plane. The
Lagoon Nebula occupies a corner of the GLIMPSE coverage. Most of our
field was observed as part of the second extension of this program
\citep[GLIMPSE II;][]{glimpse2}, but not quite all of it. Part of our
field was also observed as part of GLIMPSE3D \citep{glimpse3d}, which
was targeting (among other things) the Galactic Center. The net result
is complete coverage of our region by aggregate GLIMPSE catalogs.
All of the
GLIMPSE-affiliated observations here used the Infrared Array Camera
\citep[IRAC;][]{2004ApJS..154...10F}, with four channels (3.6, 4.5,
5.8, and 8 \mum). 
We used the `more complete, less reliable' catalogs
from both the GLIMPSE II and GLIMPSE 3D deliveries \citep[the
`Archive';][]{glimpse2,glimpse3d}.  We used TOPCAT
\citep{2005ASPC..347...29T} to merge these two large catalogs together
with a 1$\arcsec$ matching radius. (We made a note of the 2MASS
sources that the GLIMPSE pipeline had identified as counterparts for
comparison/debugging downstream.)  We merged the GLIMPSE data to the
2MASS data by position, with a 1$\arcsec$ radius, retaining all of 
the sources from that catalog.  (Note that there are more Spitzer 
data included and discussed below.) Spitzer/IRAC's spatial resolution 
at 3.6 \mum\ is $\sim$1.6$\arcsec$ \citep{2004ApJS..154...10F}.

\begin{figure}[htb!]
\epsscale{0.5}
\plotone{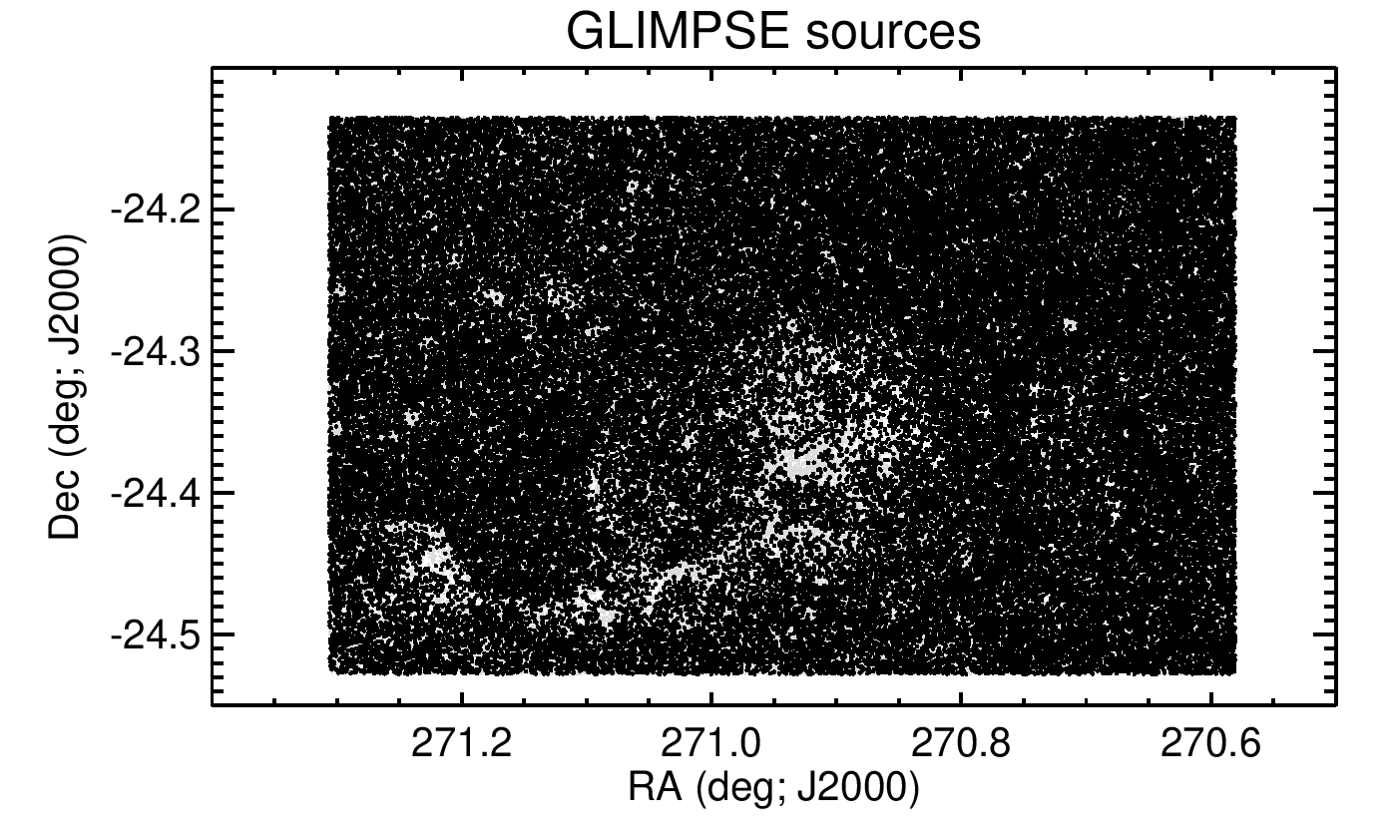}
\caption{Spitzer/GLIMPSE
\citep{glimpseprelaunch,glimpsepaper,glimpse2,glimpse3d}  sources
(black) on the entire catalog (light grey). Because there are GLIMPSE
data  densely distributed over the field, it is very hard to see the
light grey. Places where there are bright nebulosity and extinction 
(compare to optical image in Fig.~\ref{fig:where}) are evident as lower 
source densities. }
\label{fig:whereglimpse}
\end{figure}

\textbf{ Coverage and numbers.} GLIMPSE covers our entire survey
region,  with 69,080 sources. They are plotted in
Fig.~\ref{fig:whereglimpse}.   GLIMPSE's sensitivity is typically
quoted as 15.5 mag at [3.6] \citep{glimpsepaper}; a histogram of the
GLIMPSE [3.6] magnitudes in this incredibly complicated region rolls
over at $\sim$14 mags. While no YSO candidates are identified as part
of the GLIMPSE catalog itself, 2,210 GLIMPSE sources in this region
are identified in the literature as YSOs or candidates, vetted through
the analysis below. All of the sources from this catalog are retained
in the master catalog. 

\subsubsection{Spitzer/SEIP}

\begin{figure}[htb!]
\epsscale{0.5}
\plotone{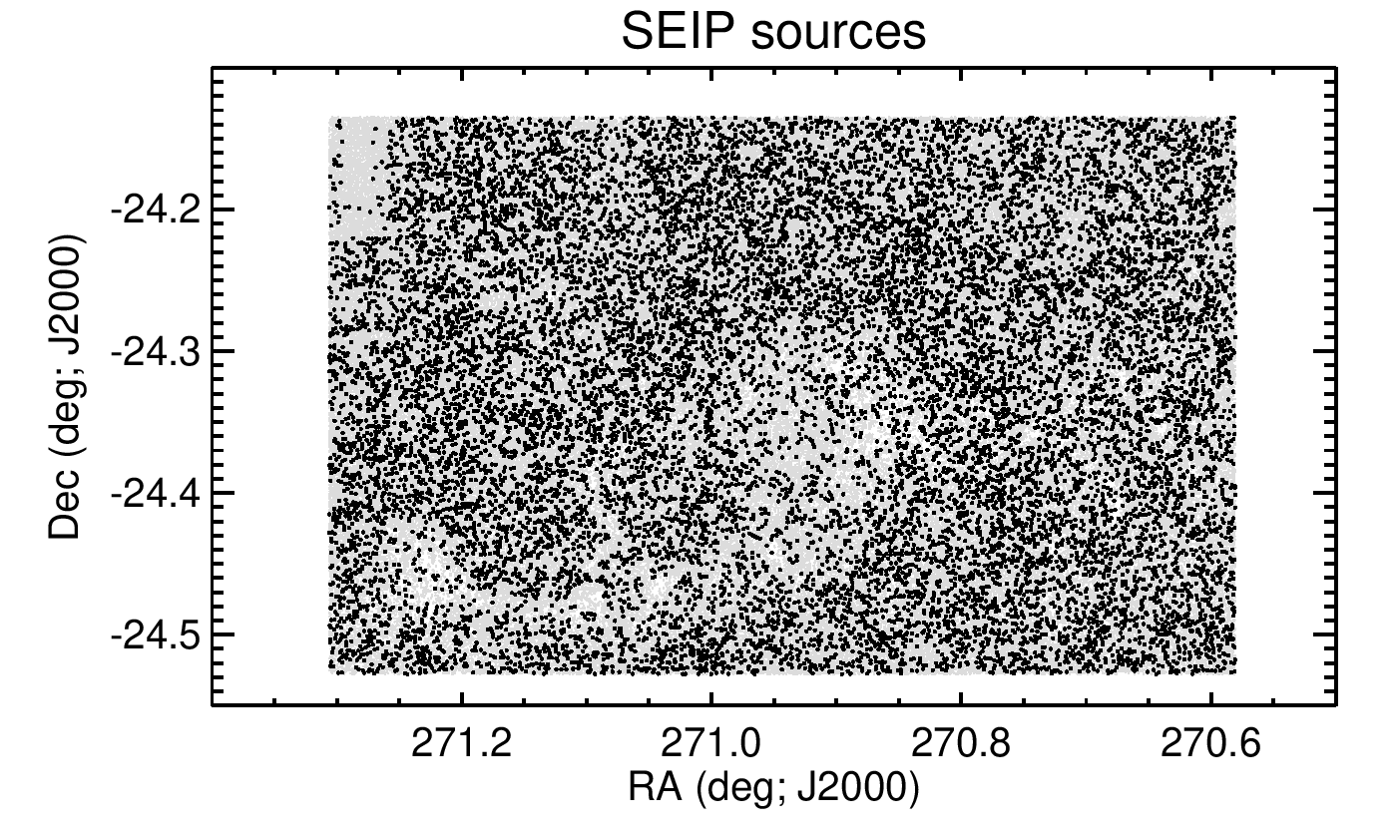}
\caption{Spitzer/SEIP \citep{seipcatalog} sources (black) on the entire 
catalog (light grey). By comparison to Fig.~\ref{fig:whereglimpse}, 
it's obvious that this catalog does not go as deep as GLIMPSE, and 
there is a mask of some sort in the upper left.}
\label{fig:whereseip}
\end{figure}

The Spitzer Enhanced Imaging Products 
(SEIP\footnote{http://irsa.ipac.caltech.edu/data/SPITZER/Enhanced/SEIP/overview.html};
\citealt{seipcatalog}) extracted sources (including bandfilled fluxes
and limits) for all four IRAC bands plus Multiband Imaging Photometer
for Spitzer \citep[MIPS;][]{2004ApJS..154...25R} 24 \mum, nominally to
a SNR=10, meaning limited to the brighter sources.  Note that SEIP is a
reanalysis of the data that compose GLIMPSE, but with generally 
shallower photometry, as can be seen in the different numbers of
sources in the catalog and in Fig.~\ref{fig:whereseip} as compared to
Fig.~\ref{fig:whereglimpse}.  We merged the SEIP data to the aggregate
catalog by position with 1$\arcsec$ radius, keeping track of limits as
distinct from detections and retaining all of the sources from that
catalog. (Note that there are more Spitzer data included and
discussed below.)

\textbf{ Coverage and numbers.} SEIP covers our entire survey region, 
with 15,778 sources. They are plotted in Fig.~\ref{fig:whereseip}. 
SEIP's sensitivity is shallower than GLIMPSE (compare
Figs.~\ref{fig:whereglimpse} and \ref{fig:whereseip}); a histogram of
the SEIP [3.6] magnitudes peaks at $\sim$12.5 and rolls over at
$\sim$13 magnitudes. YSO candidates are not identified as part of the
SEIP catalog itself.  After all of the analysis described below, 993
are identified in the literature as YSOs or candidates. All of the
sources from this catalog are retained in the master catalog.

\subsection{MYStIX}

The Massive Young Star-Forming Complex Study in Infrared and X-ray
\citep[MYStIX;][]{2013ApJS..209...26F} was a very large study that
combined Chandra X-ray Observatory \citep[CXO;][]{2002PASP..114....1W}
and  Spitzer Space Telescope \citep{mainspitzer} data (plus additional
NIR data from 2MASS and UKIRT, the United Kingdom Infrared Telescope)
to identify young stars in 20 star-forming regions, including the
Lagoon Nebula. They did extensive work to match sources across
wavelengths, a non-trivial task given the point-spread functions,
coordinate uncertainties, and large wavelength ranges involved.  The
MYStIX overview \citep{2013ApJS..209...26F} cites an average of
$\sim$1$\arcsec$ spatial resolution and a limit of \ks$\sim$14-18. Chandra's
on-axis spatial resolution is $\sim0.5\arcsec$ and varies
substantially over the field of view.  MYStIX uses a modified GLIMPSE
pipeline; Spitzer's resolution at [3.6] is $\sim1.6\arcsec$. 

We took MYStIX's NIR and Spitzer measurements, and their source
matching where it was available.  \citet{2013ApJS..209...31P}
identified IR-excess sources,  \citet{2013ApJS..209...32B} presented a
list of young stars (based on IR and X-rays), 
and \citet{2016ApJ...833..193R} extended the
analysis to include FIR and identified the youngest stars. 
\citet{2020AA...633A.155R} followed up spectroscopically to confirm
(or refute) the status of the objects they identified earlier.  
(We note for completeness that these papers also used data from 
\citealt{2013ApJS..209...28K}.)

Because the source matching was done so carefully in MYStIX, it was
important for us to inherit this matching, linking it explicitly to the
Spitzer/NIR sources, early in our process. We added in to the master
catalog sources from MYStIX that did not have Spitzer/NIR counterparts
at the positions given in the MYStIX catalog, since there is not much
more we can do for those sources. We also kept track of the sources
identified as young stars in these papers.  

\begin{figure}[htb!]
\plotone{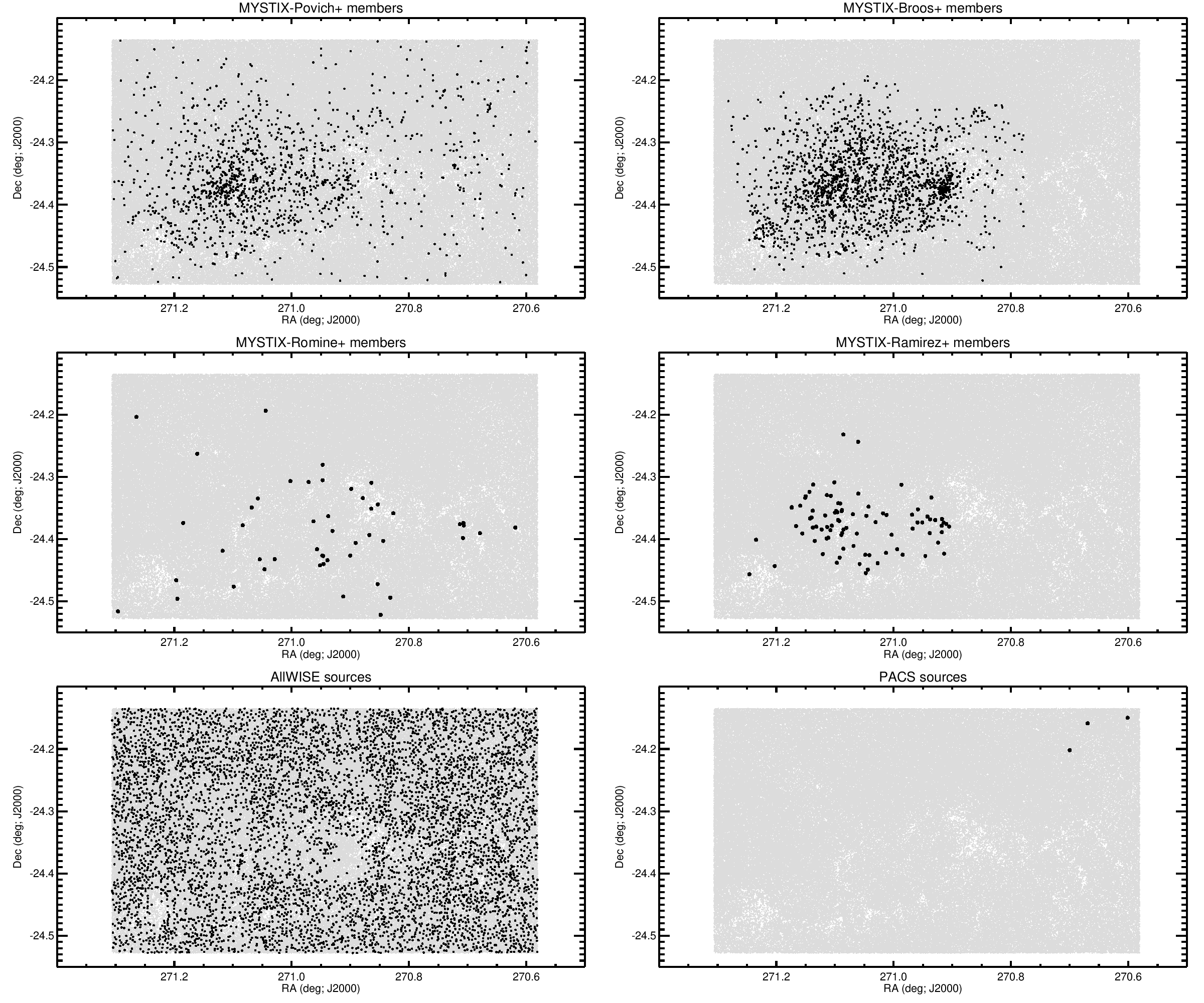}
\caption{In all cases, the indicated catalog is plotted in black and the entire catalog is shown in
light grey.
Top left: MYStIX: \citet{2013ApJS..209...31P} members. 
This is the first figure in this series where the background catalog is clearly evident.
Top right: MYStIX: \citet{2013ApJS..209...32B} members.
Middle left: MYStIX: \citet{2016ApJ...833..193R} members.
Middle right: MYStIX: \citet{2020AA...633A.155R} members.
Bottom left: WISE/AllWISE \citep{wisepaper,allwisedata} sources. 
Bottom right: Herschel/HPDP PSC 70 and 160 \citep{martonexplansupp,pacspsc70,pacspsc160} 
sources. None of the Herschel/PACS sources are identified here as YSOs or candidates.
}
\label{fig:wherepanel2}
\end{figure}

\textbf{ Coverage and numbers.} MYStIX covers our entire survey
region, but in a complicated fashion. The Spitzer data they used cover
our entire region, but the Chandra data they used do not, nor do some
of the other ancillary data they incorporated in their later work. 
The MYStIX catalogs (particularly those from the latter papers) are
often just  the YSOs, and as such the histograms are not necessarily
as easy to characterize as those for entire surveys. Therefore, we
characterize them all together; a histogram of all the MYStIX \ks\
mags used here peaks at $\sim$13 mag, and that for [3.6] peaks at
$\sim$12 mag.  \citet{2013ApJS..209...31P} reports 1,327 sources
(plotted in Fig.~\ref{fig:wherepanel2}), 1,235 of which end up on our
YSO list after all of the analysis below;  \citet{2013ApJS..209...32B}
lists 1,914 YSO candidates (plotted in Fig.~\ref{fig:wherepanel2});
\citet{2016ApJ...833..193R} describes 48 YSOs (plotted in
Fig.~\ref{fig:wherepanel2}); and \citet{2020AA...633A.155R} works with
86 YSOs (plotted in Fig.~\ref{fig:wherepanel2}). The difference
between the Spitzer and Chandra coverage can be  seen in comparing the
\citet{2013ApJS..209...31P}  and  \citet{2013ApJS..209...32B}  plots,
where the latter requires Chandra. Comparing the 1,327 sources from 
\citet{2013ApJS..209...31P} and the  1,914 sources from
\citet{2013ApJS..209...32B}, there are 962 that appear in both papers.
The 48 sources from \citet{2016ApJ...833..193R} include 14 that also
appear in \citet{2013ApJS..209...31P} and 36 that also appear in 
\citet{2013ApJS..209...32B}. Finally, the 86 from
\citet{2020AA...633A.155R} are entirely encompassed by
\citet{2013ApJS..209...32B}; 59 also appear in
\citet{2013ApJS..209...31P}.  All of the sources reported by these
papers are retained in our master catalog.

\subsection{Lower-Resolution Catalogs}

\subsubsection{WISE/AllWISE}

The Widefield Infrared Survey Explorer \citep[WISE;][]{wisepaper,allwisedata},
like 2MASS, is an all-sky survey; like Spitzer, it is tied to
the 2MASS coordinate system, so it provides a good anchor to which we
can tie other catalogs. WISE data were obtained in 3.4, 4.6, 12, and
22 \mum. We primarily used the AllWISE release \citep{allwisedata}, 
which sums up all available data prior to 2011 February.  In general,
this catalog reaches much fainter objects in 3.4 and 4.6 \mum\
than in 12 and 22 \mum; the nominal completeness limits are 
17.1, 15.7, 11.5, and 7.7 magnitudes for the four WISE bands, respectively. 
The spatial resolution is low compared to optical or even Spitzer 
-- $\sim$6.1, 6.4, 6.5, and 12$\arcsec$ for the four channels,
respectively. (Spitzer is $\sim1.6\arcsec-2\arcsec$ between 3.6 and 8 \mum, 
and $\sim6\arcsec$ at 24 \mum; see Table~\ref{tab:literature2} for 
spatial resolutions of other studies incorporated here.)

We have several long-wavelength catalogs that are lower spatial 
resolution and less sensitive. As a whole, these catalogs 
should have counterparts, if they exist at all, within the AllWISE
catalog. That is, we can use the AllWISE catalog to correct the
astrometry of the smaller, lower-resolution, longer-wavelength catalogs.

\textbf{ Coverage and numbers.} WISE/AllWISE covers our entire survey
region,  with 6,006 sources. They are plotted in
Fig.~\ref{fig:wherepanel2}.  As mentioned above, [W1] spatial
resolution  is $\sim$6$\arcsec$; a histogram of [3.4] in this
complicated region peaks at $\sim$11 mag. YSO candidates are not
identified in the AllWISE catalog itself.  After all of the analysis
described below, 284 are identified in the  literature as YSOs or
candidates. All of the sources from this catalog are  retained in the
master catalog.

\subsubsection{Herschel}

The Herschel Space Observatory \citep{herschelpaper} observed
portions of this region with both the Photodetector Array 
Camera and Spectrometer 
\citep[PACS, 70-160 \mum;][]{2010AA...518L...2P,martonexplansupp}
and the Spectral and Photometric Imaging Receiver 
\citep[SPIRE, 250-500 \mum;][]{2010AA...518L...3G}.  
The Herschel High Level Images 
(HHLI\footnote{https://irsa.ipac.caltech.edu/data/Herschel/HHLI/overview.html};
\citealt{hhli}) provide reduced images, and the PACS and SPIRE Point
Source Catalogs \citep{martonexplansupp} provide catalogs. However,
the region in which we are working here is largely off the edge of the
Herschel data. Only a few extracted PACS sources overlap the extreme
northwestern edge of our region, none of which correspond to literature
YSOs. We did not retain in our catalog Herschel sources that had no
AllWISE counterpart, on the assumption that those sources were
unlikely to have counterparts of interest in any of the other catalogs
under consideration.

\textbf{ Coverage and numbers.} Herschel/PACS covers only a tiny portion of 
our survey region. The only sources that were retained from this catalog 
are the three sources that had counterparts in AllWISE. They are 
plotted in Fig.~\ref{fig:wherepanel2}. 
YSO candidates are not identified in the Herschel/PACS catalog itself. 
None are identified in the literature considered here as YSOs or candidates.

\subsubsection{AKARI}

AKARI \citep{akaripaper,akariirc,akarifis} was an all-sky Japanese
mission, but was not as sensitive as Spitzer or WISE.  We included
data from the Infrared Camera (IRC) at 9 and 18 \mum\ and the
Far-Infrared Surveyor (FIS) at 65, 90, 140, and 160 \mum. The published
spatial resolution of IRC is $\sim$5$\arcsec$ \citep{akaripaper}; that for
FIS is 1.5$\arcsec$ to 1$\arcmin$ \citep{2015PASJ...67...50D}. 
AKARI IRC
data required a 3$\arcsec$ matching radius to find counterparts, and
the FIS data required a much larger 15$\arcsec$ matching radius. The 
reason for this probably has its origin in the fact that there are
fewer point sources detected in IRC and fewer still in FIS, so there
are fewer sources per unit area that can be tied to, e.g., astrometry 
from WISE or 2MASS.  As a result, there are larger position uncertainties in 
any given AKARI source position. However, any legitimate AKARI source 
is extremely likely to also have been detected in WISE; we have
found empirically (not just here but in other clusters as well) 
that extending the search radius to find the 
nearest WISE source is often the correct choice, and that is the origin
of the matching radii we used.  AKARI data are  
limited to the brighter sources -- nominally
at 9 \mum, all sources brighter than $\sim$0.2 Jy are detected and 
at 65 \mum, all sources brighter than $\sim$5 Jy are detected.

Fewer than 30 IRC sources correspond to literature YSOs discussed
here,  and just one FIS source has a literature YSO counterpart. We
did not retain in our catalog AKARI sources that had no AllWISE
counterpart, on the assumption that, given the other data we have
merged to this point, those sources were somewhat less likely to be
legitimate sources.

\begin{figure}[htb!]
\plotone{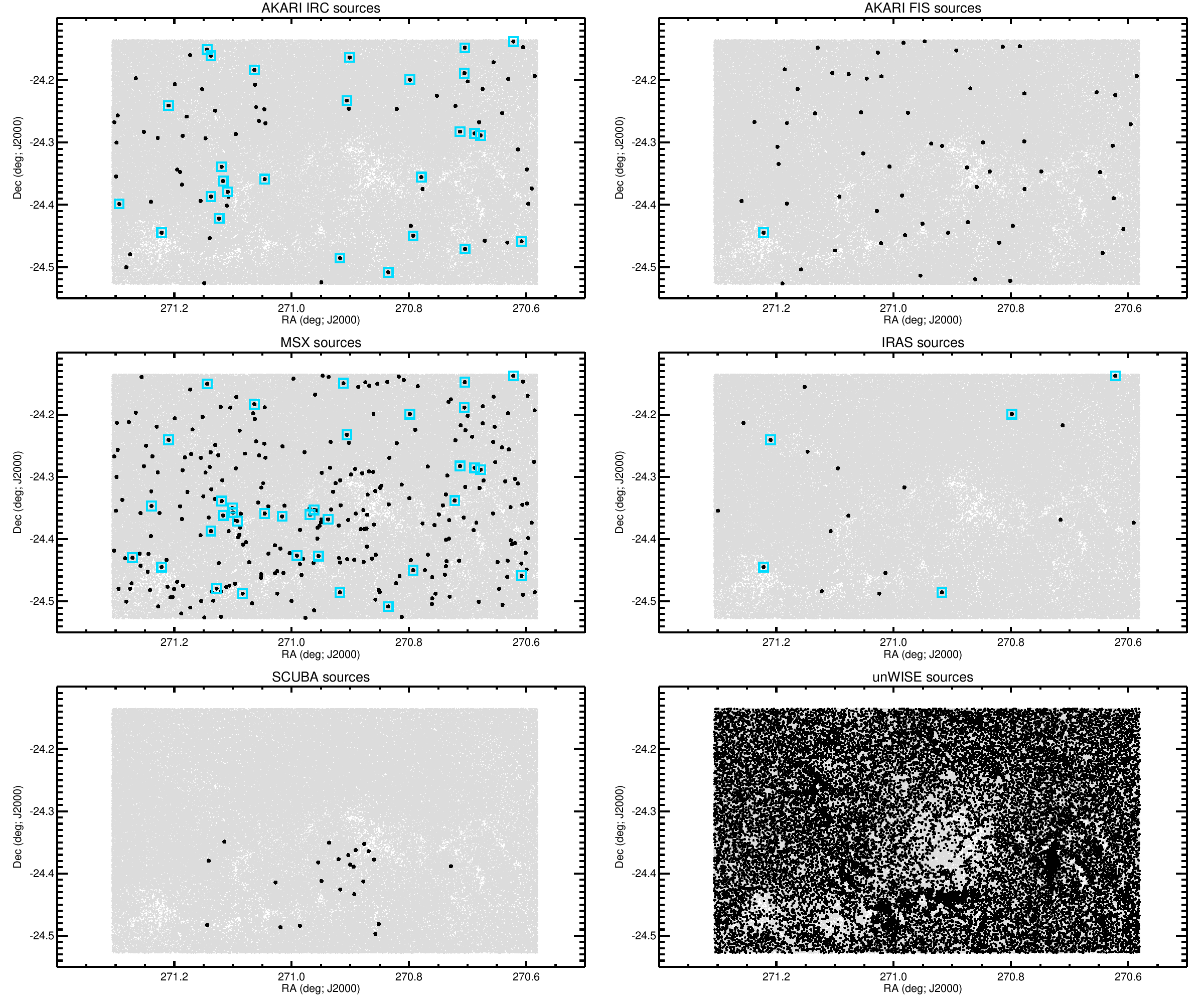}
\caption{In all cases, the indicated catalog is plotted in black and
the entire catalog is shown in light grey. If there are additional
cyan squares, these denote stars that are identified as YSOs or
candidate YSOs later in the process.
Top left: AKARI IRC \citep{akaripaper,akariirc} sources.
Top right: AKARI FIS \citep{akaripaper,akarifis} sources. 
Middle left: MSX \citep{Eganmsx,msxdata} sources. 
Middle right: IRAS PSC \citep{iraspaper1984,irasdata2019} sources. 
Bottom left: SCUBA \citep{2008ApJS..175..277D,2009yCat..21750277D} sources. 
None of the SCUBA sources are apparently counterparts of literature YSOs considered here.
Bottom right: WISE/unWISE \citep{2014AJ....147..108L,2017AJ....153...38M,2017AJ....154..161M,2019ApJS..240...30S} 
sources. Pile-ups of sources at edges of nebulosity are apparent. } 
\label{fig:wherepanel3}
\end{figure}

\textbf{ Coverage and numbers.} AKARI covers our entire survey region
with both instruments.   The only sources that were retained from this
catalog are the sources that had counterparts in AllWISE, which are 
79 IRC sources, and 61 FIS sources.  They are plotted in 
Fig.~\ref{fig:wherepanel3}.  YSO candidates are not identified in
either catalog.  After all of the analysis described below, 27 IRC
sources and 1 FIS source are identified in the literature as YSOs or
candidates.   AKARI sources in our region are bright -- the faintest
AKARI 9 \mum\  source is $\sim$0.1 Jy and the faintest  AKARI 65 \mum\
source is $\sim$0.75 Jy.

\subsubsection{MSX}

The Midcourse Space Experiment \citep[MSX;][]{Eganmsx} was an Air
Force mission that surveyed the Galactic Plane, including our region
in the Lagoon nebula, though it's not very sensitive (completeness is
reached at about $\sim$0.1 Jy) and relatively low spatial resolution
($\sim$20$\arcsec$).  They published catalogs at 4.29, 4.35, 8.28,
12.13, 14.65, and 21.3 \mum\ \citep{msxdata} but only published images
between 8 and 21 \mum.  There are nominally $\sim$300 sources in
our region. Using a 10$\arcsec$ search radius to match sources by
position, just 35 match to literature YSO candidates considered here. In
general, we did not retain MSX-only sources, but in the brightest
heart of the region (the Hourglass Nebula), it may be that MSX is the
highest spatial resolution instrument at these wavelengths that is not
saturated, so sources within 3$\arcmin$ of  18:03:37 -24:23:12 were
retained even if they were MSX-only. 

\textbf{ Coverage and numbers.} MSX covers our entire survey region. 
Sources with counterparts were retained, except in the Hourglass, 
where sources without counterparts were retained. MSX sources (a 
total of 292) are plotted in  Fig.~\ref{fig:wherepanel3}.  YSO candidates
are not identified in this catalog.  After all of the analysis
described below, 35 MSX sources are identified in the literature as
YSOs or candidates. MSX sources in our region are bright -- the
faintest MSX band A (4.29 \mum) source is $\sim$0.07 Jy.

\subsubsection{IRAS}

The Infrared Astronomical Satellite
\citep[IRAS;][]{iraspaper1984,irasdata2019}  was the first space-based
all-sky infrared survey, conducted at 12, 25, 60, and 100 \mum, with
spatial resolution of $\sim$1-3$\arcmin$, far worse than most other
data sets included here. We used 15$\arcsec$ for positional matching. 
The completeness limits for the Point Source Catalog (PSC) are 0.4-1 Jy. 
We adopted uncertainties of 20\% of the flux in all four IRAS bands.

\textbf{ Coverage and numbers.} IRAS covers our entire survey region. 
Nineteen sources with counterparts were retained, and are plotted in 
Fig.~\ref{fig:wherepanel3}. The faintest IRAS-1 flux in our region is
$\sim$1.2 Jy. YSO candidates are not identified in this catalog. 
After all of the analysis described below, just 5 IRAS sources are
identified in the literature as YSOs or candidates.

\subsubsection{SCUBA}

The Submillimetre Common User Bolometer Array (SCUBA) Legacy
Catalogues \citep{2008ApJS..175..277D,2009yCat..21750277D} provide
catalogs of point sources at 450 and 850 \mum. There are SCUBA data
covering the heart of the region, from the center of NGC 6523 to M8E,
plus three smaller pointings in the area.  Given the 33 sources
overlapping our region, we looked for counterparts within 15$\arcsec$.
Again, given the bright heart of the Hourglass Nebula, sources within
3 arcminutes of 18:03:37 -24:23:12 were retained even if they were
SCUBA-only. 

\textbf{ Coverage and numbers.} SCUBA data do not cover our entire 
survey region. All 33 sources were retained, and are
plotted in  Fig.~\ref{fig:wherepanel3}.  YSO candidates are not
identified in this catalog.  After all of the analysis described
below, none of the SCUBA sources are apparently counterparts of
literature YSOs considered here.

\subsubsection{WISE/unWISE and CatWISE}

Finally, after the other lower-resolution data, we explored the 
remaining WISE data.  The WISE satellite continued to take data
between 2013 Dec and 2024 Aug, with the mission renamed NEOWISE. 
However, the project itself was not funded to create ensemble data
products. CatWISE 
\citep{2021ApJS..253....8M,2020ApJS..247...69E,catwise} and
unWISE \citep{2014AJ....147..108L,2017AJ....153...38M,2017AJ....154..161M,2019ApJS..240...30S,unwisecat} 
were both efforts to include these more recent data; both used images
created by unWISE but obtain independent photometry. We included
catalogs from both CatWISE and unWISE, matching by position with a
1$\arcsec$ radius; we retained all sources detected only by CatWISE
and/or unWISE. The reason we didn't merge these right after the
AllWISE merge above is that we found empirically that all of the 
intervening catalogs were more likely to find a correct counterpart 
within AllWISE, and any fainter sources found in CatWISE and/or unWISE
are more likely to be decoys to the true match.  The spatial resolution
is still $\sim$6$\arcsec$; both claim sensitivities at [3.4] down to 
slightly fainter than 17th magnitude.

\begin{figure}[htb!]
\plotone{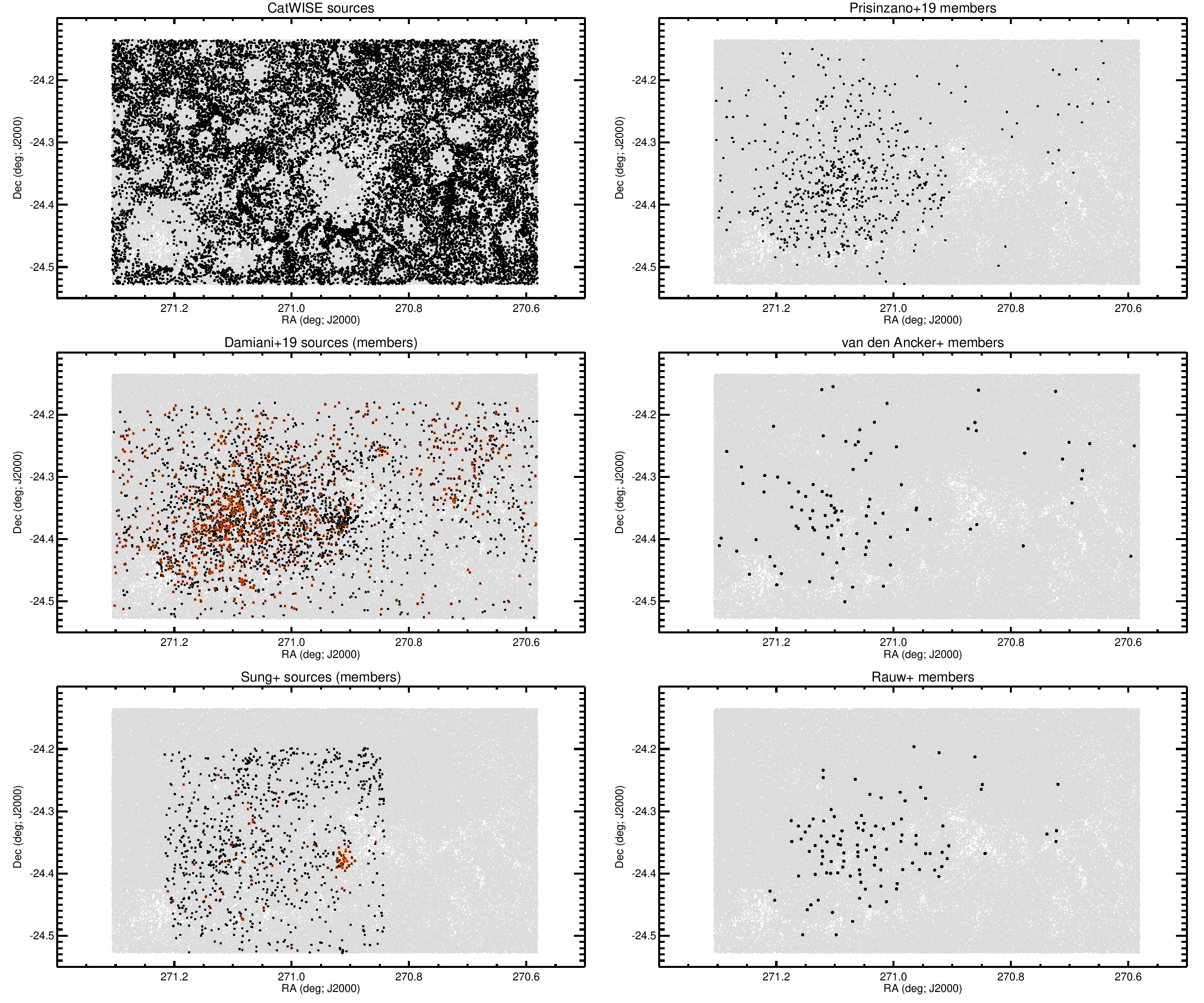}
\caption{In all cases, the indicated catalog is plotted in black and the entire catalog is shown in
light grey. In two cases here, the cited article provided a list of members distinct 
from a list of sources; those members are indicated in red.
Top left: CatWISE \citep{2021ApJS..253....8M,2020ApJS..247...69E,catwise} sources.  Remnants of source
masking are readily apparent.
Top right: \citet{2019AA...623A.159P} members. 
Middle left: \citet{2019AA...623A..25D} sources and members. 
Middle right: \citet{1997AAS..123...63V} members.
Bottom left: \citet{2000AJ....120..333S} sources and members.
Bottom right: \citet{2002AA...395..499R} members.}
\label{fig:wherepanel4}
\end{figure}

\textbf{ Coverage and numbers.} NEOWISE data cover our entire region. The
$\sim$25,000 sources from unWISE are plotted in
Fig.~\ref{fig:wherepanel3},  and the $\sim$17,000 sources from CatWISE
are plotted in Fig.~\ref{fig:wherepanel4}. Pile-ups of sources at
edges of nebulosity are apparent in unWISE, and remnants of source
masking are readily apparent in CatWISE.  Histograms of [3.4] from both 
these catalogs in this region are broad but peak at about 12.5 mag, and 
are clearly rolling over by 13. CatWISE has a longer fainter tail.
Neither of these catalogs
identified YSOs. Downstream from this point, 628
stars from unWISE and 447 stars from CatWISE were identified as YSOs
or candidates.

\subsection{Smaller Literature Catalogs}

The next catalogs we ingested were published in various articles
focused on the Lagoon region, often on NGC~6530.  Most of these
catalogs identify YSOs and publish tables of just the YSOs (or
candidates), as opposed to large catalogs of everything detected in
the region. For the most part, we retain the identification of YSOs
(or candidates), but when new photometry (or other relevant 
information) is available, we retain that as well. In some cases, this
means that the only thing that is retained from a given paper is the
fact that the star was listed as a YSO.  In all cases (except where
specified below), we retained the cross-ids as well as sources that
appeared solely in that catalog (e.g., without matches to other
catalogs).

We first work with two large, relatively recent catalogs:
\citet{2019AA...623A.159P}  and \citet{2019AA...623A..25D}.  Having
established a reasonably reliable catalog with these fairly large
catalogs, we then backtracked to older catalogs to attempt to merge
those. In some cases, we used the same technique as established in,
e.g., \citet{1333catalogpaper} where we take a relatively old
catalog, compare the sources to 2MASS to get improved coordinates, and
then use the improved 2MASS coordinates for those sources to merge
into the master catalog. For other cases, we empirically determined
the best radius for source matching -- for each object in a new
catalog, we calculate the distance to the nearest match in the master
catalog, and then make a histogram of these distances. Usually this
distribution is sharply peaked and the majority of the sources fall
$<1\arcsec$. (See individual discussions below.) 

Because these catalogs are small and they vary a lot in wavelength
coverage, in each one's ``coverage and numbers'' paragraph (and in
Table~\ref{tab:literature2}), the peak \ks\ mag of the sample is used
to characterize each sample for easy comparison to each other.

\subsubsection{\citet{2019AA...623A.159P}}

\citet{2019AA...623A.159P} explored the age spread in NGC~6530 using 
Gaia \cite[specifically Gaia DR2 data;][]{gaiadr2} and Gaia-ESO survey
\citep{2022A&A...666A.120G} data. This paper used several sources of
photometric data in the literature (all incorporated from their
origins above or below),  but the data we retained from their data
table included $V$, \ic, and \vsini. They only listed members in their
data table, so we also retained the mere fact that the stars were
listed in their paper as a member.  

\textbf{ Coverage and numbers.} \citet{2019AA...623A.159P} data cover our 
entire region but build upon data that are more concentrated on the East side;
see Fig.~\ref{fig:wherepanel4} for the locations of the 616 members. 
A histogram of \ks\ for stars from this paper peaks at $\sim$13.2 mag.

\subsubsection{\citet{2019AA...623A..25D}}

\citet{2019AA...623A..25D} also worked with NGC~6530, where their
scientific goals were to study its global properties. They used data
published by this same group (see below) and other data we
incorporated above or below, including those from MYStIX, VPHAS+ 
\citep[see Sec.~\ref{sec:vphas}
below;][]{2014MNRAS.440.2036D,2016yCat.2341....0D}, and other data.
While essentially all of the photometry they used should be absorbed
by us from their respective origins (see Table~\ref{tab:literature}),
the identification of cluster members is unique to this paper. We
retained cross-ids and photometry for later comparison, as well as all
of the flags they used for identifying possible members. 

Incorporating the information from this article was somewhat
challenging because we had to correct considerable astrometric confusion.
The baseline for their catalogs was the output of the VPHAS+ 
\citep{2014MNRAS.440.2036D,2016yCat.2341....0D} pipeline, which, at
least as of the time that it was run here, left residual sources in
the catalog. (For example, the spatial resolution of VPHAS+ should be
$\sim1\arcsec$, so all sources found within $\sim0.4\arcsec$ of each
other are likely duplicates.)  In the end, we accepted only the stars
identified by these authors as Gaia members as the best possible
members. We merged this catalog to \citet{2019AA...623A.159P} by
position with a 1$\arcsec$ radius.

\textbf{ Coverage and numbers.} \citet{2019AA...623A..25D} cover most
but not all of our region; see Fig.~\ref{fig:wherepanel4}. 
There are 2,901 sources incorporated into our master 
catalog, and 2,031 are retained as possible members. 
A histogram of \ks\ for stars from this paper peaks at $\sim$13.6 mag.

\subsubsection{\citet{1997AAS..123...63V}}

This is the first of the older, small catalogs that we merged in.
\citet{1997AAS..123...63V} presents $UBV$ photometry for
stars they thought were YSOs, as well as spectral types and
cross-identifications for those stars from \citet{1972AA....20..425V}, 
\citet{1978MNRAS.184..467S}, and \citet{1957ApJ...125..636W}.  
(The observations reported in \citet{1997AAS..123...63V} were 
probably all obtained at $\sim15-16\arcsec$ spatial resolution, 
though in one instance in the text, $\sim16\arcmin$ is mentioned.) 

Sources from \citet{1997AAS..123...63V} were retained with their 
improved 2MASS coordinates. Once the coordinates were updated to 
2MASS, we could match by position with a 1$\arcsec$ radius.

\textbf{ Coverage and numbers.} \citet{1997AAS..123...63V} covers most of 
our area. The 96 members appear in Fig.~\ref{fig:wherepanel4}. 
A histogram of \ks\ for counterparts to stars from this paper peaks at $\sim$11 mag.

\subsubsection{\citet{2000AJ....120..333S}}

\citet{2000AJ....120..333S} used H$\alpha$ and $UBVRI$ photometry to
find 37 YSOs in NGC~6530. Helpfully, they published their entire
catalog, not just the catalog of YSOs. We absorbed all of their
photometry as well as their indication of whether or not they thought
the star was a YSO. They report seeing was $\sim$2-3$\arcsec$. 
We matched by position using a 2$\arcsec$ radius,
which is an empirically-determined distance.  Most of the astrometry
did not require correction to find the appropriate
counterpart; a few sources needed some shifting by less than
2$\arcsec$ to ensure that the correct counterpart was matched. 

\textbf{ Coverage and numbers.} \citet{2000AJ....120..333S} did not cover 
our whole region; see Fig.~\ref{fig:wherepanel4} which shows the 900 sources
from that catalog as well as the 446 members we carried forward. 
A histogram of \ks\ for counterparts to stars from this paper peaks 
at $\sim$12.0 mag.

\subsubsection{\citet{2002AA...395..499R}}

\citet{2002AA...395..499R} is the only X-ray study we incorporate that
does not use data from Chandra;  it analyzes an  X-ray Multi-Mirror
Mission \citep[XMM-Newton;][]{2001A&A...365L...1J} observation of this
region.  They report on sources they detect that match uniquely to
optical sources from \citet{2000AJ....120..333S}, sources that match
ambiguously to \citet{2000AJ....120..333S} sources, and sources that
have no match in  \citeauthor{2000AJ....120..333S}. We took the
positions they reported, compared the positions to 2MASS sources, and
largely but not always adopted the 2MASS counterpart's position.
\citet{2002AA...395..499R} took all of their detections as likely
YSOs, so we followed suit and identified all of the
\citeauthor{2002AA...395..499R} sources as  implicit literature YSO
candidates. XMM's nominal resolution is $6\arcsec$.  We matched by
position using a 2$\arcsec$ radius, again an empirically determined
distance.

\textbf{ Coverage and numbers.} \citet{2002AA...395..499R} did not cover
our whole region. Fig.~\ref{fig:wherepanel4} shows all 105 sources.
A histogram of \ks\ for counterparts to stars from this paper peaks 
at $\sim$12.0 mag.

\subsubsection{\citet{2004ApJ...608..781D}}

\citet{2004ApJ...608..781D} used Chandra to study NGC~6530. They
presented matches to 2MASS counterparts in most cases. We absorbed
their catalog of X-ray detections, taking the fact that they detected
it as an indicator of likely youth. If they indicated that they found
a 2MASS counterpart, we took that as the match. If they listed more
than one 2MASS counterpart, we took the closest one. If they did not
list a counterpart, we took the position they listed for the source.
We merged these sources with a 2$\arcsec$ radius.

\begin{figure}[htb!]
\plotone{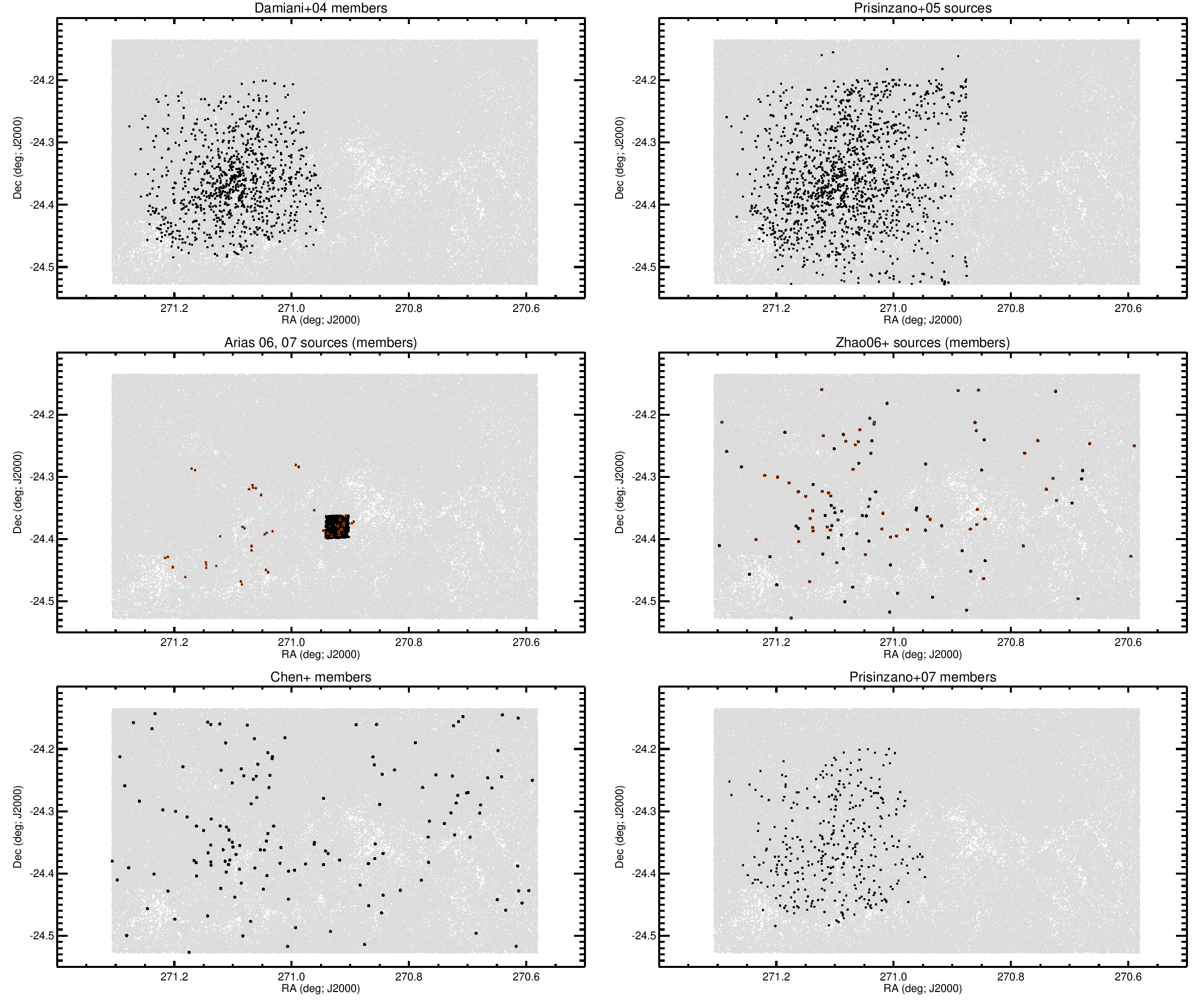}
\caption{In all cases, the indicated catalog is plotted in black and the entire catalog is shown in
light grey. In two cases here, the cited article provided a list of members distinct from a list of sources;
the members are indicated in red.
Top left: \citet{2004ApJ...608..781D} members.
Top right: Surviving \citet{2005AA...430..941P} members (black) on the entire catalog (light grey.
Middle left: \citet{2006MNRAS.366..739A} and \citet{2007MNRAS.374.1253A} sources 
 and members. 
Middle right: \citet{2006ChJAA...6..435Z} sources and members.
Bottom left: \citet{2007AJ....134.1368C} members.
Bottom right: \citet{2007AA...462..123P} members. }
\label{fig:wherepanel5}
\end{figure}

\textbf{ Coverage and numbers.} \citet{2004ApJ...608..781D} did not cover
our whole region; all 884 sources are shown in Fig.~\ref{fig:wherepanel5}.
A histogram of \ks\ for counterparts to stars from this paper peaks at $\sim$13.2 mag.

\subsubsection{\citet{2005AA...430..941P}}

\citet{2005AA...430..941P} presented ground-based $BV$\ic\ photometry
of NGC~6530, and used these data in conjunction with the X-ray data
from \citet{2004ApJ...608..781D} and 2MASS data to assess the age,
distance, and initial mass function.  We retained $BV$\ic, the
cross-identifications with \citet{2000AJ....120..333S}, 
\citet{1957ApJ...125..636W}, and \citet{2004ApJ...608..781D}. The best
seeing in the paper is described as $\sim0.8\arcsec$, but there are
several sources that are separated by $\leq0.5\arcsec$. We
consolidated these sources by hand where necessary. Aside from that,
we took the counterparts they list for \citet{2000AJ....120..333S}
and \citet{2004ApJ...608..781D}.  The objects in this paper are
described as being at least 90\% members.  We found that $\sim$70\%
have counterparts that are also identified as YSOs or YSO candidates
by at least one other study here, but the remaining $\sim$30\%, which
is $\sim$400 sources, are identified as YSO candidates solely by
\citet{2005AA...430..941P}. Based on that, and the fact that 
several subsequent papers written by many people within this 
same collaboration have reported more refined lists of YSO 
candidates, we retained the cross-ids from this paper but did
not carry along the YSO identification for any of the stars identified
as YSO candidates solely by this paper.

\textbf{ Coverage and numbers.} Data from this paper did not cover our
whole region. There were 1329 sources in the catalog from
this paper originally, but 926 are carried forward as unique, real 
members, and those are what is shown in Fig.~\ref{fig:wherepanel5}.
A histogram of \ks\ for counterparts to final YSOs or candidates 
from this paper peaks at $\sim$13.2 mag.

\subsubsection{Arias \etal}

\citet{2006MNRAS.366..739A} studied the Hourglass Nebula in the heart
of M8 using ground- and space-based NIR data.
\citet{2007MNRAS.374.1253A} continued work in this region, conducting
follow-up spectroscopy of several candidate YSOs. We retained the
$JH$\ks\ photometry from the first paper, but much of the photometry
reported in the second paper is archival data that we have absorbed (or will
below) from the original catalogs. Most of the stars in the 2006
paper may not be YSOs, but we took all of the stars in the 2007 paper
to be literature YSOs. We also retained the spectral types from the
2007 paper. The ground-based data from 2006 had seeing $<1\arcsec$. 

\textbf{ Coverage and numbers.} The data from these two papers covered 
only a very tiny portion of our field, with the latter extending over a much
larger region than the former; see Fig.~\ref{fig:wherepanel5}.
\citet{2006MNRAS.366..739A} reports on 744 targets, 127 of which are likely
members. \citet{2007MNRAS.374.1253A} has just 44 members. 
A histogram of \ks\ for counterparts to stars from \citet{2006MNRAS.366..739A} 
peaks at $\sim$15.7 mag. For counterparts to stars from \citet{2007MNRAS.374.1253A},
the histogram peaks at $\sim$13.

\subsubsection{\citet{2006ChJAA...6..435Z}}

\citet{2006ChJAA...6..435Z} identified members from proper motions
with a time baseline of 87 years. They used photographic plates with a baseline
of nearly 100 years. We retained the cross-ids and the
member probability, taking as members those with $>$90\% probability.
Neither spatial resolution nor seeing is mentioned in the article, but given
that the mean positional error is considerably less than an arcsecond, 
one assumes that the seeing was at least comparable to an arcsecond.
We used a matching radius of 1$\arcsec$.

\textbf{ Coverage and numbers.} Data from \citet{2006ChJAA...6..435Z} cover 
most of our region. There are 112 targets in their catalog, 59 of which 
they report as having a membership probability of $>$90\%, and these
are the ones highlighted in Fig.~\ref{fig:wherepanel5}. However, all 112
are ultimately identified as YSOs in one way or another. 
A histogram of \ks\ for counterparts to stars from this paper
is very broad but peaks at $\sim$10.5 mag.

\subsubsection{\citet{2007AJ....134.1368C}}

\citet{2007AJ....134.1368C} studied mass segregation in NGC~6530 and 
one other cluster using photographic plates. We retained the $B$ magnitude 
for everything 
reported, plus the identification of objects that were tagged as
likely members.  The seeing for these data was not reported in the
article. We used an empirically-determined matching radius of
2$\arcsec$.

\textbf{ Coverage and numbers.} This catalog covers most of our region. All
147 stars are shown in Fig.~\ref{fig:wherepanel5}.
A histogram of \ks\ for counterparts to stars from this paper
peaks at $\sim$10.7 mag.

\subsubsection{\citet{2007AA...462..123P}}

\citet{2007AA...462..123P} followed up some of the work from their
2005 paper \citep{2005AA...430..941P} with spectroscopy to identify
members. The information we absorbed included the photometry,
\vsini, whether or not they identified the star as having a disk, and
whether or not they identified the star as being a member.  We used an
empirically-determined matching radius of 2$\arcsec$; the overwhelming
majority of sources could be matched within 1$\arcsec$, but 3\% needed
the larger matching radius. 

\textbf{ Coverage and numbers.} Because it builds on \citet{2005AA...430..941P}, 
these data did not cover our whole region; see Fig.~\ref{fig:wherepanel5},
where 332 stars are shown.
A histogram of \ks\ for counterparts to stars from this paper
peaks at $\sim$12.4 mag.

\subsubsection{\citet{2010MNRAS.407.1170K}}

\citet{2010MNRAS.407.1170K} studied M8 with Spitzer/IRAC,
incorporating $JH$\ks\ data from 2MASS and
\citet{2006MNRAS.366..739A,2007MNRAS.374.1253A}. They published
photometry for the YSO candidates they identified using the 
\citet{2009ApJS..184...18G,2010ApJS..189..352G}  approach. We matched
by position with a 1$\arcsec$ radius and retained the IRAC photometry
as well as the fact that they are listed as YSOs.
\citet{2010MNRAS.407.1170K} do not report photometric uncertainties;
where our best available IRAC value (see Sec.\ \ref{sec:bestphot}) 
comes from them, we assigned an uncertainty of 0.3 mag.

\begin{figure}[htb!]
\plotone{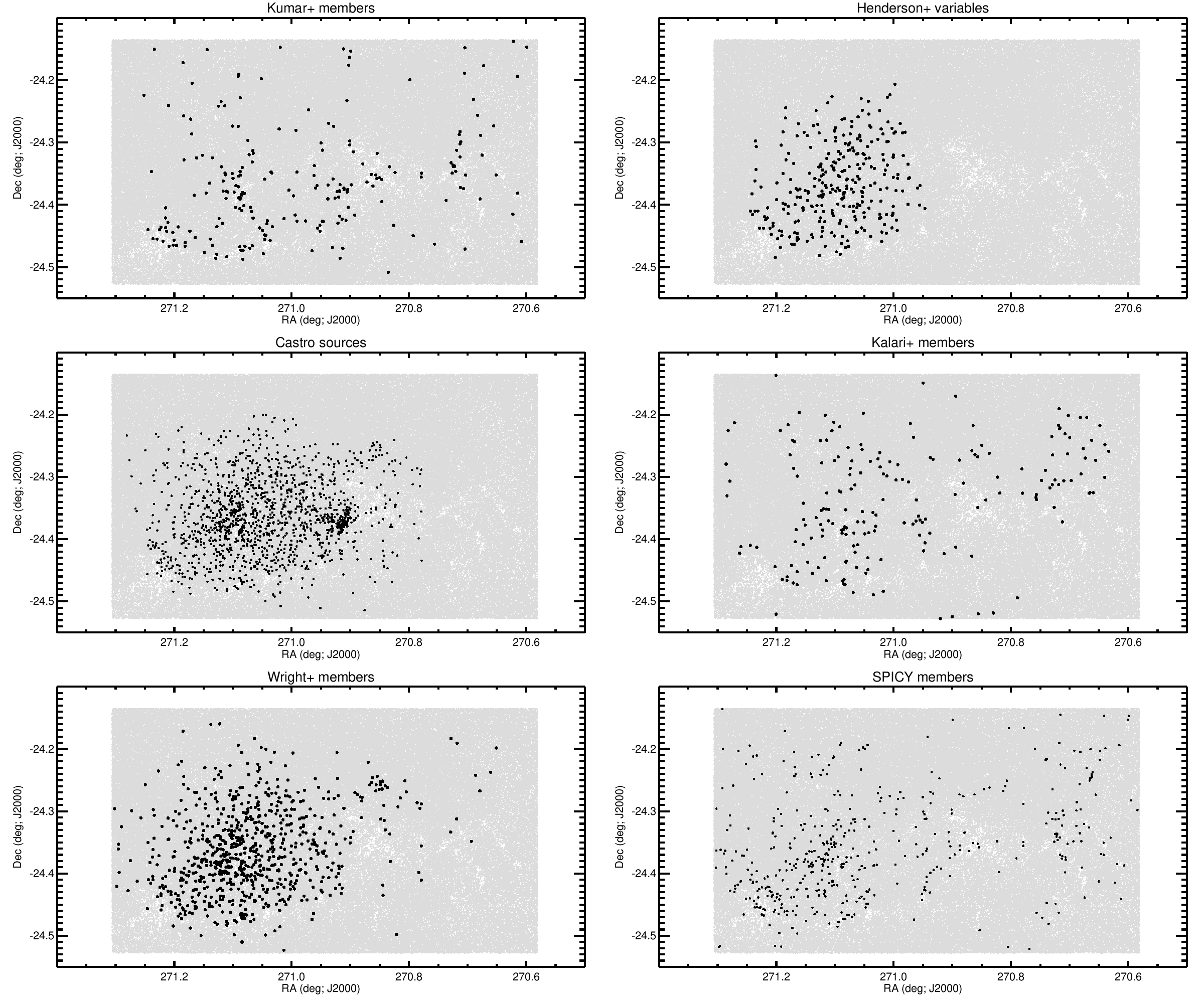}
\caption{In all cases, the indicated catalog is plotted in black and the entire catalog is shown in
light grey. 
Top left: \citet{2010MNRAS.407.1170K} members. 
Top right: \citet{2012ApJ...747...51H} variables.
Middle left: \citet{2013PhDT........36C} sources.
Middle right: \citet{2015MNRAS.453.1026K} members.
Bottom left: \citet{2019MNRAS.486.2477W} members.
Bottom right: SPICY: \citet{2021ApJS..254...33K,2021ipac.data.I556K} members.}
\label{fig:wherepanel6}
\end{figure}

\textbf{ Coverage and numbers.} The 221 targets reported by \citet{2010MNRAS.407.1170K}
are distributed across most of our area; see Fig.~\ref{fig:wherepanel6}.
A histogram of \ks\ for stars from this paper peaks at $\sim$11.6 mag.

\subsubsection{\citet{2012ApJ...747...51H}}

\citet{2012ApJ...747...51H} presented rotation periods for 290 stars
in the Lagoon Nebula, to which they add literature data, all of which
has been incorporated above. The spatial resolution of their data is not
described in their paper, but the seeing at Cerro Tololo is 
often $\sim1-1.5\arcsec$. We used an empirically determined
matching radius of 2$\arcsec$; we retained the reported period and
tagged all of these reported variable stars as implicit candidate YSOs.

\textbf{ Coverage and numbers.} \citet{2012ApJ...747...51H} focused their
attention on a subset of our field. The 290 targets for which they
reported periods appear in Fig.~\ref{fig:wherepanel6}.
A histogram of \ks\ for stars from this paper peaks at $\sim$13.2 mag.

\subsubsection{\citet{2013PhDT........36C}}

\citet{2013PhDT........36C} worked with some Chandra observations of
the Lagoon Nebula that were not included in the works by Damiani,
Prisinzano, \etal\ above, but were included in the MYStIX analysis. A
copy of the relevant table from his thesis was obtained via private
communication in 2015 with the intention of being incorporated into
our catalog. 
Sources were not (could not be) matched to counterparts
in as sophisticated a fashion as was done within the MYStIX program. 
There are 1482 sources from the \citet{2013PhDT........36C} catalog. 
Treating the \citet{2013PhDT........36C} catalog just as all the other
independent catalogs here and merging it (matching by position within
2$\arcsec$) in to the master catalog,  84\% have counterparts within
MYStIX. Out of those that don't match MYStIX, 64 sources in the final
catalog have positions that don't match any other source from any
other catalog used here; given the wealth of other data we have, it is
unusual that these sources would both be legitimate additional sources
and have no other counterparts. There are also 91 sources that are
tagged as possible YSOs based solely on their presence in the Castro
catalog (all the remaining Castro sources were also identified as YSOs
by at least one other catalog here). Based on the entirety of this
assessment, we left the Castro catalog merged in for cross-ids, but
did not add in to the master catalog those sources found only in 
\citet{2013PhDT........36C}, and did not tag sources found only in
Castro as YSO candidates. (This is the same treatment as for
\citealt{2005AA...430..941P} above.)

\textbf{ Coverage and numbers.} The data from \citet{2013PhDT........36C}
do not cover our whole region; see Fig.~\ref{fig:wherepanel6}. All 
1,350 sources are shown in the Figure, though only 1,303 end up
as YSO candidates after the analysis here. 
A histogram of \ks\ for counterparts to stars from this paper
peaks at $\sim$13.5 mag.

\subsubsection{\citet{2015MNRAS.453.1026K}}

\citet{2015MNRAS.453.1026K} used VPHAS+ (see Sec.~\ref{sec:vphas})
data to estimate accretion rates for 235 YSOs in the Lagoon; they used
both H$\alpha$ and $u$-band and selected YSO candidates. We matched by
position with a search radius of 2$\arcsec$, and retained their
cross-id (with it their implicit YSO tag) and their estimate of
accretion rate.

\textbf{ Coverage and numbers.} Data from \citet{2015MNRAS.453.1026K} 
cover most of our region; see 
Fig.~\ref{fig:wherepanel6}. All 195 targets are included as YSOs. 
A histogram of \ks\ for counterparts to stars from this paper
is very broad but peaks at $\sim$12.7 mag.

\subsubsection{\citet{2019MNRAS.486.2477W}}

\citet{2019MNRAS.486.2477W} worked with the Gaia-ESO data set to
explore the 3D structure of NGC 6530. All of the ancillary
(non-Gaia-ESO) data they worked with come from several different 
resources we have already incorporated above or below, 
so the only thing we retained is the membership
flag (their table in \citealt{2019yCat..74862477W}). 

\textbf{ Coverage and numbers.} Data from \citet{2019MNRAS.486.2477W} 
basically cover our region, but are biased towards the east; see 
Fig.~\ref{fig:wherepanel6}. All 776 targets are included as YSOs. 
A histogram of \ks\ for counterparts to stars from this paper
is broad and flat but is turning over by $\sim$13.6 mag.

\subsubsection{SPICY: \citet{2021ApJS..254...33K}}

\citet{2021ApJS..254...33K} worked with machine learning to identify
YSOs over the Galactic plane using data we have incorporated above
or below, and used MYStIX data as the training set. We pulled data
from this catalog, called SPICY \citep[Spitzer/IRAC Candidate
YSO;][]{2021ipac.data.I556K}, in our
region, matched to the master catalog within $1\arcsec$, and
identified them as YSO candidates in our catalog.

\textbf{ Coverage and numbers.} Sources from SPICY cover our region; see 
Fig.~\ref{fig:wherepanel6}. All 466 targets are included as YSOs. 
A histogram of \ks\ from these stars peaks at $\sim$13.0 mag.

\subsubsection{\citet{2022MNRAS.517.6191O}}

\citet{2022MNRAS.517.6191O} studied the variability of YSOs in this
region in \ks. They identified their YSOs based on a subset of the
literature we have also considered above. However, the region on the
sky that they considered is larger than the region we have considered,
so not all of their sources are in our catalog. We matched by position
within 1$\arcsec$ and retained  their mean \ks, rotation period, and
variability class. We also tagged their variable sources as implicit 
YSO candidates.

\begin{figure}[htb!]
\plotone{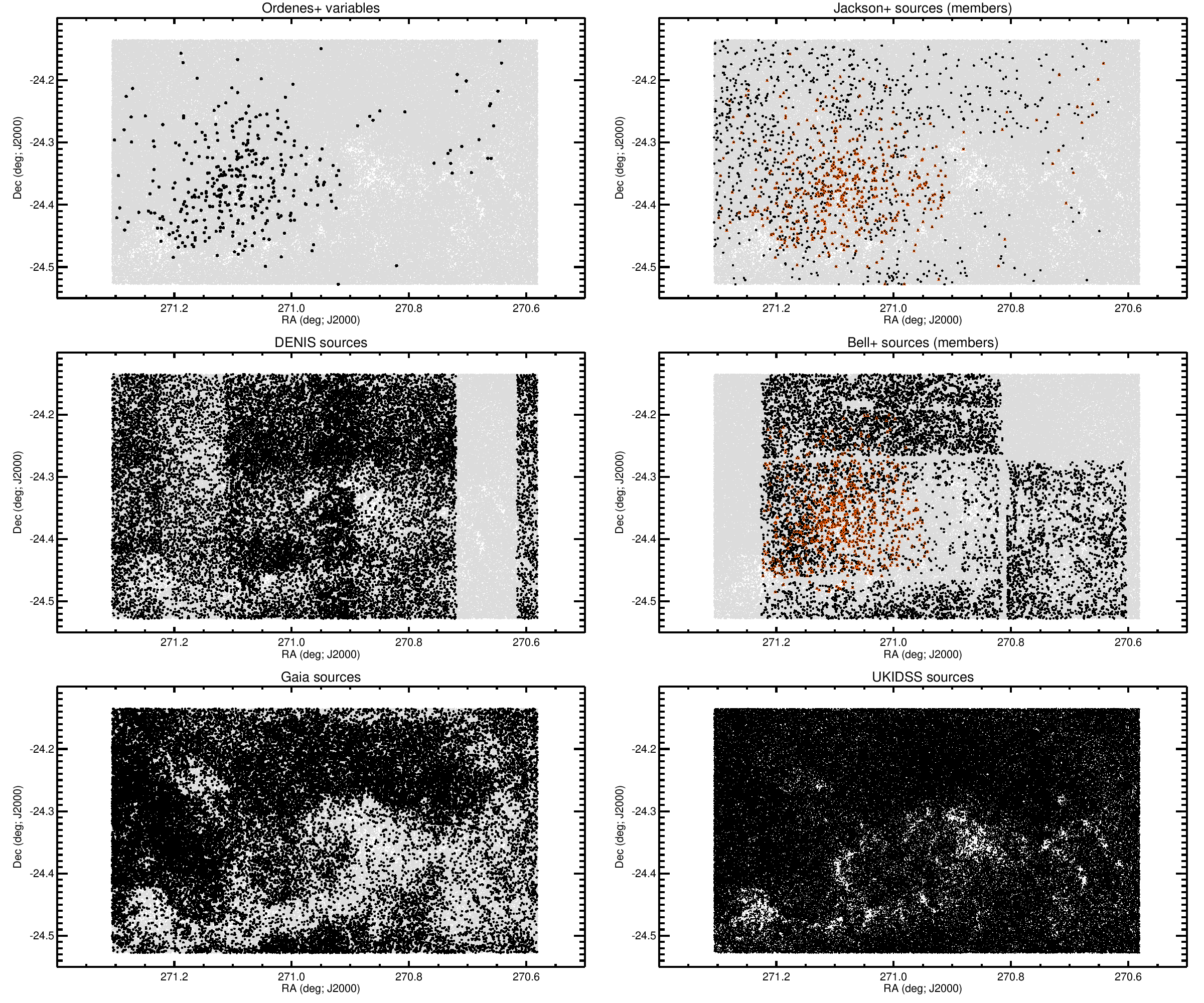}
\caption{In all cases, the indicated catalog is plotted in black and
the entire catalog is shown in light grey. In two cases here, the
cited article provided a list of members distinct from a list of
sources; the members are indicated in red.
Top left: \citet{2022MNRAS.517.6191O} variables. 
Top right: \citet{2022MNRAS.509.1664J} catalog and members.
Middle left: DENIS \citep{denispaper,denisdata} sources. 
Tile boundaries and margins are immediately apparent.
Middle right: \citet{2013MNRAS.434..806B} sources and members. 
Bottom left: Gaia DR3 \citep{gaiadr3,gaiadr3data} sources. 
The additional extinction from dust is immediately obvious.
Bottom right: UKIDSS \citep{ukidsspaper,ukidssdata} sources. }
\label{fig:wherepanel7}
\end{figure}

\textbf{ Coverage and numbers.} Data from \citet{2022MNRAS.517.6191O} 
basically cover our region, but are biased towards the east; see 
Fig.~\ref{fig:wherepanel7}. All 361 variables are shown.
A histogram of \ks\ for counterparts to stars from this paper
is at $\sim$13.2 mag.

\subsubsection{\citet{2022MNRAS.509.1664J}}
\label{sec:jackson}

\citet{2022MNRAS.509.1664J} explored membership of NGC~6530 (among many
other clusters) using Gaia-ESO spectra. We matched their catalog to
ours based on Gaia DR3 ID -- we did not match by position, so the actual
merging of this catalog took place after the Gaia merge below, but we
list this catalog here in the text because it is another small
catalog. We took as members those where their $P_{3D}\geq0.90$ (retaining 
that number in our catalog) and
identified those stars as YSO candidates in our catalog.
Stars with $P_{3D}=-1$ are in their sample but
were excluded from their maximum-likelihood membership analysis, because
of incomplete Gaia data, because they were identified as likely
background giants or distant background stars, or because their radial
or tangential velocity was more than 75 km~s$^{-1}$ from the cluster
median \citep[][their Sections 2.4, 3.1, and 4.3]{2022MNRAS.509.1664J}.

\textbf{ Coverage and numbers.} Data from \citet{2022MNRAS.509.1664J}
 cover our region. There are 1458 targets, 735 of which are in our
 final catalog and 439 of which we took as members tagged by this paper; see
 Fig.~\ref{fig:wherepanel7}.
A histogram of \ks\ from counterparts of these stars peaks at $\sim$13.2 mag.

\subsection{Larger Catalogs}

We waited until this point in the process to merge the larger catalogs
that are more likely to have false sources, or sources that would be
more likely to provide decoy matches to YSO candidates in the above
catalogs. For example, a bright IR source is reasonably likely to
break into several optical pieces (if for no other reason than the
optical surveys tend towards higher spatial resolution), but the
bright IR source is also more likely to be the YSO, so by merging the
large optical catalogs later in the process, we are more likely to 
match the correct sources by simple position matching. 

Some of these large catalogs report a magnitude with an uncertainty of
zero, or with no uncertainty at all. Rather than discard these
measurements, we assigned them an uncertainty of 0.3 mag, both so that
they can be used in the SEDs and to flag them as suspect for later
investigators. The affected bands are noted in each subsection below.

\subsubsection{DENIS}

The Deep Near Infrared Survey of the Southern Sky
\citep[DENIS;][]{denispaper,denisdata}  covered our region in $I$,
$J$, and \ks. Nominally, the depth of this survey is $K_D\sim18.5$,
and the spatial resolution is $\sim1-3\arcsec$.
We matched sources from this catalog with a $2\arcsec$
radius. We incorporated into the catalog sources that appeared for the
first time in DENIS. 

\textbf{ Coverage and numbers.} DENIS nominally covers the whole
region, but as can be seen in Fig.~\ref{fig:wherepanel7}, there is a
stripe of missing  coverage, a stripe of lesser coverage, and stripes
of deeper coverage (presumably where tiles overlapped). There are
24,550 DENIS sources in the master catalog, 1,550 of which get tagged
as YSOs by any other literature considered here.  A histogram of $K_D$
for sources in our region is clearly rolling over by $\sim$13 mag.

\subsubsection{\citet{2013MNRAS.434..806B}}

\citet{2013MNRAS.434..806B} published their entire large catalog as
part of their analysis  of 13 star forming regions. We retained their
catalog number, $UgriZ$, and their membership flag.  The spatial resolution
is not described in their paper but the seeing is likely $\sim1.5\arcsec$.
We matched sources from this catalog with a $2\arcsec$ radius. 

\textbf{ Coverage and numbers.} \citet{2013MNRAS.434..806B} has complicated
tile boundaries and edges; see Fig.~\ref{fig:wherepanel7}. All 8,567 sources
in the catalog are shown, along with the 936 sources identified as members.
A histogram of \ks\ from counterparts of these stars peaks at $\sim$15.0 mag.

\subsubsection{Gaia}
The next set of these very large catalogs provided additional challenges 
due to their size. We used TOPCAT to merge all of these with a 1$\arcsec$
radius. 

We first merged Gaia DR3 \citep{gaiadr3,gaiadr3data} data for
this region with geometric distances from \citet{2021AJ....161..147B}  
(via VizieR; \citealt{2021yCat.1352....0B}). We
retained optical photometric data in $G_{BP},G,G_{RP}$ as well as
parallax and proper motion, plus other diagnostics like renormalised 
unit weight error (RUWE), astrometric excess noise, and duplicated 
source flags. We merged those two catalogs together
within TOPCAT, and then carried that merged catalog forward.

As a space-based astrometric mission, the spatial resolution is 
at least $\sim$0.5$\arcsec$ and it reaches at least $G\sim21$ mag,
which is better in both resolution and sensitivity than anything 
else we have incorporated. However, since it is working in the 
optical, it is limited by the dust in this region, so it doesn't
reach those depths everywhere here in practice.

\textbf{ Coverage and numbers.} Gaia covers our whole region, and is
limited by dust; see Fig.~\ref{fig:wherepanel7}. There are 36,939
sources in our master catalog, 2,374 of which are identified by other
studies here as YSOs. For the ones that also have distances from 
\citet{2021AJ....161..147B}, there are 27,693 in the master catalog,
2,202 of which are identified by other studies here as YSOs.
A histogram of $G$ in this region peaks at $\sim$20.7 mag. 
A histogram of \ks\ from counterparts of these stars peaks at 
$\sim$15.0 mag, but that is limited by the existence of \ks\ 
counterparts, not Gaia.

\subsubsection{UKIDSS}

The next large catalog we used was that from the UKIRT (United
Kingdom Infrared Telescope) Infrared Deep Sky Survey (UKIDSS)
Galactic Plane Survey \citep{ukidsspaper,ukidssdata}, which observed
in $JHK_s$.  Nominally, this survey reaches \ks$\sim$18 mag, with a
resolution of  $\sim0.8\arcsec$ \citep{ukidsspaper}. 
Where UKIDSS reports an uncertainty of zero, we assigned 0.3 mag (see above).

\textbf{ Coverage and numbers.}  UKIDSS covers our whole region. 
There are 157,995 sources from UKIDSS 
(see Fig.~\ref{fig:wherepanel7}), 2,580 of which are tagged as YSOs by the 
literature explored here.  A histogram of the UKIDSS \ks\ for all 
sources in this region is broad but rolls over by \ks$\sim$16 mag.

\subsubsection{VVV}

VISTA variables in the V\'ia L\'actea
\citep[VVV;][]{2012AA...537A.107S,2017yCat.2348....0M} was another
large catalog we used, specifically their DR2, which includes
single-epoch observations. They observed in $ZYJH$\ks; the magnitudes
are on the Vega system, calibrated to 2MASS \citep{2012AA...537A.107S}. 
The spatial resolution is $\sim1\arcsec$ and nominally the \ks\ limit
is between 16 and 18, depending on location in the sky 
\citep{2012AA...537A.107S}.
Where VVV reports an uncertainty of zero (only in $Z$ and $Y$, and only
for a handful of YSOs), we assigned 0.3 mag.

\begin{figure}[htb!]
\plotone{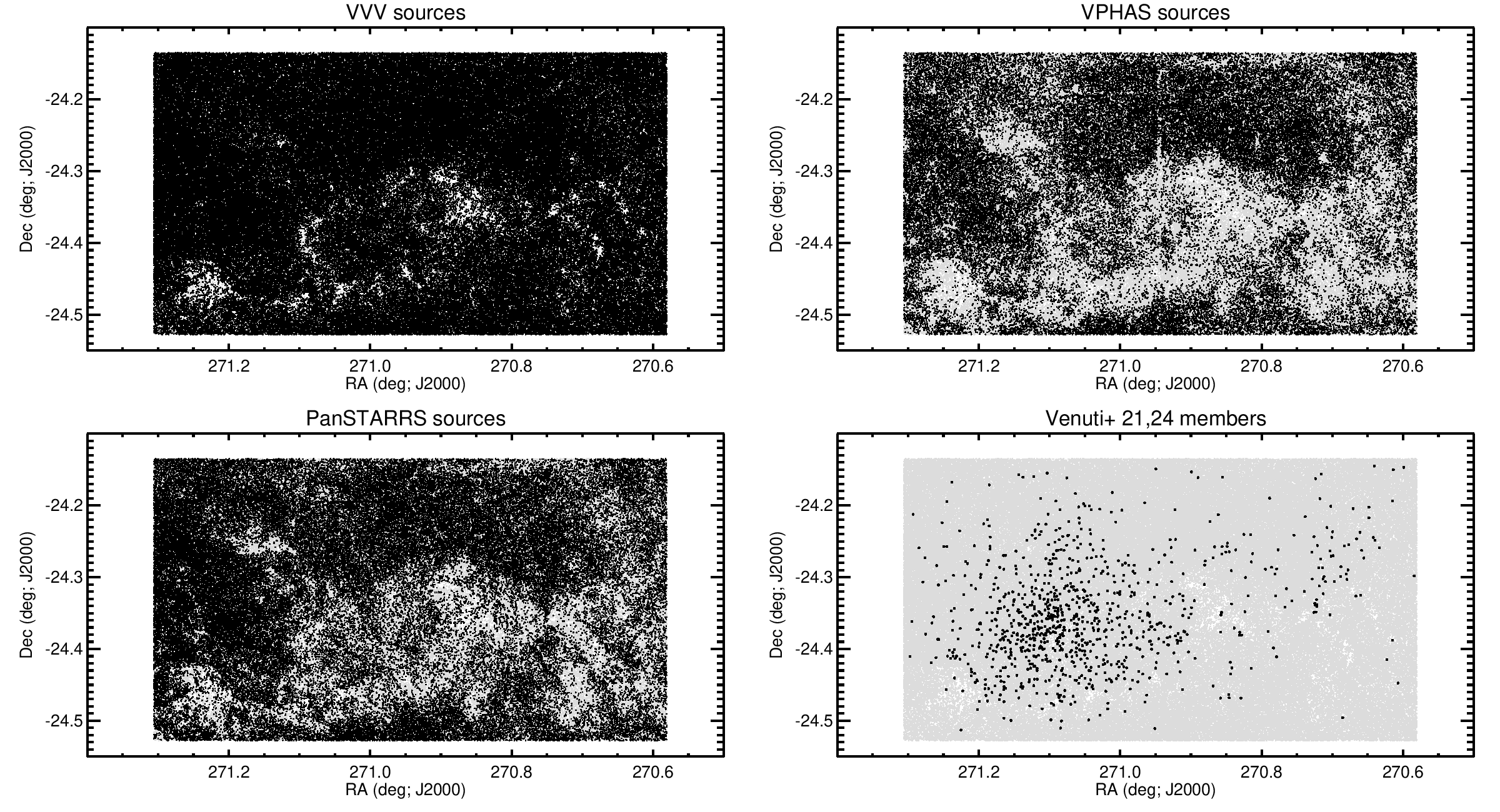}
\caption{In all cases, the indicated catalog is plotted in black and 
the entire catalog is shown in light grey. 
Top left: VVV \citep{2012AA...537A.107S,2017yCat.2348....0M} sources. 
Top right: VPHAS+ \citep{2014MNRAS.440.2036D,2016yCat.2341....0D} sources.  
This survey, being further into the optical
than UKIDSS or VVV, is more impacted by dust.
Bottom left: PanSTARRS \citep{panstarrs,2017yCat.2349....0C} sources.
Bottom right: \citet{venuti2021} and \citet{venuti2024} members.}
\label{fig:wherepanel8}
\end{figure}

\textbf{ Coverage and numbers.} VVV covers our whole region. There are
168,338 sources from VVV  in our master catalog (see
Fig.~\ref{fig:wherepanel8}), 2,665 of which are tagged as YSOs by the 
literature explored here.  The histogram of VVV  \ks\ for sources in
this region is broad and textured, but peaks at $\sim$15.7 and rolls
over by $\sim$16.5 mag.

\subsubsection{VPHAS+}
\label{sec:vphas}

The VST Photometric Halpha Survey of the Southern Galactic Plane and
Bulge \citep[VPHAS+;][]{2014MNRAS.440.2036D,2016yCat.2341....0D}  was
carried out in $ugri$ and H$\alpha$.  This survey, being further into
the optical than UKIDSS or VVV, is more impacted by dust; see
Fig.~\ref{fig:wherepanel8}. The spatial resolution is $\sim1\arcsec$
and the target depth is 20th mag (on the Vega system) at all bands
\citep{2014MNRAS.440.2036D}. 
Where VPHAS+ reports an uncertainty of zero, we assigned 0.3 mag.

\textbf{ Coverage and numbers.} VPHAS+ covers our whole region; see
Fig.~\ref{fig:wherepanel8}.  There are 59,473 sources from VPHAS+ in our
master catalog, 2,280  of which are tagged as YSOs by the literature
explored here. A histogram of VPHAS+ $i$ magnitudes for sources in this region
peaks at $\sim$21, better than designed.

\subsubsection{Pan-STARRS}

The last large catalog was Panoramic Survey Telescope and Rapid Response System
\citep[Pan-STARRS;][]{panstarrs,2017yCat.2349....0C}. These data are $grizy$ 
mean PSF magnitudes, averaged over the
single-epoch detections. The impact of the dust can be seen in Fig.~\ref{fig:wherepanel8}.
The quoted spatial resolution is $\sim1\arcsec$ and depth is $\sim$20 mag at 
$y$ \citep{panstarrs}.
Where Pan-STARRS reports no uncertainty, or an uncertainty of zero, we
assigned 0.3 mag.

\textbf{ Coverage and numbers.} Pan-STARRS covers our whole region. 
There are 84,118 stars from PanSTARRS in our master catalog, 2,494 of which 
are tagged as YSOs by the literature explored here. A histogram of 
$y$ magnitudes for targets in this region peaks at $\sim$19.0 mag.

\subsubsection{Combining These Big Catalogs}

We took these five catalogs (Gaia plus the merged distances, UKIDSS, 
VVV, VPHAS+, and Pan-STARRS) and carried out a 5-way position merge
with a 1$\arcsec$ radius in TOPCAT, retaining all sources whether or
not they had a counterpart.  We took all of the sources as-is from the
catalogs of origin, and merged them into the master catalog with a
1$\arcsec$ radius.  Further weeding of phantom sources from these
catalogs was identified downstream from this point.

\subsection{The Last Literature Catalogs: Venuti \etal}

Finally, we merged in the catalogs from \citet{venuti2021,venuti2024}.
Both of these projects started from earlier versions of our catalog,
so it made sense to add these catalogs near the end of our process.
Both of these catalogs, however, include sources from a region larger
than our current catalog.  Out of 1018 unique sources found in the
combined catalogs from both Venuti papers, 72 are outside of the
region we defined for this paper, leaving  946; 275 from
\citet{venuti2021} have counterparts in our current catalog, and 943
from \citet{venuti2024} have counterparts in our current catalog. We
retained the average $ugri$H$\alpha$, spectral types, and mass accretion rates, as well
as catalog ID (and implicit identification as a YSO), from these
catalogs. The photometry is taken from \citet{venuti2024}, or from
\citet{venuti2021} for bands missing from the later catalog; where both
catalogs report a value, they agree to within 0.005 mag. We assigned an
uncertainty of 0.15 mag to all of the Venuti \etal\ photometry.
Some source confusion present in our early catalog version
has been corrected here, but this only affects 15 stars, and none of
these issues affect the conclusions of \citet{venuti2021} or
\citet{venuti2024}.

\textbf{ Coverage and numbers.} Our region is covered by both
\citet{venuti2021} and \citet{venuti2024}. The 271 surviving
counterparts from \citet{venuti2021} and the 928 surviving
counterparts from \citet{venuti2024} are shown in
Fig.~\ref{fig:wherepanel8}.  A histogram of \ks\ from counterparts of
these stars peaks just below $\sim$13th mag.

\subsection{Patching Spitzer Photometry}

By the time we reached this point in the process, most of the
literature-identified YSO candidates had Spitzer IRAC and MIPS-24
measurements. For those that did not have measurements, we performed
aperture photometry. Following guidelines from the
IRAC Instrument Handbook \citep{irachandbook}, for the IRAC data, we
used an aperture of 3 native px, an annulus of 3-7 px, and aperture
corrections 1.124, 1.127, 1.143, and 1.234 for the four bands,
respectively. Similarly, from the MIPS Instrument Handbook
\citep{mipshandbook}, for MIPS-24,  we used an aperture of 2.857 px,
an annulus of 2.857-5.306 px, and an aperture correction of 2.05. This
bandfilling is only necessary for targets where somehow all the
existing Spitzer photometry attempts (primarily SEIP, GLIMPSE, and
MYStIX) missed them -- meaning that they are likely very faint (or
very bright), and/or with a very complicated background (near a bright
source, artifacts, or nebulosity). In those cases, aperture
photometry is likely sub-optimal, but it is unclear if more advanced
photometric techniques would be useful.  As a result, when we
inspected SEDs (see Sec.~\ref{sec:sedimginsp} below), if this new 
photometry was inconsistent with the rest of the SED, particularly 
for MIPS, we simply dropped it. 

\subsection{Best Available Photometry}
\label{sec:bestphot}

For some bands, it's useful to have a `best available' value.

The best available $JH$\ks\ photometry were populated first from 
2MASS, then UKIDSS, VVV, DENIS (just $J$\ks), MYStIX, and finally
\citet{2006MNRAS.366..739A}.  The catalog in
Appendix~\ref{app:datatable} (Table~\ref{tab:bigdata}) has the best
available data as well as all of the individual $JH$\ks\ values from
all the surveys. The individual $JH$\ks\ values from all
the surveys appear in the SEDs.
The best available values carry the uncertainties of the catalog they
were taken from, including the 0.3 mag uncertainties assigned where the
original catalog gave none (see the individual subsections above).

The best available IRAC and MIPS photometry  were populated in this
priority order: SEIP,  MYStIX, GLIMPSE, \citet{2010MNRAS.407.1170K},
and then patched photometry (as described above). The catalog in
Appendix~\ref{app:datatable} (Table~\ref{tab:bigdata}) has the best
available data as well as all of the individual IRAC and MIPS values from
all the individual surveys.  Because all of these studies are working
with the same data, the best available data are plotted in the
SEDs.

\subsection{SED and Image Inspection}
\label{sec:sedimginsp}

Because the source surface density is so high, the potential for
catalog mismatches is high. As an integral part of the catalog
matching process, we constructed SEDs for each literature-identified
YSO and overlaid catalogs on as many images as possible and relevant
(typically at least 2MASS, WISE, Spitzer, POSS,  Pan-STARRS, and/or
VVV), attempting to resolve source confusion issues using this
additional information. 

For the list of literature YSO candidates, we iterated among SEDs, 
images, remerging catalogs as we moved through the list multiple
times, looking in detail at each source. If the SED was contiguous
throughout all available wavelengths and the star seemed point-like
and relatively isolated in all images in IRSA's Finder Chart tool
\citep{finderchart}, then we assumed it was correctly merged across
catalogs.  These SED and image inspections also primarily revealed :

\begin{itemize}
    \item discontinuous SEDs, suggesting a source mismatch or incorrect photometry; 
    \item SEDs where the net result looks like two sources are plotted on the same SED, suggesting source confusion or a source mismatch in at least one band; 
    \item catalogs with artifacts that draw matches of real detections away from a real source; 
    \item `detections' from the same catalog that are much closer together than the 
    stated spatial resolution of the survey of origin; 
    \item `detections' that are not real -- no source is apparent in the corresponding image and the flux as plotted in the SED is inconsistent with the rest of the SED; 
    \item sources that are single in some low-spatial-resolution bands but that are multiple in higher-spatial-resolution bands; 
    \item X-ray sources that have no known counterparts at any other bands here.  
\end{itemize}
If we could shift matches around within 1-1.5$\arcsec$ to create two
(or occasionally three) continuous, internally-consistent SEDs where
before there was a single garbled one, we retained the cross-matches
that resulted in the two (or three) good SEDs. If one source from one
catalog could be identified as a likely artifact, we dropped it and
shifted any spurious matches to the nearest real source. If a logical
solution wasn't obvious even if the SED was a mess, we left it as is.
If a flux was inconsistent with the rest of the SED, and the source
wasn't apparent in the image, that flux was removed or converted to a
limit.  If there was information from Gaia that could help (RUWE, 
astrometric excess noise, duplicated source flags), we incorporated
that information in our assessment. Many sources were carefully
tweaked, consolidated, unhooked, removed, or otherwise clarified 
during this time-consuming manual process, resulting in the final
cleaned-up catalog presented here. Some X-ray sources obstinately
persisted in having no known counterparts at any of the bands
assembled here, and at this time, there is not much more we can do
about exploring those sources, other than listing them in the
catalog\footnote{These sources can be  identified in
Table~\ref{tab:bigdata} by looking for sources with npts\_sed=0.}.

\subsection{Example SEDs}

\begin{figure}[htb!]
\plotone{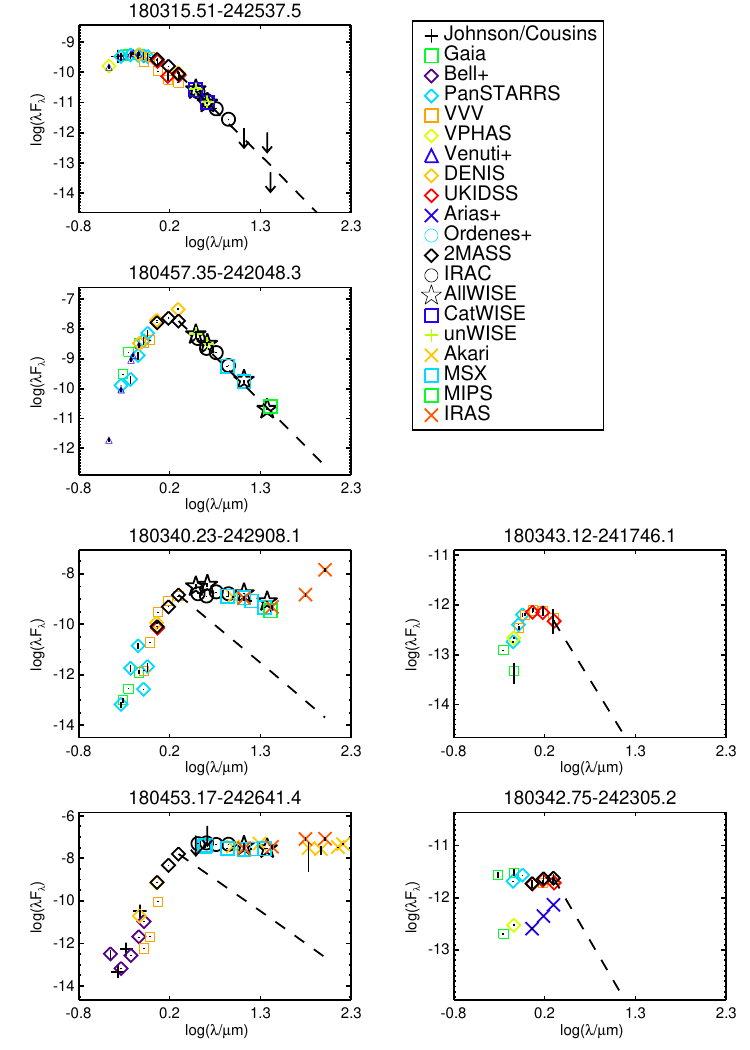}
\caption{Example SEDs, part 1.  These use cgs units for the $\lambda
F_{\lambda}$ axis, erg sec$^{-1}$ cm$^{-2}$, and the wavelength axis
is in microns.  Error bars are vertical black lines, most obvious here
in just a few of the points.  The dashed line is the expected flux
density from photosphere assuming \ks\ is on the photosphere.
Coordinate-based names appear above each plot. Sample SEDs show (first
row) a likely earlier type, well-populated SED with no likely IR
excess to 8 \mum\ and upper limits at 12, 22, and 24 \mum; (second
row) a likely later type, well-populated SED with no IR excess; 
(third row, left) well-populated SED with substantial IR excess (and
detections at all four IRAS bands); (third row, right) minimally
populated SED with no points longer than 2 \mum; (fourth row, left)
well-populated SED of embedded object and large IR excess, with
detections at several IRAS, MSX, and Akari bands; (fourth row, right)
minimally populated, likely irretrievably confused SED with no points
longer than 2 \mum.}
\label{fig:exampleseds1}
\end{figure}

Figure~\ref{fig:exampleseds1} shows some of the final SEDs that
emerged from this entire process. In that figure, there are two sample
SEDs  where there is likely to be no IR excess and two SEDs where
there is a very large IR excess and detections out past 50 \mum.   The
first thing that is immediately apparent from the SEDs is just how
many different catalogs have been combined.  The different catalogs
are largely in very good agreement, such that it is hard to
distinguish the individual symbols in the figure, especially in the
optical.  Some of the residual scatter in the SED is likely from
intrinsic variability, particularly in the optical for the stars
with a significant IR excess. 

Figure~\ref{fig:exampleseds1} also has an example SED of an object
about which we know nothing past \ks, so we cannot assess whether or
not it has an IR excess. And, finally, Figure~\ref{fig:exampleseds1}
includes an object for which we could not resolve the evident source
confusion that is apparent from the SED. Additional data will be needed
to untangle this source and the few others like it in our final catalog.

\section{Disks}
\label{sec:disks}

\subsection{Identifying IR Excesses: Initial Placement}

We have diverse data for the YSO candidates in our collection, so it
is difficult to come up with a single prescription for identifying disks 
across the entire sample. To attempt to take into account all the available
data and still identify the YSOs with IR excesses, we used a quantitative 
approach for an initial classification and then manually inspected each
SED multiple times to place the object in an appropriate bin. We
now describe the quantitative approach for the initial classification.

As discussed in, e.g., \citet{2012AJ....144..135M} or \citet{2008ApJ...674.1086T},
one way of calculating the significance of an IR excess 
is the following for two bands, $B1$ and $B2$, where $\sigma$ is the uncertainty :
\begin{equation} \chi_{\rm B1-B2} = \frac{([B1]-[B2])_{\rm
observed} - ([B1]-[B2])_{\rm predicted}}{\sigma_{([B1]-[B2])}}
\end{equation} 
We have many bands to compare, but comparing across different instruments 
can be complicated (see below re: systematic IRAC offset). We also
don't have spectral types for most of our stars. For most FGK and early M 
stars, if the shorter band is on the photosphere, 
the expected IR color is 0, but for later M stars,
it is not 0. Fortunately, given that this target is at $\sim$1250 pc, there
are not many M stars in our YSO list, so we assumed that the predicted
photospheric IR colors were 0 for all of our YSOs. For situations where
the shorter band is not on the photosphere, in most cases, the SED is 
well-populated enough that the excess is apparent from the SED alone, and 
any $\chi$ calculations are less important. 

We calculated $\chi$  within Spitzer (for colors [I1]$-$[I2],
[I1]$-$[I3], [I1]$-$[I4], [I1]$-$[M24], using the best available
values) and within WISE  (for colors [W1]$-$[W2], [W1]$-$[W3],
[W1]$-$[W4], using AllWISE); and then comparing \ks\ with  Spitzer
(for colors [\ks]$-$[I2], [\ks]$-$[I3], [\ks]$-$[I4], [\ks]$-$[M24],
using the best available values) and WISE (for colors [\ks]$-$[W2],
[\ks]$-$[W3], [\ks]$-$[W4], using the best available \ks\ and
AllWISE).  Because the Spitzer data are the highest resolution and
sensitivity IR data we have,  the $\chi$ values for those should be
the most sensitive. The $\chi$ within WISE should be the next most
sensitive, followed by the ones that combine \ks\ and Spitzer, and
then \ks\ and WISE.  The distribution of these $\chi$ values were
inspected and used to set some limits.  We fit Gaussians to each
distribution and looked at how far out we needed to go to reach
$\sim3\sigma$ from the mean; this is easier where there are more
$\chi$ values.  There are few $\chi_{\rm WISE}$ in general, but even
within those, there are very few involving [W3] or [W4], just because
there are so few with detections at those bands. Based on the
distributions, for each $\chi$, a corresponding threshold was set for
identifying an IR excess: $\chi>4$ for any of the combinations within
Spitzer,  $\chi>5$ within WISE, $\chi>8$ for the combination of \ks\
and Spitzer, and $\chi>9$ for the combination of \ks\ and WISE.

For each source, if there are data past 2 \mum, and everything that
has detections also has the corresponding $\chi$ above those
thresholds, then it has a significant IR excess. If it is all below
those thresholds, then it does not have a significant IR excess. It
is common, however, that there are mixed results.  More weight was
put on longer wavelengths than shorter. An excess that starts at $>$5
\mum\ can still be a legitimate excess. A significant excess at 24 or
8 \mum\ with no detections longer than that, but no excess shorter
than that, is still a significant IR excess.  An apparently significant 
excess at 3 or 4 \mum, but photospheric fluxes between 5 and 25 \mum, is 
probably not a real excess.  More weight was placed
on Spitzer than WISE data, because of the difference in sensitivity
and spatial resolution. If there was no significant excess within
Spitzer or WISE, but there was a significant excess when comparing 
those to \ks, it's possible that there is a small excess. 

\subsection{IR Excesses: Final Placement}
After automatic placement in the initial, provisional excess bins,
each object was individually visually inspected and their placement adjusted
if necessary. Patterns within the final disk categories were noted and
similar stars were grouped together; see next section. 

We emphasize that the quantitative use of the $\chi$ values is for an 
initial disk placement, but all stars' SEDs are inspected and verified 
individually to be sure that they are properly characterized before 
their final disk placement.

\subsection{IR Excess Categories}

There are three easy categories into which we can place SEDs.  One is
``no data'' -- there are no data past \ks, so there is no information
as to whether or not there is an IR excess (18\% of YSO sample); an 
example appears in the third row of Figure~\ref{fig:exampleseds1}. 
``Unambiguous no excess'' (``no'') means that there are data at \ks\
and longer bands, and everything up to and including the longest
wavelength available is consistent with photospheric emission (30\% of
YSO sample); two examples appear in the first two rows of 
Figure~\ref{fig:exampleseds1}. This includes situations where only
the longest band has an excess but it is not significant because there
is also also a big error bar.  ``Unambiguous excess'' (``yes'') means
that there is more than one band past \ks\ involved in a many-sigma IR
excess -- that is, the SED could be in a textbook as an example SED
(like the two in the lower left of Fig.~\ref{fig:exampleseds1});
there is no ambiguity as to whether or not there is an excess (26\% of
YSO sample). Most (72\%) of the YSO sample are in these
straightforward bins.

We found it necessary to create additional categories beyond these 
`easy' categories. There are cases we have tagged ``no?'' (5\%) --
there are data past \ks, the longest wavelength available seems like
it might be consistent with the photosphere, but maybe there is a
small excess (Figure~\ref{fig:exampleseds2}c).  There are stars where
the SED is photospheric except for one band, the longest band, which
seems to be significantly above the photosphere (``1-band''; 4\%;
Figure~\ref{fig:exampleseds2}e).  There are stars where there seems to
be an excess, but it is only manifesting significantly in one band,
and the adjacent shorter-wavelength point is offset from the
photosphere, though not statistically significantly given the $\chi$
value (``1+-band''; 2\%; Figure~\ref{fig:exampleseds2}f).  There are
stars where there is marginal evidence for a disk (``yes?''; 8\%;
Figure~\ref{fig:exampleseds3}) -- most often that means that 2 bands
seem to have a slight excess. 

In some cases ($\sim$7\% of the YSOs), there is an apparently
systematic offset of the IRAC measurements. That is, all of the
measured IRAC bands (all four, or, more commonly, just the shortest
two where the longest 2 are non-detections) appear on a line parallel
to the Rayleigh-Jeans line but offset slightly high. There is no
significant excess apparent within the IRAC bands (\eg,
[I1]$-$[I2]$\sim$0, [I1]$-$[I4]$\sim$0), but there is a small,
apparently significant excess when comparing to \ks.   Examples of
these stars are in Figure~\ref{fig:exampleseds2},  panels a and b,
where a has all four IRAC bands offset and b has just IRAC 3 and 4
offset. We spent considerable time investigating the origin of this
offset to no avail. This does not appear to be an error in data
reduction. It could be a result of source confusion at the IRAC bands
(additional contributing flux at IRAC), or it could be real.
We have identified these sources in the tables below, so that they can
be followed up (or dropped) as later investigators prefer. We have
generally lumped these in with the `no excess' stars because the most
sensitive excess metric is $\chi$ within IRAC, and that is usually
$<$4 (our threshold), often $<<$4, for these stars.  Combining these
with the other ``no disks'' means that 36\% of the sample have no
disks. 

In some cases, as with the IRAC offset stars, we have added an extra
qualitative subgrouping to help identify similar sources within the SED
collection. Some targets have more than one subgroup. These
subcategories are:
\begin{itemize}
    \item fewpoints -- SED that has very few points in it. So many of our targets 
    have such well-defined SEDs that it is worth calling out those targets whose placement 
    is less secure specifically because there are so few points in the SED.
    (Note that this is different from those with {\em no} points in the SED, which
    are the X-ray-only sources; those have npts\_sed = 0 in the data table in 
    Appendix~\ref{app:datatable}.)
    \item offset/offsetx -- SED that has an IRAC offset (as described above), 
    but an additional `x' means it appears to also have a legitimate IR excess. 
    (Figure~\ref{fig:exampleseds2}ab)
    \item vshape -- SED that seems to have an excess but where the excess points 
    rise abruptly into a `V' shape that is almost implausibly steep, though the points that
    create that steep rise seem to be legitimate detections. (Figure~\ref{fig:exampleseds2}d)
    Within this class, vshape2, vshape3, and vshape4 identify three subgroups of SEDs
    that more closely resemble each other than they do the rest of the vshape SEDs;
    these subgroups are qualitative and are intended only to group like with like.
    \item reddened -- SED that is highly reddened and/or embedded such that the 
    highest rollover point is 
    longer than \ks. (In these cases, $\chi$ values are less meaningful.)
    \item scatter/scatterx -- SED with a lot of scatter; additional `x' means 
    it appears to also have a legitimate IR excess. This group encompasses those sources
    where the scatter could be due to intrinsic variability or source confusion, but some of 
    the scatter could also be photometric problems; we currently have no way to know. 
    Note that a star could have no data past \ks\ and still be tagged `scatter' in order
    to include these source confusion/variability cases. 
    (Figure~\ref{fig:exampleseds1}, lower right)
    \item straightup -- SED that rises more or less straight up, very steeply. These 
    could be deeply embedded YSOs or they could be some other kind of source. 
    (Figure~\ref{fig:exampleseds2}h)
    \item what/whatx -- SED is strange, and just doesn't look like a YSO(?); 
    additional `x' means it appears to also have a legitimate IR excess. 
    (Figure~\ref{fig:exampleseds2}g)
    \item weird/weirdx -- SED is, well, weird; additional `x' means it appears 
    to also have a legitimate IR excess.
\end{itemize}
A `?' appended to any of these (e.g., vshape?) means the source possibly belongs
to that subgroup. Where more than one subgroup applies, they are separated by `/'
(e.g., offsetx/vshape). The secondary category can also contain one of the primary
categories (offset, offseti3i4, 1band, 1+band) when a second one applies.

On the whole, out of 3019 stars, 537 (18\%) have no data past \ks, 
1097 (36\%) have unambiguously no disk (no IR excess), 160 (5\%)
probably have no disk, 193 (6\%) likely have a small excess at the
longest available wavelength,  252 (8\%) likely have a disk, and 780
(26\%) have an unambiguous IR excess (disk). 

All of these disk classifications are included in the data table in 
Appendix~\ref{app:datatable}.

\subsection{Example SEDs}
 
\begin{figure}[htb!]
\plotone{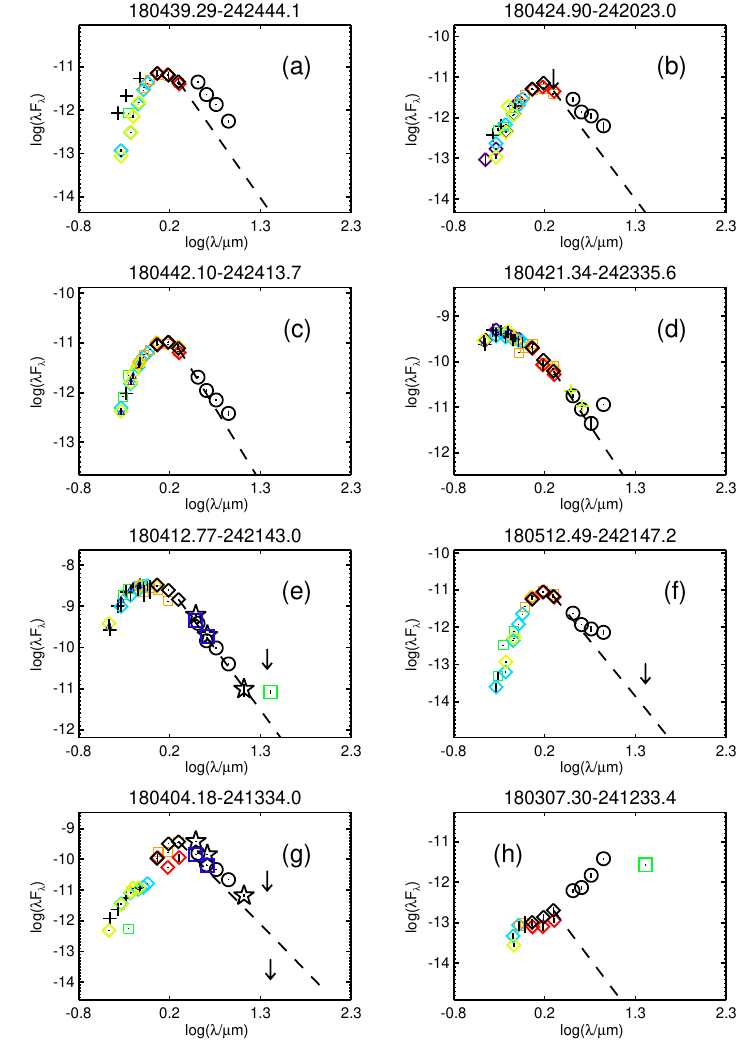}
\caption{Example SEDs, part 2.  These use cgs units for the $\lambda
F_{\lambda}$ axis, erg sec$^{-1}$ cm$^{-2}$, and the wavelength axis
is in microns.  Legend is as in Fig.~\ref{fig:exampleseds1}.
Coordinate-based names appear above each plot. Sample SEDs show: a-
offset IRAC; b- offset IRAC-3 and -4; c- ``no?'' disk;  d-
``no?''/v-shape; e- 1-band; f- 1+band; g- weird/scatter; h-
straightup.  See the text for more details.}
\label{fig:exampleseds2}
\end{figure}

Figure~\ref{fig:exampleseds1} includes SED examples of stars that were
unambiguously `no IR excess'  (`no') and `clear IR excess' (`yes'), 
as well as `no data past \ks' (`nodata'). 
Figure~\ref{fig:exampleseds2} has examples of more complex cases
discussed above. Panels a and b demonstrate the  offset IRAC -- panel
a has all four IRAC channels offset and panel b has primarily the
longest two bands offset, although a case could be made  that the
first two channels are also offset. The uncertainty on the points  is
large enough that there may or may not be an IR excess. In contrast,
panel c is  a case where IRAC-1 and -2 are photospheric, and the SED
is pulling away from the photosphere in IRAC-3 and -4 such that there
could be a legitimate  excess, and the excess is larger at 4 than 3,
consistent with astrophysical expectations. The error bars are small
enough that the IR excess at IRAC-3 and -4 is just over the thresholds
described above. Panel d's SED is photospheric until IRAC-4, where the
SED rises abruptly, perhaps unphysically. There is nothing obviously
wrong with the IRAC-4 photometry, except for the shape of this SED.
It's characterized here as `v-shape' as a result. As in panel d, 
panel e is photospheric until the longest band, in this case MIPS-24. 
However, here the rise from photospheric at 24 \mum\ isn't
unphysical.  This one is characterized here as `1-band.' Panel f's SED
looks like it has an IR excess at the two longest detections, IRAC-3
and -4. However, the error bar and departure from the photosphere at
IRAC-3 combine to make the IR excess insignificant given the limits we
have set. But, viewed in  context with the significant excess at
IRAC-4, the shape of the SED suggests that the IRAC-3 excess may be
real. This one is characterized as `1+band,' meaning that technically
there is only a significant IR excess at one band, but, in context,
there may be a real excess at the second-to-last detection in the
SED. The last two panels are both more unusual. Panel g is both
strangely shaped and has a lot of scatter in the SED, suggesting
either source confusion or intrinsic variability. This SED on its own
doesn't seem stellar. Panel h has an SED that more or less goes
straight up in the IR; it's characterized here as ``straightup.''
This particular SED is reasonably well-populated; many of the
``straightup'' SEDs are less-well-populated, and many have
long-wavelength counterparts (as this one does).

\begin{figure}[htb!]
\plotone{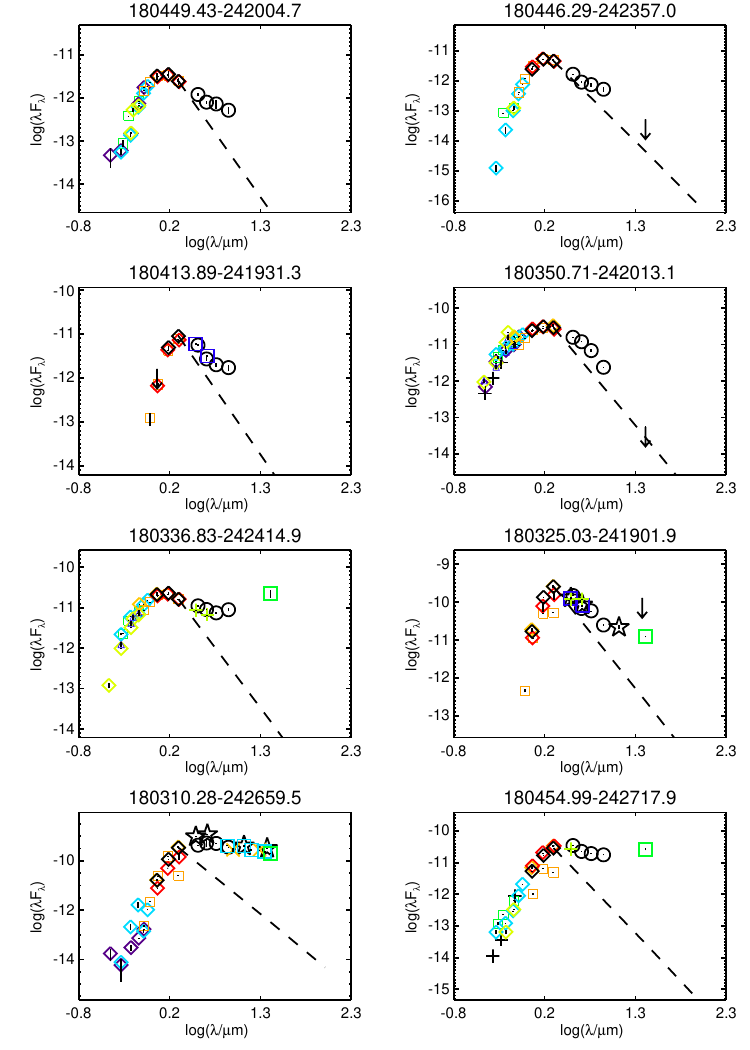}
\caption{Example SEDs, part 3.  These use cgs units for the $\lambda
F_{\lambda}$ axis, erg sec$^{-1}$ cm$^{-2}$, and the wavelength axis
is in microns.   Legend is as in Fig.~\ref{fig:exampleseds1}.
Coordinate-based names appear above each plot. Sample SEDs show
possible disks (``yes?'') in first two rows, and confident disks
(``yes'') in the last two rows.}
\label{fig:exampleseds3}
\end{figure}

Figure~\ref{fig:exampleseds3} collects four of the `yes?' disks and
four of the unambigous 'yes' disks just as examples of the diversity
of shapes and scatter and embeddedness. Plots like these will be provided
for all of the YSO candidates in our final catalog delivery to IRSA.

\subsection{Disk Reconciliation}

\citet{venuti2024} placed their YSOs in bins of disk-bearing,  disk
candidate, or disk-free. The stars that we have in common are, in most
cases, in the same bins in both analyses.  In the few cases where they
don't agree, most commonly, we now have filled in missing Spitzer
measurements (enabling classification not possible before),  or
cleaned up source confusion (which may have removed Spitzer fluxes
previously assigned to a source). For example, there are  three
sources very close together: J180414.57-242221.1, J180414.63-242221.7, 
and J180414.70-242219.5. All three are legitimate, separate sources --
Gaia,  PanSTARRS, VVV, and VPHAS all resolve three sources here.  All
three of them have multiple literature identifications as YSOs.  But
once this source confusion is taken into account, IR flux is
difficult-to-impossible to apportion properly to J180414.63-242221.7.
It previously had some flux assigned that likely is actually due to
J180414.57-242221.1, so the disk assessment was incorrect before.

\subsection{Spatial Distribution of Disks on the Sky}

\begin{figure}[htb!]
\plotone{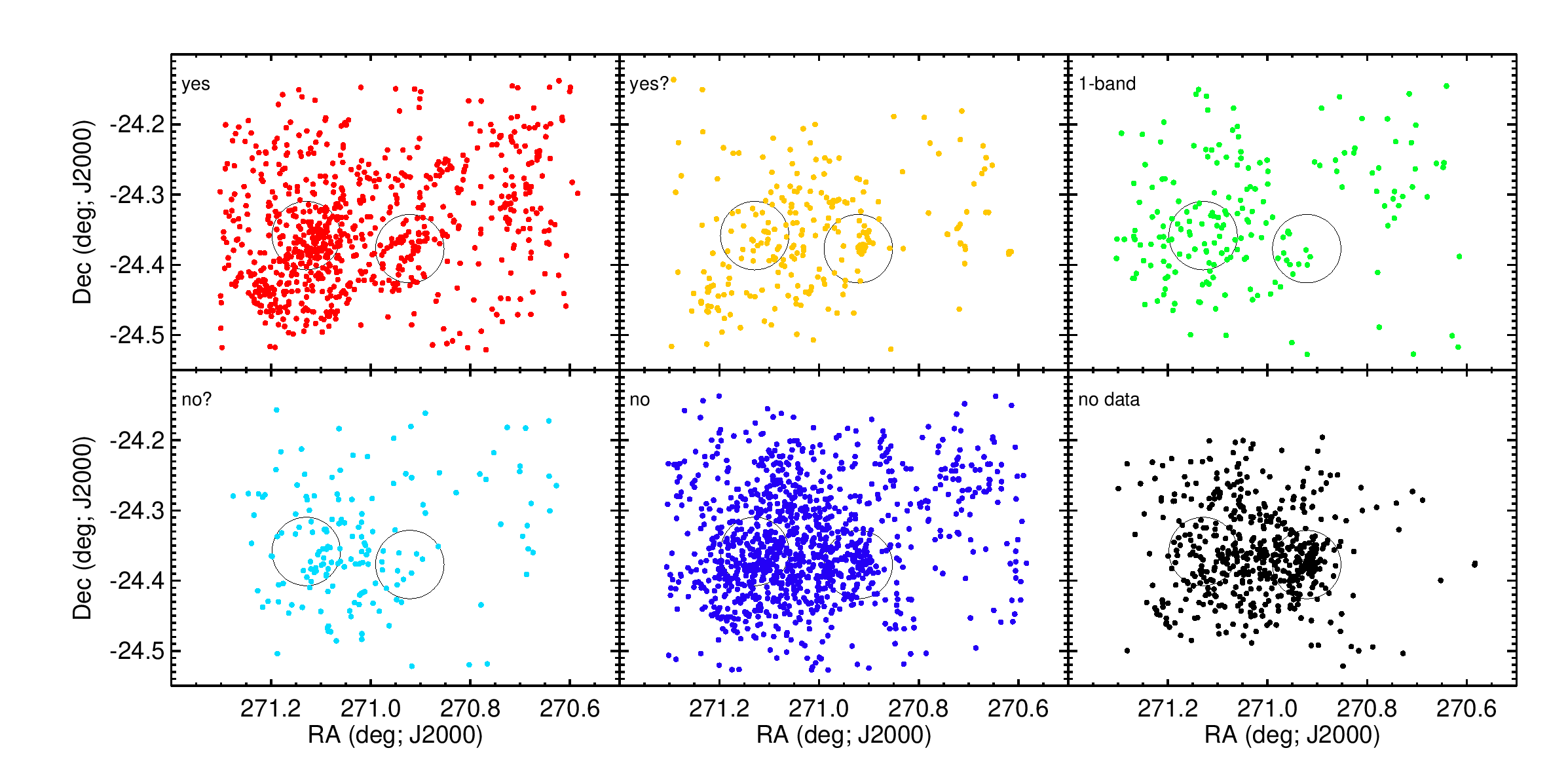}
\caption{Location on the sky (in decimal degrees, J2000) of YSOs and 
candidate YSOs, for the 6 different SED  disk categories.  From left
to right, top to bottom: confident disks (`yes'), less confident 
disks (`yes?'), possible disks (largely 1-band), less confident  no
disks (`no?'), confident no disks (`no'), and no information. Black 
circles in each panel are the approximate locations of NGC~6530 
(left) and the Hourglass Nebula (right). 
Concentrations of YSOs can be seen by eye in at least 4 of the 6
panels here, as is residual texture reflecting where there are 
more data (e.g., more disks in NGC 6530), or less data (e.g., 
more `no data' in the Hourglass). }
\label{fig:disks2}
\end{figure}

The distribution of stars with and without disks on the sky is somewhat
difficult to assess. The source density of YSOs and candidates is very
high in several areas, but fades rapidly as a function of distance from
those concentrations. Fig.~\ref{fig:disks2} plots the 6 different SED 
disk categories on the sky -- confident disks (`yes'), less confident 
disks (`yes?'), possible disks, less confident no disks (`no?'), 
confident no disks (`no'), and no information.

Concentrations of YSOs can be seen by eye in most of the panels in
Fig.~\ref{fig:disks2}. We expect that members with disks should, on 
average, be younger and therefore more clustered than stars without
disks. However, selection effects could be a complication here -- 
the limitations of, e.g., having no infrared data past
2 \mum\ in the Hourglass can be seen in the distribution of targets
with no information.

Looking at the distribution of the objects on the sky as a function 
more explicitly of SED slope (as opposed to binned like in
Fig.~\ref{fig:disks2}) reveals no new information.  There is a clear
clump of steeply positive slopes in the heart of the Hourglass, but no
other structure jumps out by eye, much like Fig.~\ref{fig:disks2}. 
Further restricting this kind of analysis to only the best-populated
SEDs or just the most confident disks/no disks similarly does not 
produce much new information.

\section{Properties of the Final Catalog}
\label{sec:properties}

\subsection{Final Catalog}
\label{sec:finalcat}

Our final, complete merged catalog of the 3019 YSOs and candidates 
is in an electronic table accompanying this article and delivered to IRSA. 
A list of the columns can be found here in the Appendix, and includes 
photometry and cross-IDs from the other catalogs merged here. 
An entire set of plotted SEDs is also in the IRSA delivery and can be 
viewed there in RA order or in disk confidence order. 

\subsection{Source Numbers}
\label{sec:sourcenumbers}

There are 3019 YSOs or YSO candidates listed in the literature in our region 
that have survived the merging process. 

To reiterate some critical points from above, as  seen in
Table~\ref{tab:literature}, there are data from $\sim$50 different
origins incorporated here. Note that not all of these original
catalogs cover our whole region (there are more covering NGC 6530, for
example),  nor are they to similar sensitivity or spatial resolution. 

Table~\ref{tab:literature2} includes the numbers of sources per origin
catalog that appear in our final merged catalog and the final list
of literature-selected YSOs and candidate YSOs as a guide to how
common or rare that catalog's information is in our final catalog of
literature YSOs.  The rows where the number of sources and the number
of YSOs are the same are the papers where their data tables consist
solely of YSOs or candidates.

\subsection{SED points}
\label{sec:sedpoints}

Table~\ref{tab:literature} also indicates that there are $\sim$80
different possible measurements that could appear in the SEDs, plus
additional measurements such as $L_x$, \vsini, or $P_{\rm rot}$. Note,
though, that this does not necessarily mean $\sim$80 distinct
wavelengths; a few surveys use exactly the same filters. For example,
AllWISE, CatWISE, and unWISE all have the same WISE-1 and WISE-2
bands; CatWISE and unWISE use more data than AllWISE, but they use the
same data as each other, so the points from those catalogs are not
entirely independent. Several other catalogs use very similar but not
exactly the same filters; Pan-STARRS $gri$ is similar but not the
same as VPHAS $gri$.

\begin{figure}[htb!]
\plottwo{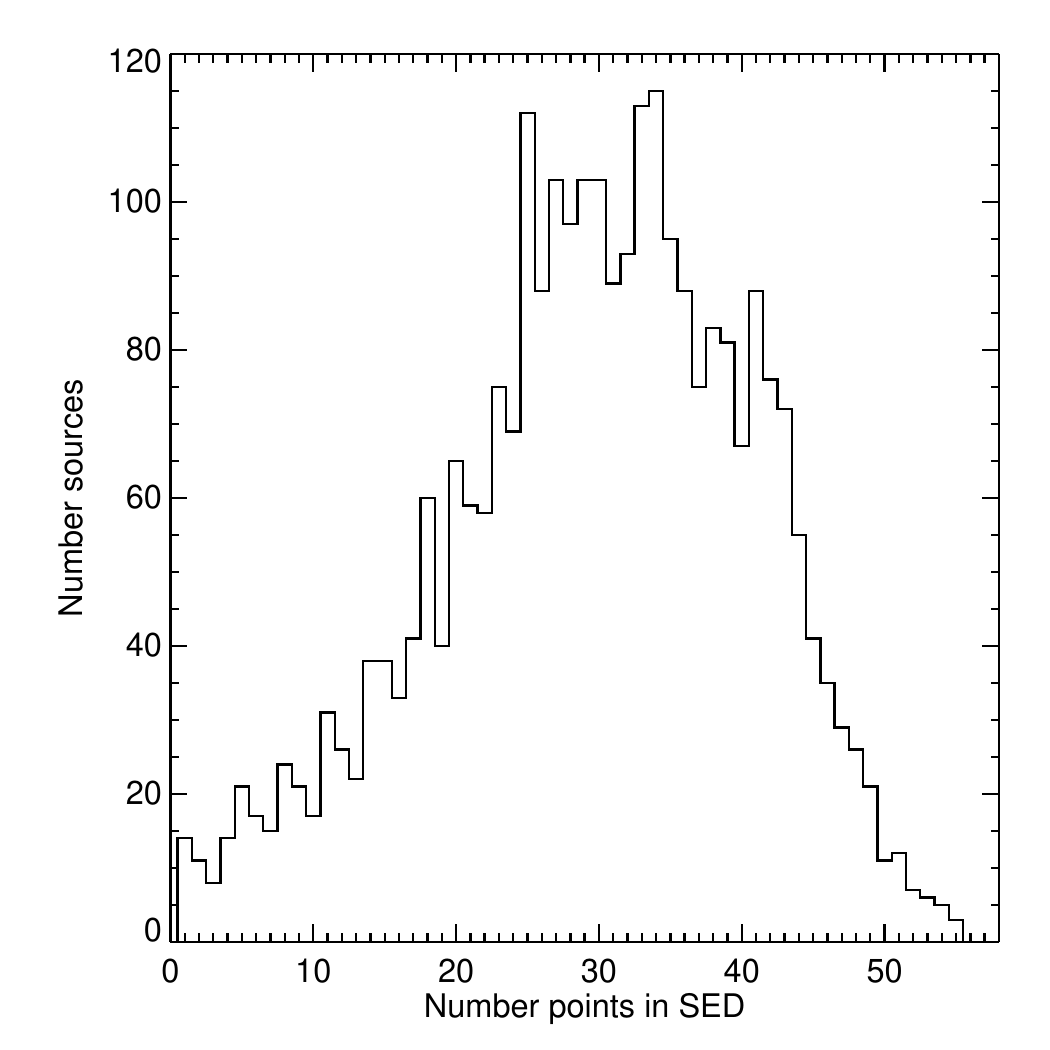}{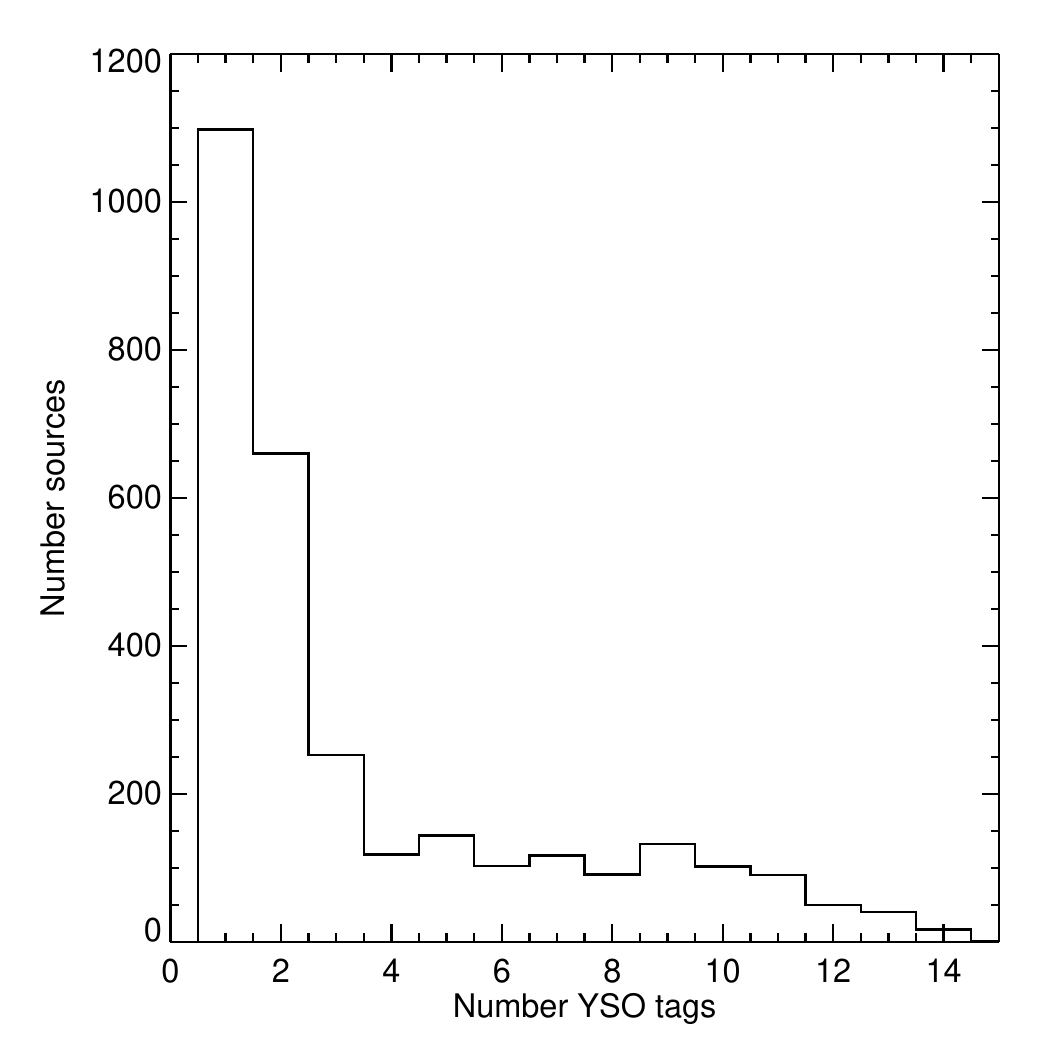}
\caption{Left: Histogram of number of points in the SED for all of the 
literature YSOs or candidate YSOs. The median is 30 points; $\sim$75\% 
have more than 20 points in the SED. At the other extreme, 210 (not shown here) 
have no points in their SED -- they are X-ray only. 
Right:  Histogram showing the number of YSO tags per source, effectively 
highlighting those that have multiple YSO tags.  Seventy-five percent have
5 or fewer YSO tags. There are 110 sources that at least 12 papers have 
identified as YSOs, and 302 sources that at least 10 papers have identified
as YSOs.}
\label{fig:sednpts}
\end{figure}

Figure~\ref{fig:sednpts} has a histogram of the number of points or 
limits in the SED for all of the literature YSOs or candidate YSOs.
The median  of this distribution is 30 points or limits in the SED,
and  the maximum number of points or  limits any star has is 55. 
About 75\% of them have more than 20 points or limits, so the SEDs in
general are very well defined.  At the other extreme, however, there
are 210 targets with no measurements in their SED at all -- they are
X-ray only sources, whose nature is unknown (but whose YSO nature is
suspected). They could, however, be active galactic nuclei (AGN),
seen through the Galaxy. 

\begin{figure}[htb!]
\plotone{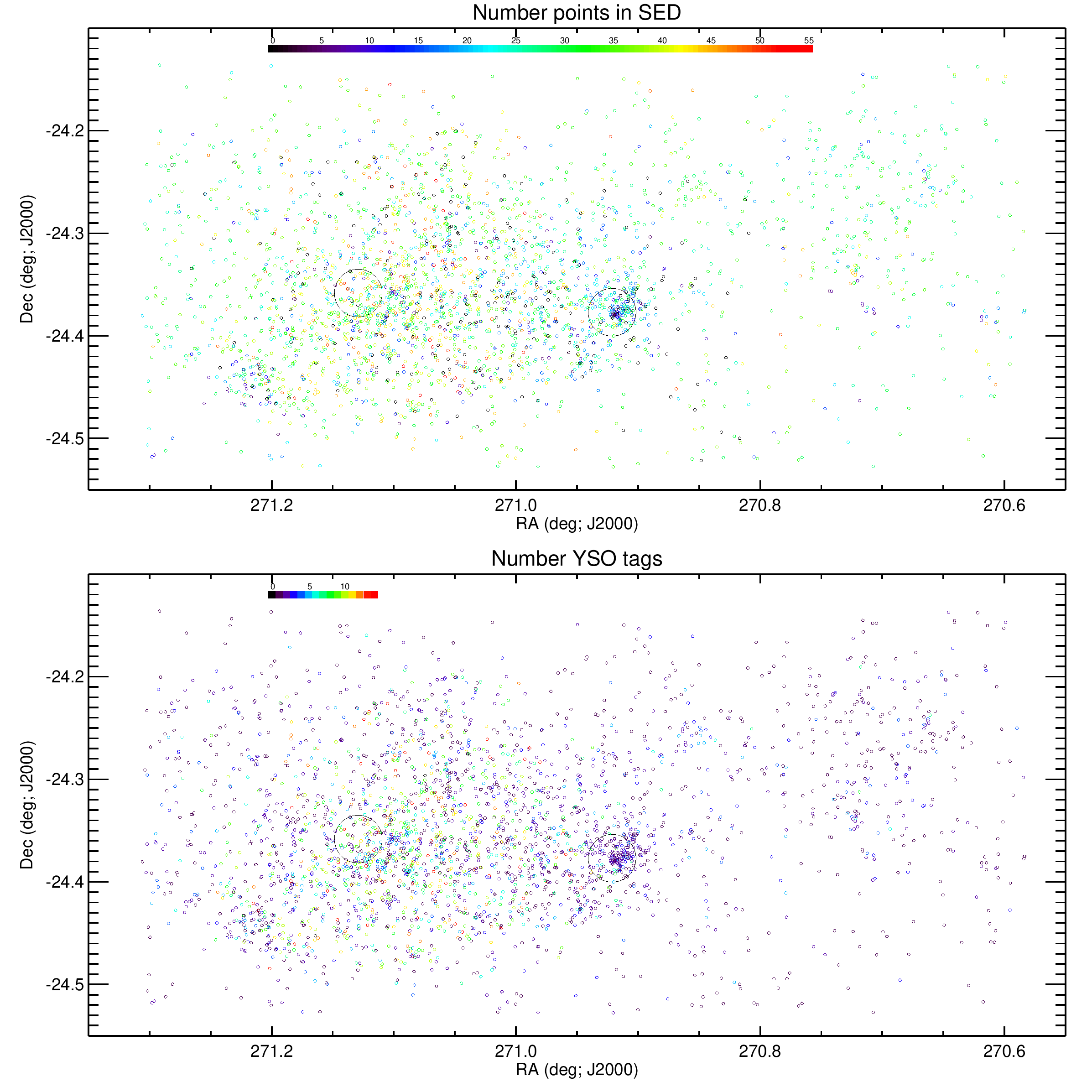}
\caption{Location on the sky (in decimal degrees, J2000) of YSOs and 
candidate YSOs, where the color of the point corresponds to the number
of points in the SED (top) or the number of YSO tags (bottom). There
are more literature data available on the East side of our region, around
NGC~6530 ($\sim$271.1$\arcdeg$,  $-$24.3$\arcdeg$, indicated by the large 
black circle on the left; see
Fig.~\ref{fig:where}), so there are generally more points per SED and
more YSO tags.  There are generally fewer points per SED and YSO tags
in the heart of the Hourglass ($\sim$270.92$\arcdeg$, -24.38$\arcdeg$, 
indicated by the large black cirle on the right;
see Fig.~\ref{fig:where}),  where high surface brightness and high
source surface density both limit catalog completeness.}
\label{fig:nptsonsky}
\end{figure}

Additional complicating factors can be seen in 
Fig.~\ref{fig:nptsonsky}. There are more literature surveys available
on the East side of our region, in NGC~6530 ($\sim$271.1$\arcdeg$, 
$-$24.3$\arcdeg$), so in general, there are more YSOs \& candidates,
and more points per SED, in the East half of our region. In regions
of very bright nebulosity and high source surface density (in the
heart of the Hourglass, $\sim$270.92$\arcdeg$, -24.38$\arcdeg$) there
are distinctly fewer points on average available in any given SED,
just because source confusion and high surface brightness both limit
catalog completeness. 

\subsection{YSO Tags}
\label{sec:ysotags}

Just looking at the literature, 25 different papers explicitly
identified YSOs (Table~\ref{tab:literature}). As described above, we 
discarded YSO candidates that were identified in only 
\citet{2005AA...430..941P} or \citet{2013PhDT........36C}. While there
are many YSOs or candidates that are identified in more than one
paper, there are also many others that are identified in only one
paper.  An approach one could take would be to simply retain only
those YSOs that were identified in more than one or two papers
as `more reliably YSOs.' There is some implicit bias in this because
as seen in Fig.~\ref{fig:nptsonsky}, there are more surveys (and
therefore more tags) in NGC~6530 and fewer tags in the Hourglass.
Additionally, different methods of searching for YSOs reach different
kinds of targets -- for example, using positions and proper motions (PPM)
prior to Gaia was limited to optically brighter stars, and searches
using Spitzer will easily reach relatively faint IR excess sources
that may be entirely hidden from existing optical surveys.

\begin{figure}[htb!]
\epsscale{0.7}
\plotone{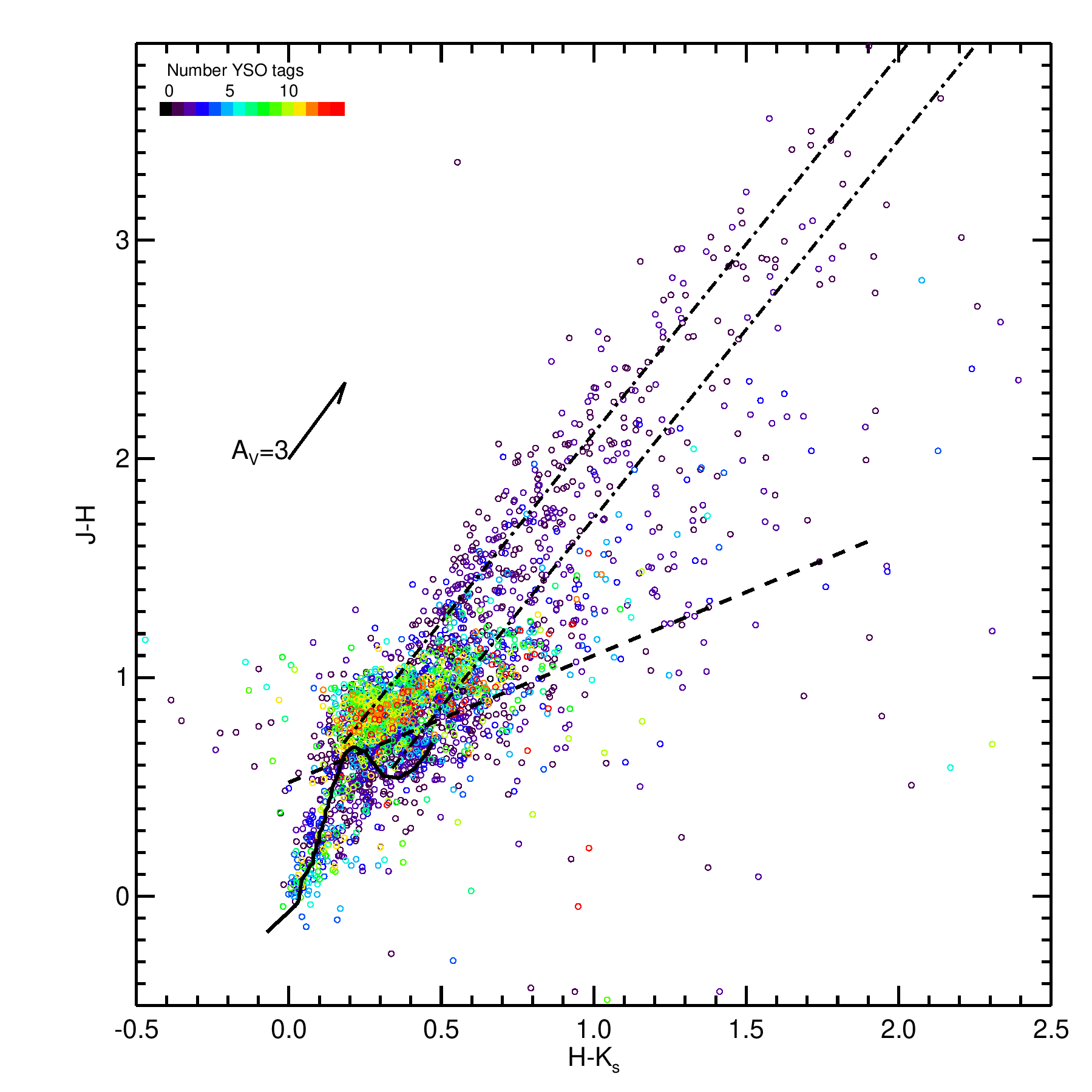}
\caption{A plot of $J-H$ vs.\ $H-K_{\rm s}$ for the YSOs. The point
colors  correspond to the number of stars in the bin, as indicated.
The thick, solid line is the ZAMS \citep{pecautmamajek}; the
dashed line is the \citet{meyerttlocus} T~Tauri locus. A reddening
vector (following the reddening law from \citet{Indebetouw}  and
\citet{mathis}) are shown, and the dash-dot lines are reddening
vectors extending roughly from the extremes of the ZAMS
relation. The YSOs with the most literature tags are clumped 
above the ZAMS. Many of the most reddened stars are identified 
in just a few papers, typically using mid-IR or X-rays to identify
YSOs. }
\label{fig:tagjhjk}
\end{figure}

\begin{figure}[htb!]
\epsscale{0.5}
\plotone{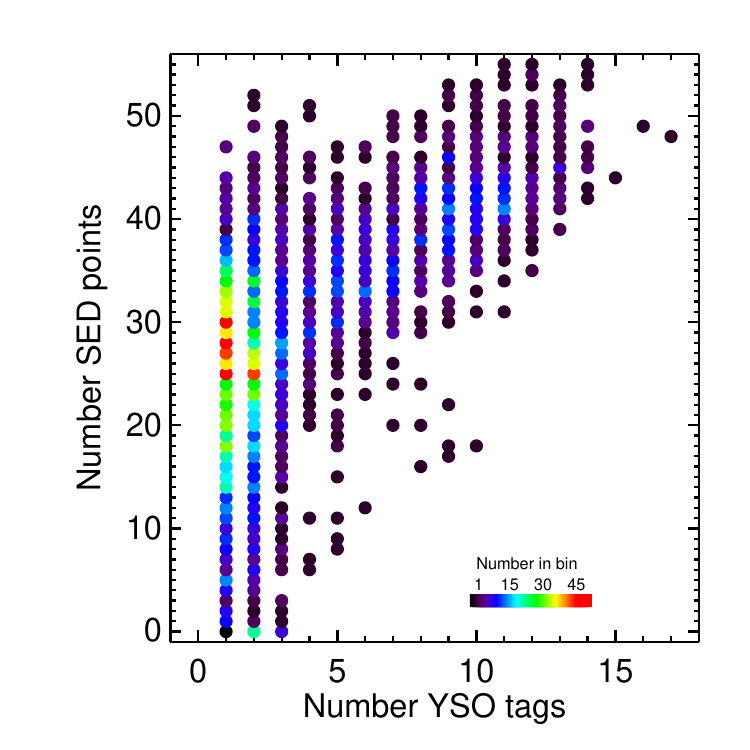}
\caption{A plot of the number of papers tagging the star as a YSO 
against the number of points in its SED. The point colors 
correspond to the number of stars in the bin, as indicated. It is
roughly the case that the more papers there are identifying the 
YSO, the more points there are in the SED, but there are a lot of 
YSOs that are only tagged in one paper whose SEDs are well-populated 
from large catalogs.}
\label{fig:ysotagssedpoints}
\end{figure}

Table~\ref{tab:ysotags} lists, for each journal article, the number of
objects that were identified as YSOs or YSO candidates (``tagged
YSOs'').  It also lists the number of those sources that were also
identified in at least one other paper, and the number of those
sources that were only identified in that paper. 
Figure~\ref{fig:sednpts} includes a histogram of the number of sources
as a function of how many times they were tagged as YSOs. There are
110 YSOs that have at least 12 papers that identify them as YSOs, and
302 that are identified in at least 10 papers, so even sight unseen,
those seem likely to be legitimate YSOs. Roughly two-thirds of the
sample have two or more YSO tags. If one assumes that identification
by two or more studies indicates YSO  legitimacy, then that two-thirds
of the list is more  likely to be legitimate YSOs. One-third have four
or more YSO tags. 

In contrast, however, one-third of the YSOs/candidate YSOs  have just
one paper identifying them as a YSO (only $\sim$15\% of which have no
points in their SED). As can be seen in  Table~\ref{tab:ysotags}, the
majority of those single-tagged YSOs  ($\sim$500) come from the
aggregate of MYStIX studies 
\citep{2013ApJS..209...31P,2013ApJS..209...32B,2016ApJ...833..193R,2020AA...633A.155R},
with another $\sim$200 from \citet{2019AA...623A..25D}. The former is
a comprehensive multi-wavelength effort; the latter uses Gaia and
Chandra data. These studies are therefore unlikely to contain many 
spurious sources. Nonetheless, these single-tagged YSOs also often
seem to be outliers in a variety of optical and infrared color-color
and color-magnitude diagrams. Some of that comes from the fact that
those studies, particularly the MYStIX ensemble, is often reaching
fainter objects than most of the other work, and so the photometry may 
be more uncertain. Further study is needed.

Figure~\ref{fig:tagjhjk} shows the $JHK_{\rm s}$ color-color diagram
for the  YSO sample with the number of YSO tags indicated.   The
empirical zero-age main sequence (ZAMS) \citep{pecautmamajek} is shown
in the figure, along with the \citet{meyerttlocus} T~Tauri locus and a
reddening vector (following the reddening law from \citet{Indebetouw} 
and \citet{mathis}). The stars with the most YSO tags appear in this
diagram in a location consistent with slightly reddened ZAMS stars. 
This doesn't mean that they are likely contaminants, 
just that their NIR colors are affected by reddening but 
not affected by IR excesses (they could have IR excesses $>$2 \mum).  
Many of the most reddened stars are identified in just a few papers,
most commonly those reaching faint sources and using mid-IR or X-rays
to identify YSOs, both of which can more  easily penetrate dust than
some of the other YSO identification methods. Many of the stars that
have near-IR excesses (those near the T~Tauri locus and those reddened
from that locus) are YSOs identified based on IR excesses. Most of the
earliest papers (particularly those using pre-Gaia positions and
proper motions) find stars near where early type stars are expected to
fall in this diagram -- between 0 and 0.5 in $H-K_s$, and 0 and 0.5 
in $J-H$. A Gaia color-magnitude diagram of these stars has 
the most-tagged YSOs clustered about a $\sim$6 Myr isochrone, with
a long tail up to the brightest stars (identified most often in 
early position and proper motion studies), and considerable scatter
among the fainter, single-tagged sources. 

Figure~\ref{fig:ysotagssedpoints} looks at the correlation between 
number of papers identifying the target as a YSO and the number of
points in the SED. There does seem to be a correlation in that the
more papers there are identifying the source as young, the more points
there are in the SED, which makes sense if one assumes that the papers
that identify YSOs also publish additional photometry. However, as can
also be seen in Table~\ref{tab:literature}, that is not necessarily
the case. The most populated bins in  Fig.~\ref{fig:ysotagssedpoints}
are, in fact, YSOs tagged in just one paper, with $\sim$25-30 points
in their SED. Many of these stars come from MYStIX or
\citet{2019AA...623A..25D}, and their SEDs are populated by large or
all-sky catalogs.

\begin{deluxetable}{p{5.5cm}cccc}
\tablecaption{Numbers of YSO tags\tablenotemark{a} \label{tab:ysotags} }
\tablewidth{0pt}
\tablehead{\colhead{(1)} & \colhead{(2)} &\colhead{(3)} &\colhead{(4)} &\colhead{(5)} \\
\colhead{Reference} &  \colhead{{Total tagged YSO}}& 
\colhead{Total tagged YSO in $\geq$2} & \colhead{Total tagged YSO in $\geq$3} &\colhead{Total tagged YSO only once}}
\startdata
\citet{1997AAS..123...63V} (PPM) & 96 & 92 & 86 & 4 \\ 
\citet{2000AJ....120..333S} (H$\alpha$) & 74 & 65 & 60 & 9 \\ 
\citet{2002AA...395..499R} (X-rays) & 105 & 87 & 84 & 18 \\ 
\citet{2004ApJ...608..781D} (X-rays) & 884 & 871 & 814 & 13 \\ 
\citet{2005AA...430..941P} (optical) & 926 & 926 & 832 & 0 (403 dropped) \\ 
\citet{2006ChJAA...6..435Z}  (PPM) & 59 & 59 & 57 & 0 \\ 
\citet{2007MNRAS.374.1253A} (NIR)& 44 & 39 & 36& 5 \\ 
\citet{2007AJ....134.1368C} (PPM) & 147 & 110 & 100 & 37 \\ 
\citet{2007AA...462..123P} (opt spec; with disks) & 90 & 90 & 90 & 0 \\ 
\citet{2007AA...462..123P} (opt spec; members) & 332 & 288 & 256 & 44 \\ 
\citet{2010MNRAS.407.1170K} (MIR) & 221 & 155 & 139 & 65 \\ 
\citet{2012ApJ...747...51H} (periods) & 290 & 290 & 290 & 0 \\ 
MYStIX: \citet{2013ApJS..209...31P,2013ApJS..209...32B,2016ApJ...833..193R,2020AA...633A.155R} (X-rays+MIR) & 2145 & 1581 & 1117 & 564 \\ 
\citet{2013MNRAS.434..806B} (optical) & 661 & 651 & 644 & 10 \\ 
\citet{2013PhDT........36C} (X-rays) & 1303 & 1303 & 1008 & 0 (47 dropped) \\ 
\citet{2015MNRAS.453.1026K} (optical) & 195 & 153 & 123 & 42 \\ 
\citet{2019AA...623A.159P} (opt spec) & 616 & 595 & 562 & 21 \\ 
\citet{2019AA...623A..25D} (PPM) & 910 & 687 & 650 & 223 \\ 
\citet{2019MNRAS.486.2477W} (PPM) & 776 & 774 & 734 & 2 \\ 
\citet{2021ApJS..254...33K} (MIR, PPM, variab) & 466 & 434 & 276 & 31 \\ 
\citet{2022MNRAS.509.1664J} (PPM) & 439 & 429 & 426 & 10 \\ 
\citet{2022MNRAS.517.6191O} (variab) & 361 & 361 & 336 & 0 \\ 
\enddata
\tablenotetext{a}{Column (1): reference publishing a list of YSOs
(or YSO candidates), followed by a brief parenthetical reminder of their primary method of
identifying YSOs. Column (2): the number of sources published by this reference
as YSOs or YSO candidates within our region of interest. 
Column (3): the number of sources from this reference that
are also tagged as YSOs in any other reference discussed here (that is, they are tagged
as YSOs by a total of 2 or more references considered here). 
Column (4): the number of sources from this reference that
are also tagged as YSOs in 3 or more other references discussed here (that is, they are tagged
as YSOs by a total of 3 or more references considered here). 
Column (5): the number of sources from this reference that are only tagged 
as YSOs once (that is, only by this reference). Example: \citet{1997AAS..123...63V} used 
positions and proper motions to identify YSOs; they published 96 YSOs in our region, 92 of
which were also identified as YSOs by at least one other paper considered in our analysis, and 86 
of which were also identified as YSOs by at least two other papers considered here. 
Four of the YSOs in \citet{1997AAS..123...63V} are listed only in \citet{1997AAS..123...63V} 
and no other place investigated in our analysis as YSOs. See the text for discussion around 
dropping from consideration the YSOs tagged only by \citet{2005AA...430..941P} or 
\citet{2013PhDT........36C}.}
\end{deluxetable}

\begin{figure}[htb!]
\epsscale{0.6}
\plotone{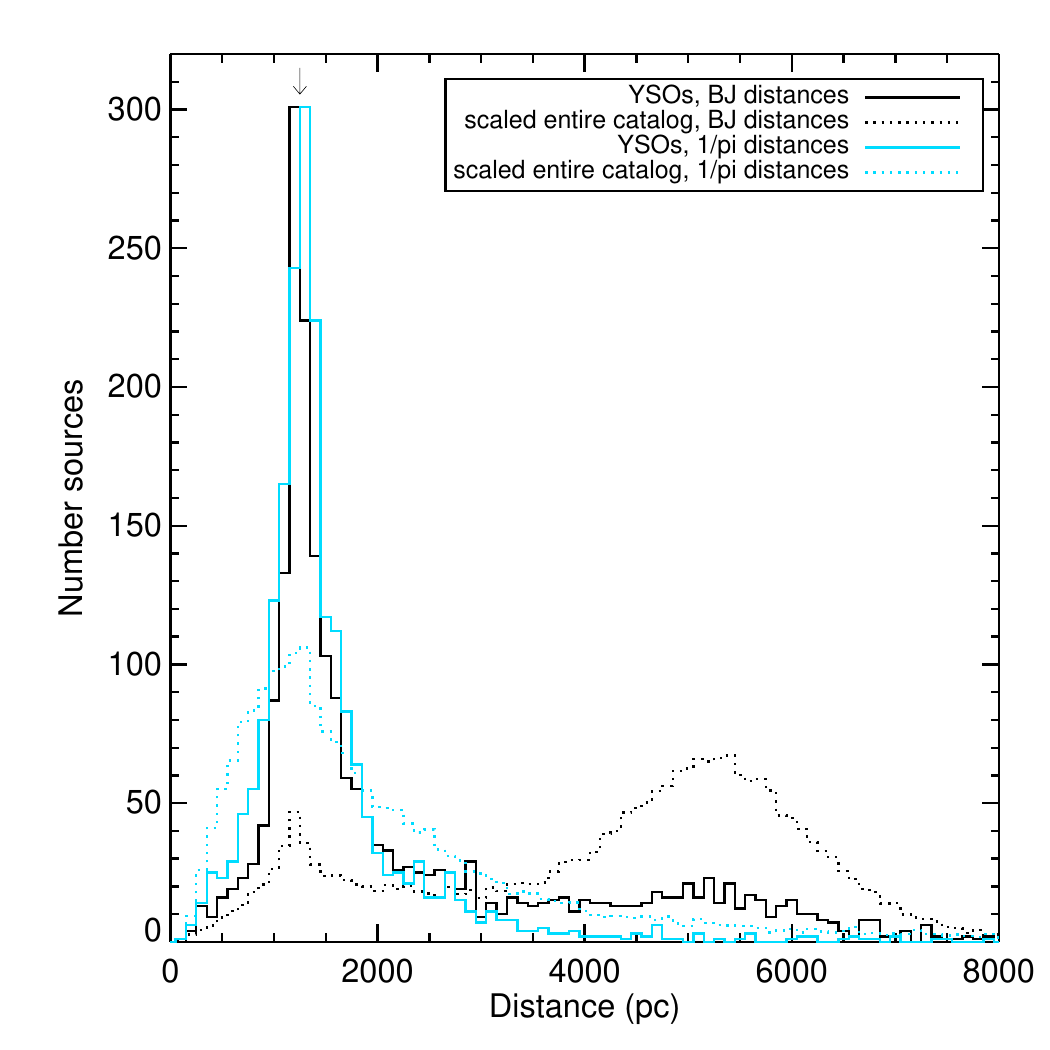}
\caption{Histogram of distances obtained from Gaia DR3, both from  
\citet{2021AJ....161..147B} (black lines) and simply inverting the
parallax (blue lines).  The solid lines are the YSOs; the dotted lines
are the entire catalog we assembled in this region, scaled as if it
had the same total number of sources as the YSO catalog (it actually
has $\sim$211,000 sources).  There is a long, flat tail out past 8000
pc. The arrow at the top is placed  at 1250 pc, the best available
distance estimate for the Lagoon Nebula.  The YSO catalog is strongly
biased towards  distances centered on $\sim$1200 pc, as expected if
they were all true members of the Lagoon.}
\label{fig:distances}
\end{figure}

\subsection{Distances}
\label{sec:distances}

Distances, even using Gaia data, can be complicated. Numerous papers discuss the
strengths and weaknesses of different Gaia-derived distances. We noted in a 
previous paper \citep{ic417paper} that particularly in the case of young stars, 
most of the Gaia distance estimates may not be terribly reliable much past 2 kpc. 
The Lagoon nebula is at $\sim$1250 pc \citep{2019AA...623A.159P}, so it should be
close enough for most if not all of the legitimate members, but separating them 
from the background could be difficult.

Figure~\ref{fig:distances} is a histogram of distances for the YSO
catalog, and for comparison, the entire assembled catalog, scaled as
if it was the same size as the YSO catalog.  Just two distance values
are shown -- the geometric distance as derived by \citet{2021AJ....161..147B},
and that obtained by inverting the (positive) parallax.  The YSO
catalog is strongly biased towards much closer sources than the 
ensemble distribution; the most common value is at $\sim$1250 pc, 
as expected, regardless of which distance one uses. 

However, the structure near $\sim$5200 pc is interesting. If that
$\sim$5200 pc peak is real, it presumably is a result of galactic
structure behind the Lagoon Nebula. In the ensemble catalog, the
Bailer-Jones distances are entirely responsible for that peak; the
distribution of YSO candidates is fairly flat between $\sim$3000 and
$\sim$6000 pc, and then falls off. The inverted parallaxes have no
such peak, and instead have a slightly broader distribution at
$\sim$1250 pc.  While there is a very long, flat tail, with a handful
of YSOs out past 8 kpc (using either distance estimate),
three-quarters of the YSOs with Bailer-Jones distances $>$4 kpc
(creating that bump) are tagged as YSOs in at most 2 articles
($\sim$45\% are tagged in just 1 paper).  If the objects with reported
distances $>$4,000 pc are really at those distances, they are
significantly less likely to be true members of the Lagoon SFR, but it
is unclear if those distances are accurate.  

We have included the distances in our final catalog, but have not 
used the distances to exclude YSOs from our final catalog. Future 
Gaia data releases may permit better distance-based exclusions.

\section{Reddening}
\label{sec:reddening}

In subsequent analyses, we will plot against \vmkz\ as a proxy for
mass, so we would like to be able to deredden stars reasonably
reliably. Reddening here is both patchy and occasionally substantial,
so this is not a small problem. Because of the rich literature, we
have spectral type estimates for 928 stars, biased towards
things that are likely YSOs, and from there we can attempt to explore
constraints on reddening.
Most of these spectral types (805) come from \citet{venuti2024}.
Another 86 come from \citet{2020AA...633A.155R}: 70 from their own
spectroscopy, and 16 OB stars for which they compiled literature
spectral types (their Table 5). Ten come from \citet{2007MNRAS.374.1253A}
and one from \citet{venuti2021}. For the remaining 26 stars, we adopted
the spectral type listed in SIMBAD \citep{2000AAS..143....9W}; these
originate in \citet{2012AA...546A...9P} (15 stars),
\citet{1939AnBos...5E...1W} (5), \citet{1961ApJ...133.1081W} (2), and
\citet{1989ApJS...69..301G}, \citet{2014ApJS..211...10S},
\citet{2020AA...633A.155R}, and \citet{2022AA...657A.131M} (one each).
The source of each adopted spectral type is given in the catalog
(column sptysrc; see Table~\ref{tab:bigdata}).

\begin{figure}[htb!]
\epsscale{0.5}
\plotone{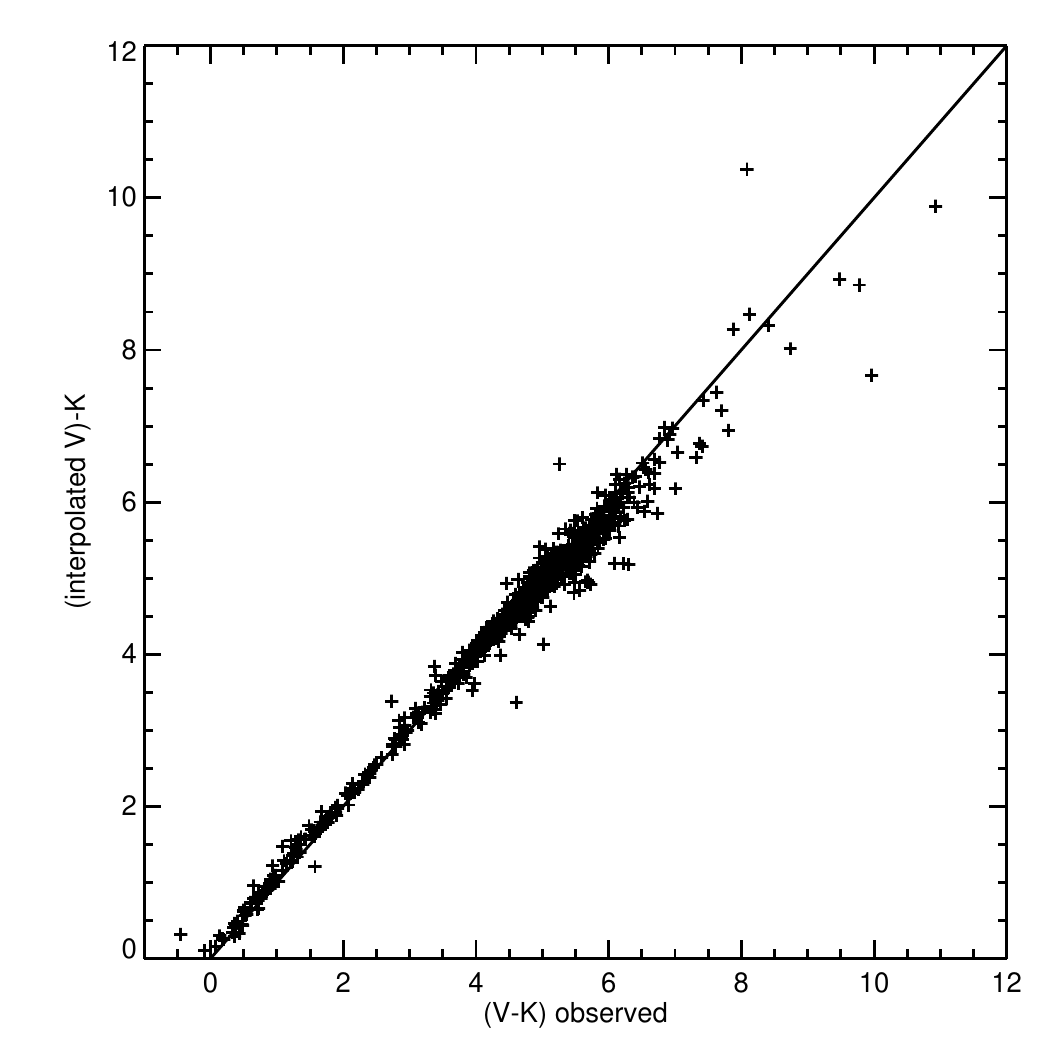}
\caption{Comparison of \vmk\ for stars where both $V$ and \ks\ are
observed explicitly ($x$-axis), and where \ks\ is observed, but where
$V$ is interpolated from the points near 0.55 \mum\ in the SED
($y$-axis). A 1:1 line is overplotted. There is, in general, good
agreement. We take this as confirmation that is is valid to
interpolate values for $V$  in cases where there is no measured $V$.}
\label{fig:comparevkobssed}
\end{figure}

Finding a measured $V$ in the literature is relatively rare, but since
these SEDs are in general very well defined in the optical, we can
with reasonable confidence interpolate a value for $V$. Therefore, as
in other of our papers \citep[\eg,][]{ucllcc}, we interpolated a value
for $V$ using points near 0.55 \mum, omitting H$\alpha$. To check the accuracy of this,
for the stars where we do have measurements at $V$, we can compare the
interpolated and measured values of $V$, and find that they are in
general in good agreement; see Figure~\ref{fig:comparevkobssed}.   We
take this as confirmation that is is valid to interpolate values for
$V$ in cases where there is no measured $V$.

\begin{figure}[htb!]
\epsscale{0.5}
\plotone{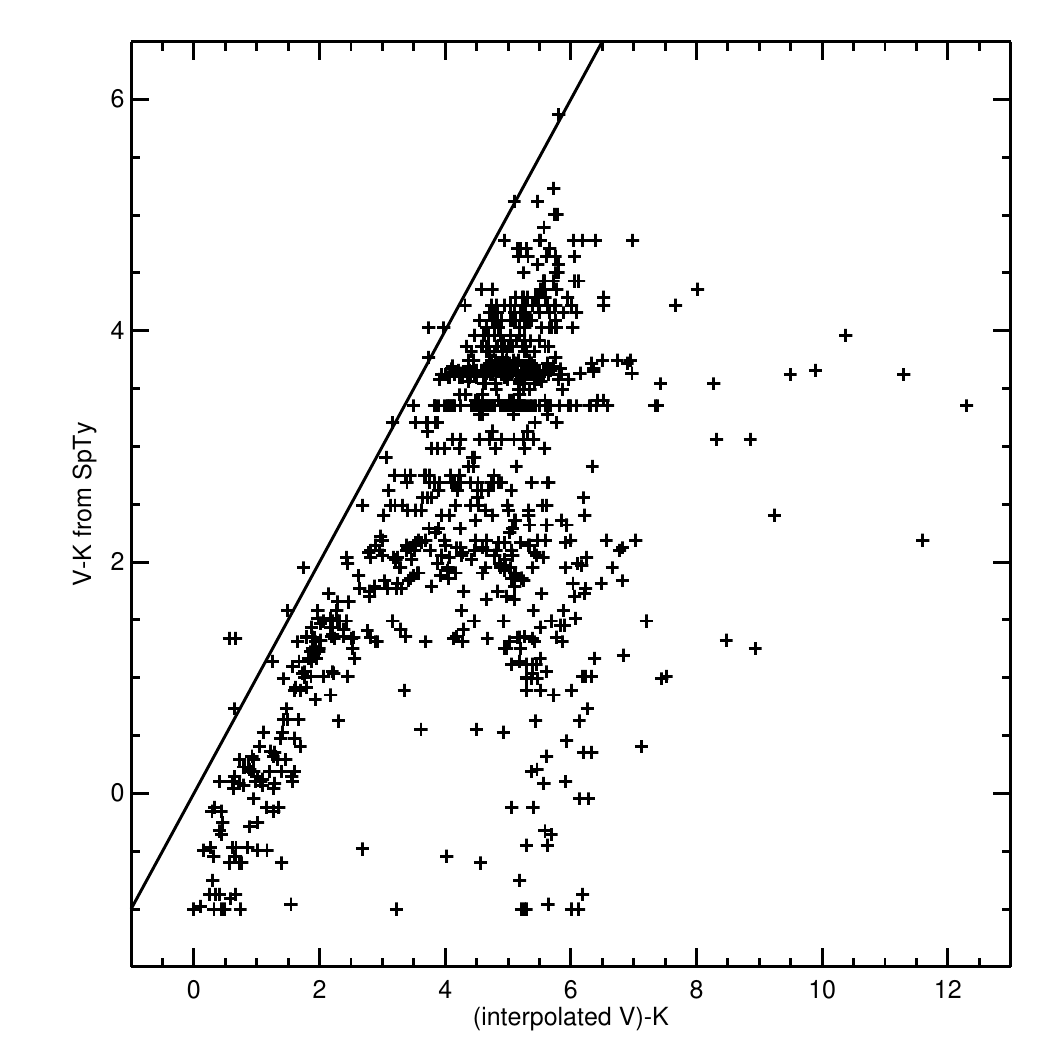}
\caption{Comparison of \vmk\ for stars where $V$ is interpolated from
the points near 0.55 \mum\ in the SED (and \ks\ is observed) and the
\vmkz\ expected based on the spectral type. A 1:1 line is overplotted.
There is typically at least
some reddening, and sometimes substantial reddening.}
\label{fig:comparevkobsspty}
\end{figure}

Next, we know that there will be a range of reddening corrections. 
Figure~\ref{fig:comparevkobsspty} compares the \vmk\ values (where
\ks\  is observed, but where $V$ is interpolated from the points near
0.55 \mum\ in the SED) to the expected \vmkz\ values  given the
spectral type or the spectral type range available in the  master
catalog assembled here, and the colors from \citet{pecautmamajek}.  
The quantization arises because the types are quantized, and thus the
colors are as well.  For the bulk of the distribution, there is
roughly a 1:1 relationship, though there is a net offset due, 
presumably, to an overall extinction contribution. There is clearly an
additional population for which there is substantial reddening, where
the observed \vmk\ is very different from the expected \vmkz\ based on
spectral type (or spectral type estimate). Stars with and without IR
excesses are scattered throughout this diagram; the `hook' structure 
on the right contains both stars with and without IR excesses, as does
the structure that runs parallel to the 1:1 relationship.

\begin{figure}[htb!]
\epsscale{0.5}
\plotone{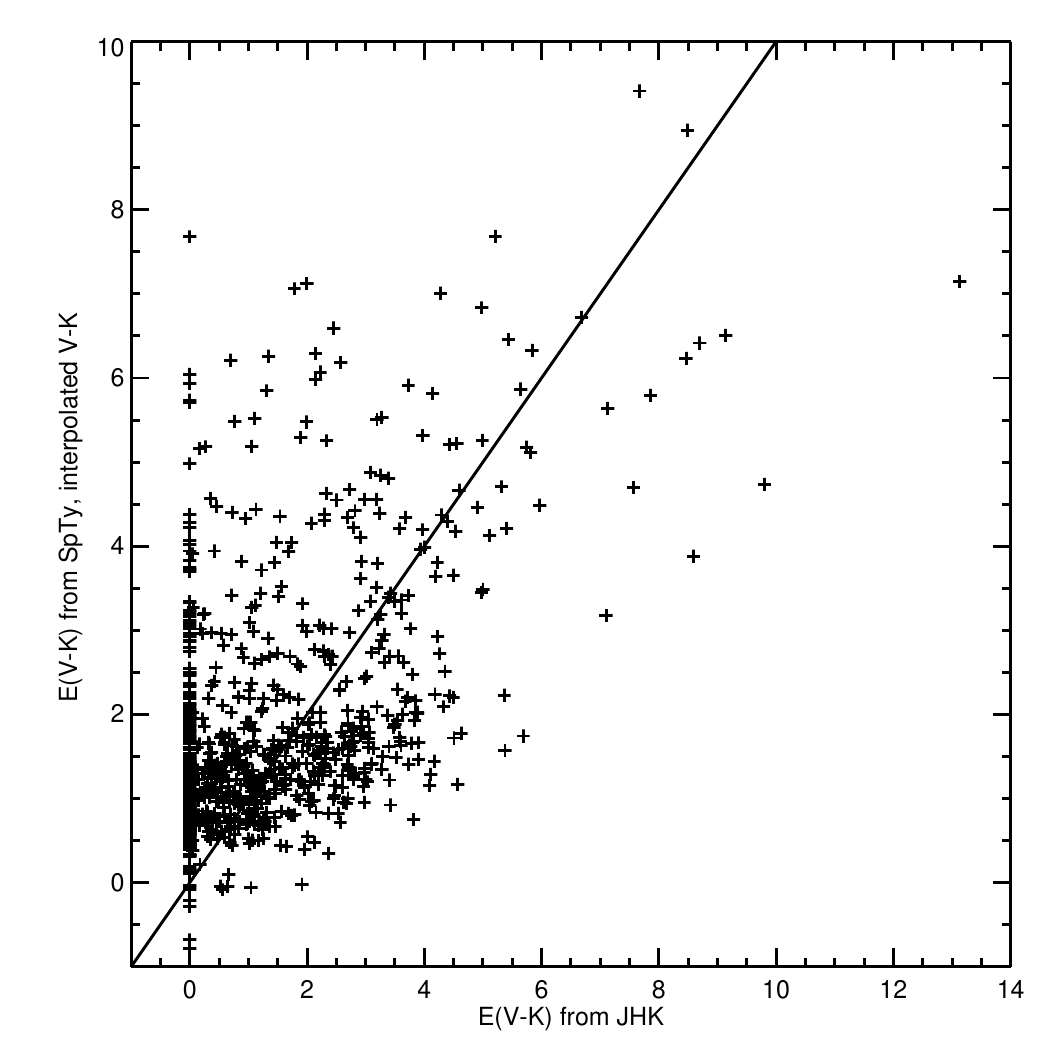}
\caption{Comparison of $E$\vmk\ for stars where the reddening is estimated
from the placement in the $JHK_s$ diagram ($x$-axis) and the $E$\vmk\
calculated from the observed \ks, interpolated $V$, and colors assumed  
from the spectral type.  A 1:1 line is overplotted.
There is somewhat of a correlation, but one would like something better.}
\label{fig:comparejhksptyav}
\end{figure}

We would like to be able to work with stars for which no spectral
type estimate is available. We can also use the placement of the stars
on a $J-H$ vs.\ $H-K_s$ diagram to estimate reddening, by shifting
them back along the reddening law derived by \citet{Indebetouw} to
the expected $JHK_s$ colors for young stars from \citet{pecautmamajek}
or the T~Tauri locus from \citet{meyerttlocus}, then converting to
$E(V-K_s)$ via $A_K = 0.114 A_V$ \citep{ccm}.  The difficulty is that,
in this region, sometimes this procedure results in a very different reddening
estimate than can be obtained by a comparison to spectral types.
Figure~\ref{fig:comparejhksptyav} compares the $E$\vmk\ obtained from
the placement in the $JHK_s$ diagram and that from the observed \ks,
interpolated $V$, and colors assumed from the spectral type. The
values are not completely uncorrelated, but we expected there to be
more of a correlation than this. There are sometimes objects where
the spectral type mandates much larger corrections than the placement
in the $JHK_s$ diagram suggests, which we expected, since disk excess
can affect $JHK_s$, but there are also some stars where the
$JHK_s$ placement would be bluer (suggests a much larger correction)
than the spectral type mandates.

\begin{figure}[htb!]
\plotone{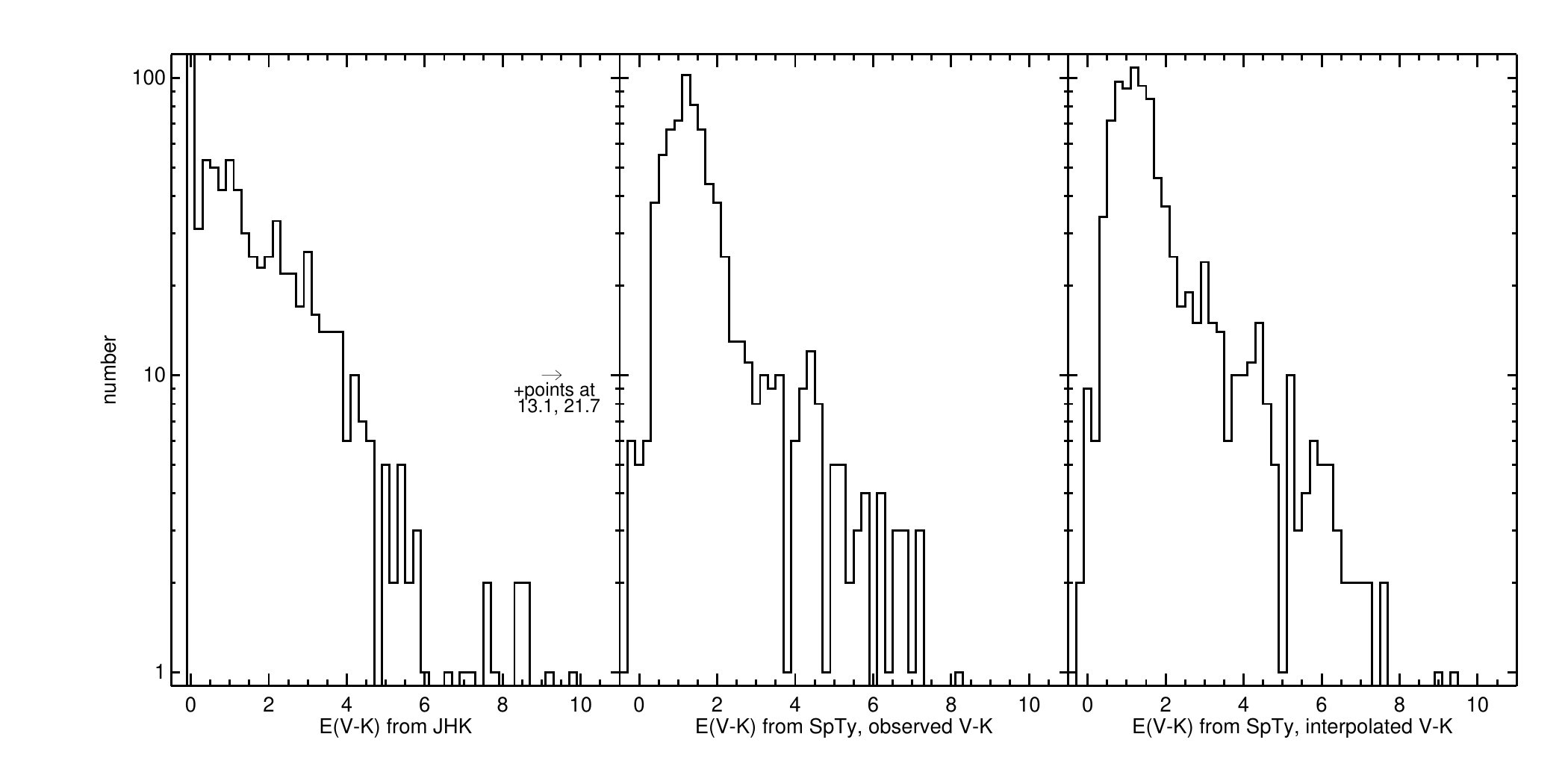}
\caption{Comparison of $E$\vmk\ for stars where the reddening is
estimated from the placement in the $JHK_s$ diagram (left), the
$E$\vmk\ calculated from the observed $V$ and \ks\ (middle) and colors
assumed   from the spectral type, and  the $E$\vmk\ calculated from
the interpolated $V$ and colors assumed   from the spectral type
(right). The left hand plot $x$-axis is truncated for ease of
comparison with the other two plots; there are two stars that  would
be far to the right in this histogram, at $E$\vmk=13.1 and 21.7. See
the text for more discussion.}
\label{fig:evkhistos}
\end{figure}

Another way to compare these methods can be found in 
Figure~\ref{fig:evkhistos}. The left histogram has the $E$\vmk\ for
stars where the reddening is estimated from the placement in the
$JHK_s$ diagram; this can be done for a very large fraction of the
entire catalog, but here is only shown for the stars with spectral
types.  There is no clear peak to this distribution, unless one counts
the peak at 0, which is what happens when the routine fails to
converge at all, or fails to converge on a positive \av. The other two
panels are $E$\vmk\ for stars where the reddening comes from a
comparison to expected colors for their spectral type (or spectral
type range).  The distribution for those where both $V$ and \ks\ are
observed,  the middle panel, has a mode at $E$\vmk$\sim$1.25. The
distribution in which \ks\ is observed but $V$ is interpolated from
the SED is about the same, but may be slightly higher. The median for 
$E$\vmk\ (observed $V$) is $\sim$1.39 and the median for $E$\vmk\
(interpolated $V$) is $\sim$1.40. Given all the uncertainties here,
these are probably identical.

Note that there are two stars that would be far to the right in the
$JHK_s$ histogram, at $E$\vmk=13.1 and 21.7. These stars are
J180419.08-241545.1 and J180422.23-242215.7, respectively.  The latter,
J180422.23-242215.7, is slightly easier to understand; its spectral
type is M5-M7III from \citet{2020AA...633A.155R}. Its SED suggests it
is very embedded, with a substantial IR excess; the near-IR could
be affected by IR excess, all of which likely explains why the $E$\vmk\ from
$JHK_s$ is so large.  
The former, J180419.08-241545.1, is confusing. \citet{venuti2024}
says it is an F7.5 star. It has very different $JHK_s$ 
measurements from 2MASS and UKIDSS; 2MASS seems most consistent with
the rest of the SED. The optical measurements from  more than one
catalog (PanSTARRS, VVV, VPHAS, Gaia, \citet{2000AJ....120..333S})
all agree with each other and pull away from what would be a
Wein-shaped extension from the 2MASS peak of the SED. Assuming that
\ks\ is on the photosphere (which may not be a good assumption), there
appears to be an IR excess at 3-4 \mum; if 3-4 \mum\ is instead on the
photosphere, there is an IR excess by 8 \mum\ that extends to at least
24 \mum, the longest wavelength point we have for this star.  There is an
apparently MYStIX-only source within 3\arcsec, so source confusion 
may be important too. More information is needed to understand what is
happening with this object. 

In general, for stars in this catalog that need to be dereddened for
this or our other future analysis, we took the spectral type
correction where we could, and fell back to that from the $JHK_s$
diagram when we had no other choice.

\section{YSO Reliability and Completeness}
\label{sec:reliability}

Because of the wildly inhomogeneous constituent catalogs that we have 
merged to create this final catalog, it's very hard if not impossible to 
assess completeness in a statistically rigorous fashion in terms of
brightness, wavelength, or even spatial coverage.  Only some of the 
catalogs we merged cover our entire field. However, our catalog is 
complete in terms of published literature between $\sim$1997 and 
$\sim$2024, which is what we set out to do. It combines data over 
$\sim$4 orders of magnitude in wavelength to create well-populated 
SEDs.

Specifically because we have so many points delineating the SEDs
after all the catalog merging, we had hoped that we would be able to
make a better assessment in many cases of the YSO nature of these
sources. For example, if a source was identified as a YSO based only
on, say, near- and mid-IR (2MASS+IRAC) data, contamination from,
e.g., background asymptotic giant branch (AGB) stars or star-forming
background galaxies is far more likely than if we have optical+NIR+MIR
data to constrain the SED. 

It turns out, however, that (as discussed above) discarding the YSO
candidates that  were identified in only \citet{2005AA...430..941P} or
\citet{2013PhDT........36C} likely was among the most significant
things we could have done to clean the list of the most questionable
YSOs.  

The simplest filter that could be easily imposed to make the catalog 
more reliable but less complete is to restrict the list to have at
least 2 or 3 YSO tags before accepting it as a YSO.
Table~\ref{tab:subsamples} shows the sample sizes for several
different subsets of the data. If one drops the YSOs from the list
that have only one YSO tag, then one loses 36\% of the catalog (1098
sources). If one drops those with just one or two YSO tags, then one
loses 1098+661 sources, 58\% of the sample. These that would be so
dropped include, e.g., X-ray sources without any other counterparts,
sources that have well-populated but stellar SEDs (no IR excess), and
sources with SEDs that honestly are quite strange looking (may be
stars, may be something else), all of which sounds like it could be 
appropriate. However, it also drops several hundred that have very 
well-populated stellar SEDs with clear IR excesses, including 
some IRAS sources, suggesting that such a filter would also 
discard obvious YSOs.

\begin{deluxetable}{p{5.5cm}cp{5.5cm}}
\tablecaption{Subsamples and Fractions \label{tab:subsamples} }
\tablewidth{0pt}
\tablehead{
\colhead{Description} &  \colhead{Number}& 
\colhead{Notes} }
\startdata
YSOs in catalog & 3019 \\
YSOs tagged in just 1 paper & 1098 & 36\% of catalog \\
YSOs tagged in just 2 papers & 661 & 22\% of catalog\\
YSOs tagged in just 3 papers & 253 & 8\% of catalog \\
YSOs tagged in just 4 papers & 118 & 4\% of catalog \\
YSOs tagged in just 5 papers & 144 & 5\% of catalog \\
YSOs tagged in 6 or more papers & 745 & 25\% of catalog\\
\hline
SED points $\leq$10 & 371 & 12\% of catalog\\
SED points $>$10 and $\leq$20 & 395 & 13\% of catalog \\
SED points $>$20 and $\leq$30 & 867 & 29\% of catalog\\
SED points $>$30 and $\leq$40 & 899 & 30\% of catalog\\
SED points $>$40 & 487 & 16\% of catalog\\
\hline
IR excess: no data & 537 & 18\% of catalog\\
IR excess: unambiguous no disk (1.x or ``no'') & 1097 & 36\% of catalog\\
IR excess: possibly no disk (1.x or ``no?'') & 160 & 5\% of catalog\\
IR excess: 1 band (3.x)  & 193 & 6\% of catalog\\
IR excess: possibly disk (4.x or ``yes?'') & 252 & 8\% of catalog \\
IR excess: unambiguous disk (5.x or ``yes'') & 780 & 26\% of catalog; (252+780)/3019 is 34\% of catalog\\
\hline
Bailer-Jones distance exists & 2202 & 73\% of catalog \\
Bailer-Jones distance $<$2 kpc & 1406 & 64\% of those with distances\\
Bailer-Jones distance $<$2 kpc \& 3 YSO tags & 914 & 42\% of those with distances\\
Bailer-Jones distance $<$2 kpc \& IR excess 4 or 5 & 479 & 22\% of those with distances\\
Bailer-Jones distance $<$2 kpc, 3 YSO tags, \&  IR excess 4 or 5 & 358 & 16\% of those with distances\\
\hline
YSOs selected via PPM & 1582 & 52\% of sample; 1029 (65\% of the PPM sample) have more than 3 YSO tags \\
YSOs selected via H$\alpha$ & 74 & 2\% of sample; 60 (81\% of the H$\alpha$ sample) have more than 3 YSO tags\\
YSOs selected via other optical methods & 1281 & 42\% of sample; 1010 (79\% of the `other optical' sample) have more than 3 YSO tags\\
YSOs selected via X-rays & 2284 & 76\% of sample; 1169 (51\% of the X-ray sample) have more than 3 YSO tags\\
YSOs selected via IR & 2280 & 76\% of sample; 1131 (50\% of the IR sample) have more than 3 YSO tags\\
YSOs selected via variability & 815 & 27\% of sample; 600 (74\% of the variability sample) have more than 3 YSO tags\\
\enddata
\end{deluxetable}

There are so many catalogs in this merged collection that
adding a requirement for a certain minimum number of points 
in the SED is not a very stringent requirement, but it does exclude
the X-ray-only objects. Only 12\% of the catalog has 10
or fewer points in the SED, leaving 88\% of the catalog intact. 

Assuming a goal of having a very reliable catalog of YSOs, one could
also impose a requirement of having an IR excess. This necessarily 
excludes those YSOs which no longer have disks, but it does ensure that
the YSOs still in the catalog are more likely to be young. If one selects
only the most reliable IR excess stars, or the likely IR excess stars, 
then one retains just 34\% of the full catalog. If one requires detection at
any band redward of 2 \mum, such that any estimate of the photosphere is 
possible, regardless of whether or not there is a significant excess,
then 18\% of the catalog is dropped. 

Distances are complicated, as discussed above, but for the sake of
argument, if one requires a distance from
\citet{2021AJ....161..147B},  then one is left with 73\% of the
catalog, necessarily the less embedded of the sample (though it also
drops the X-ray sources without counterparts, of course). A further
restriction down to $<$2 kpc leaves 64\% of those with distances (47\%
of the whole catalog).  Including additional restrictions of at least
3 YSO tags and an IR excess leaves just 358 YSOs, but at least
nominally, they could be the most reliable YSOs.

YSOs selected by PPM prior to Gaia tend to be the optically brighter 
and higher mass stars. There are not many YSOs selected via H$\alpha$ 
but those also tend to have an IR excess. MYStIX used combinations 
of X-rays and IR excess to identify YSOs, and they are largely responsible
for the large numbers of X-ray and IR-selected YSOs in 
Table~\ref{tab:subsamples}. Unsurprisingly, variability seems to 
be a good way to find YSOs; most of the YSOs identified via variability 
are also tagged as young via other methods. 

\section{Final Thoughts}
\label{sec:theend}

We have assembled a reasonably complete, cleaned catalog of literature-identified
YSOs in the Lagoon Nebula. Future surveys such as the Roman Galactic Plane
Survey \citep{2025arXiv251107494G} will observe this region with high
spatial resolution and sensitivity, and the Lagoon Nebula is also a prime 
target of interest for time-domain young star science with Vera C. Rubin 
Observatory's Legacy Survey of Space and Time (Rubin LSST; \citealp{bonito2023}); 
rather than being limited to analysis 
using only the data from within a given survey, this catalog will enable 
future work to be more easily linked to prior analyses. 

This catalog enables us to better exploit the time series data from this 
region and investigate periodic variability in a companion paper
\citep{lagoonrotation}. The catalog and disk classifications 
compiled here are also being used in another work by our team (Venuti et al., 
in preparation) that aims to determine how the inner disk properties and their 
evolutionary timescales in the Lagoon Nebula are impacted by the presence of 
stellar companions, with information on stellar multiplicity being extracted 
from Gaia astrometric and Gemini speckle imaging data.



\begin{acknowledgments}
Tremendous thanks to the NITARP (https://nitarp.ipac.caltech.edu) 
2019 IDYL team (D.~Strasburger, M.~Bechtel, D.~Mattern, and R.~Swanson and their students) 
for their work on an early version of this catalog.
\begin{itemize}
\item This research has made use of the NASA/ IPAC Infrared Science Archive (IRSA), which is operated by the Jet Propulsion Laboratory, California Institute of Technology, under contract with the National Aeronautics and Space Administration.
\item This publication makes use of data products from the Wide-field Infrared Survey Explorer (WISE), which is a joint project of the University of California, Los Angeles, and the Jet Propulsion Laboratory/California Institute of Technology, funded by the National Aeronautics and Space Administration.
\item This work is based in part on archival data obtained with the Spitzer Space Telescope, which was operated by the Jet Propulsion Laboratory, California Institute of Technology under a contract with NASA.
\item This publication makes use of data products from the Two Micron All Sky Survey (2MASS), which is a joint project of the University of Massachusetts and the Infrared Processing and Analysis Center/California Institute of Technology, funded by the National Aeronautics and Space Administration and the National Science Foundation.
\item Herschel is an ESA space observatory with science instruments provided by European-led Principal Investigator consortia and with important participation from NASA.
\item This research made use of data products from the Midcourse Space Experiment (MSX). Processing of the data was funded by the Ballistic Missile Defense Organization with additional support from NASA Office of Space Science. 
\item The Pan-STARRS1 Surveys (PS1) and the PS1 public science archive have been made possible through contributions by the Institute for Astronomy, the University of Hawaii, the Pan-STARRS Project Office, the Max-Planck Society and its participating institutes, the Max Planck Institute for Astronomy, Heidelberg and the Max Planck Institute for Extraterrestrial Physics, Garching, The Johns Hopkins University, Durham University, the University of Edinburgh, the Queen's University Belfast, the Harvard-Smithsonian Center for Astrophysics, the Las Cumbres Observatory Global Telescope Network Incorporated, the National Central University of Taiwan, the Space Telescope Science Institute, the National Aeronautics and Space Administration under Grant No. NNX08AR22G issued through the Planetary Science Division of the NASA Science Mission Directorate, the National Science Foundation Grant No. AST–1238877, the University of Maryland, Eotvos Lorand University (ELTE), the Los Alamos National Laboratory, and the Gordon and Betty Moore Foundation.
\item This work is based in part on data obtained as part of the UKIRT Infrared Deep Sky Survey (UKIDSS).
\item This work has made use of data from the European Space Agency (ESA) mission
{\it Gaia} (\url{https://www.cosmos.esa.int/gaia}), processed by the {\it Gaia}
Data Processing and Analysis Consortium (DPAC,
\url{https://www.cosmos.esa.int/web/gaia/dpac/consortium}). Funding for the DPAC
has been provided by national institutions, in particular the institutions
participating in the {\it Gaia} Multilateral Agreement.
\item The VVV is based on data products from observations made with ESO
    Telescopes at the La Silla Paranal Observatory under programme ID
    179.B-2002
\item The VPHAS catalogue is based on data products from observations
made with ESO Telescopes at the La Silla Paranal Observatory under public survey programme ID, 177.D-3023
\item This research has made use of the SIMBAD database, operated at CDS,
Strasbourg, France \citep{2000AAS..143....9W}.
\item This work was partially supported by the National Aeronautics 
and Space Administration (NASA) under grant No. 80NSSC21K0633 
issued through the NNH20ZDA001N Astrophysics Data Analysis Program (ADAP).
\item Generative AI tools were used in preparing this work. Gemini (Google) was used 
to suggest rewording of individual paragraphs. Claude (Anthropic; Claude Opus 5.5), 
used through the Claude desktop app (Cowork), was used to debug and cross-check the 
IDL code used to assemble the catalog, to check the catalog against the pipeline 
outputs, to format the machine-readable table, to locate bibliographic records, 
and to draft or suggest wording for parts of the text (e.g., the provenance of 
literature spectral types and descriptions of catalog columns). All AI-assisted 
text, code changes, and table contents were reviewed, verified, and revised 
by the authors, who take full responsibility for the content of this paper.
\end{itemize}
\end{acknowledgments}

\begin{contribution}
Author Rebull worked with her NITARP
(https://nitarp.ipac.caltech.edu) 2019 team to assemble an initial
version of this catalog. Author Rebull then continued to redo, re-merge,
re-work, and refine the catalog several times, then did the analysis
and wrote the paper.  Co-authors Cody and Venuti helped resolve source
confusion issues with the catalogs used in their Lagoon papers, and
commented on the manuscript. 
\end{contribution}

\facility{IRSA}

\bibliography{newbiblio}{}
\bibliographystyle{aasjournalv7}

\appendix
\section{Data Table}
\label{app:datatable}

The full primary data table appears as a machine-readable version 
in the electronic version of the journal and in more than one format in
the delivery to IRSA. The table here describes the form and content of
the data table. 

\startlongtable
\begin{deluxetable}{cp{10cm}}
\tabletypesize{\scriptsize}
\tablecaption{Contents of Table: Literature YSOs and YSO Candidates in the Lagoon Nebula \tablenotemark{a}\label{tab:bigdata}}
\tablewidth{0pt}
\tablehead{\colhead{Column} & \colhead{Contents}}
\startdata
\cutinhead{Primary Identifications }
cat\_num & Position-based catalog number -- assembled here to be compliant with IAU nomenclature rules, based on the best J2000 RA and Dec we have for the object. All names should start with `J' when used in the text (as per \citet{nedbestpractices}), but are not listed as such here just for space considerations.  \\
name & Alternate name (from Simbad) \\
ra & Right ascension (J2000; decimal degrees)\\
dec & Declination (J2000; decimal degrees)\\
ysolit & All the YSO tags from the literature, separated by semicolons (;). Strings should largely be self-
explanatory; an additional \# means that the corresponding reference assigned it a number, implying it was 
a YSO, e.g., ``spicy\#'' means that the SPICY catalog gave it a number, so by extension that means SPICY tagged
it as a YSO. \\
\cutinhead{Basic High-Level Descriptors }
ysotags & Number of references tagging this star as a YSO.\\
npts\_sed & Number of points in the SED assembled here.\\
longestwave & Wavelength in microns of the longest wavelength detection here.\\
sedclass & SED class of this star, where the class comes from a fit to all detections (not limits) between
2 and 25 \mum.\\
sedslope & SED slope for this star, which is a fit to all detections (not limits) between 2 and 25 \mum.\\
finaldisk1 & Final primary disk category (see the text for more information). \\
finaldisk2  & Final secondary disk category, if any (subgroups defined in Section~\ref{sec:disks}; `?' = possibly; multiple subgroups separated by `/') \\
finaldiskcode & Final disk code: 0 = no data past \ks\ (noinfo); 1 = no IR excess (no); 1.1 = no significant excess, but all IRAC bands offset slightly above the photosphere relative to \ks\ (offset); 1.2 = as 1.1, but only IRAC-3 and IRAC-4 offset (offseti3i4); 2 = probably no excess (no?); 3 = significant excess in the longest-wavelength band only (1band); 3.1 = excess significant in one band, with the adjacent shorter-wavelength band also offset (1+band); 4 = likely excess (yes?); 5 = unambiguous excess (yes). \\
spty & Adopted spectral type \\
sptysrc & Origin of adopted spectral type (see Section~\ref{sec:reddening})\\
\cutinhead{Optical Photometry \& Associated Information}
Umag & $U$ magnitude (Vega mag; all errors taken to be 0.1 mag)\\
Usrc & Origin of $U$ magnitude (string that should be self-explanatory)\\
Bmag & $B$ magnitude (Vega mag; all errors taken to be 0.1 mag)\\
Bsrc & Origin of $B$ magnitude (string that should be self-explanatory)\\
Vmag & $V$ magnitude (Vega mag; all errors taken to be 0.1 mag)\\
Vsrc & Origin of $V$ magnitude (string that should be self-explanatory)\\
Rcmag & $R_c$ magnitude (Vega mag; all errors taken to be 0.1 mag)\\
Rcsrc & Origin of $R_c$ magnitude (string that should be self-explanatory)\\
Icmag & $I_c$ magnitude (Vega mag; all errors taken to be 0.1 mag)\\
Icsrc & Origin of $I_c$ magnitude (string that should be self-explanatory)\\
Gaia3ID & Identifier from Gaia DR3 catalog\\
gaia3gmag & Gaia DR3 $G$ magnitude (Vega mag) \\
gaia3gmerr & Gaia DR3 $G$ magnitude error (Vega mag)\\
gaia3bpmag & Gaia DR3 $G_{BP}$ magnitude (Vega mag) \\
gaia3bpmerr & Gaia DR3 $G_{BP}$ magnitude error (Vega mag)\\
gaia3rpmag & Gaia DR3 $G_{RP}$ magnitude (Vega mag) \\
gaia3rpmerr & Gaia DR3 $G_{RP}$ magnitude error (Vega mag)\\
gaia3plx & Gaia DR3 parallax (mas) \\
gaia3eplx & Gaia DR3 error on parallax (mas) \\
gaia3dist & Gaia DR3 distance (pc) \\
gaia3bjdist & Geometric distance ($r_{\rm geo}$) from \citet{2021AJ....161..147B}, based on Gaia EDR3, in pc\\
gaia3bjdistup & 84th percentile of the geometric distance from \citet{2021AJ....161..147B}, in pc\\
gaia3bjdistdwn & 16th percentile of the geometric distance from \citet{2021AJ....161..147B}, in pc\\
PanID & Identifier from PanSTARRS catalog\\
pangmag & PanSTARRS $g$ magnitude (AB mag)\\
pangmerr & PanSTARRS $g$ magnitude error (AB mag); 0.3 mag where assigned (see text)\\
panrmag & PanSTARRS $r$ magnitude (AB mag)\\
panrmerr & PanSTARRS $r$ magnitude error (AB mag); 0.3 mag where assigned (see text)\\
panimag & PanSTARRS $i$ magnitude (AB mag)\\
panimerr & PanSTARRS $i$ magnitude error (AB mag); 0.3 mag where assigned (see text)\\
panzmag & PanSTARRS $z$ magnitude (AB mag)\\
panzmerr & PanSTARRS $z$ magnitude error (AB mag); 0.3 mag where assigned (see text)\\
panymag & PanSTARRS $y$ magnitude (AB mag)\\
panymerr & PanSTARRS $y$ magnitude error (AB mag); 0.3 mag where assigned (see text)\\
VPHASnum & identifier from VPHAS catalog\\
vphasumag & VPHAS $u$ magnitude (Vega mag)\\
vphasumerr & VPHAS $u$ magnitude error (Vega mag); 0.3 mag where assigned (see text)\\
vphasgmag & VPHAS $g$ magnitude (Vega mag)\\
vphasgmerr & VPHAS $g$ magnitude error (Vega mag); 0.3 mag where assigned (see text)\\
vphasrmag & VPHAS $r$ magnitude (Vega mag)\\
vphasrmerr & VPHAS $r$ magnitude error (Vega mag); 0.3 mag where assigned (see text)\\
vphasimag & VPHAS $i$ magnitude (Vega mag)\\
vphasimerr & VPHAS $i$ magnitude error (Vega mag); 0.3 mag where assigned (see text)\\
vphashamag & VPHAS H$\alpha$ magnitude (Vega mag)\\
vphashamerr & VPHAS H$\alpha$ magnitude error (Vega mag); 0.3 mag where assigned (see text)\\
bellnum & Identifier from \citet{2013MNRAS.434..806B}\\
bellumag & \citet{2013MNRAS.434..806B} $U$ magnitude (AB mag)\\
bellumerr & \citet{2013MNRAS.434..806B} $U$ magnitude error (AB mag)\\
bellgmag & \citet{2013MNRAS.434..806B} $g$ magnitude (AB mag)\\
bellgmerr & \citet{2013MNRAS.434..806B} $g$ magnitude error (AB mag)\\
bellrmag & \citet{2013MNRAS.434..806B} $r$ magnitude (AB mag)\\
bellrmerr & \citet{2013MNRAS.434..806B} $r$ magnitude error (AB mag)\\
bellimag & \citet{2013MNRAS.434..806B} $i$ magnitude (AB mag)\\
bellimerr & \citet{2013MNRAS.434..806B} $i$ magnitude error (AB mag)\\
bellzmag & \citet{2013MNRAS.434..806B} $Z$ magnitude (AB mag)\\
bellzmerr & \citet{2013MNRAS.434..806B} $Z$ magnitude error (AB mag)\\
bellmemb & Set to 1 if \citet{2013MNRAS.434..806B} indicates it is a member.\\
venuti21num & Identifier from \citet{venuti2021}.\\
venuti21spty  & Spectral type from \citet{venuti2021}.\\
venutiumag	& $u$ magnitude from \citet{venuti2024} or \citet{venuti2021} (AB mag)\\
venutiumerr	& $u$ magnitude error (assigned, 0.15 mag) (AB mag)\\
venutigmag	& $g$ magnitude from \citet{venuti2024} or \citet{venuti2021} (AB mag)\\
venutigmerr & $g$ magnitude error (assigned, 0.15 mag) (AB mag)\\
venutirmag	& $r$ magnitude from \citet{venuti2024} or \citet{venuti2021} (AB mag)\\
venutirmerr	& $r$ magnitude error (assigned, 0.15 mag) (AB mag)\\
venutiimag	& $i$ magnitude from \citet{venuti2024} or \citet{venuti2021} (AB mag)\\
venutiimerr & $i$ magnitude error (assigned, 0.15 mag) (AB mag)\\
venutihamag	 & H$\alpha$ magnitude from \citet{venuti2024} or \citet{venuti2021} (AB mag)\\
venutihamerr& H$\alpha$ magnitude error (assigned, 0.15 mag) (AB mag)\\
sungnum & Identifier from \citet{2000AJ....120..333S} (optical data from this paper are included
in the $UBVRI$ columns above).\\
sungysoflag & Set to 1 if \citet{2000AJ....120..333S} indicates it is a YSO \\
\cutinhead{NIR Photometry \& Associated Information}
bestjmag	& Best available $J$ magnitude (Vega mags)\\
bestjmerr	& Best available $J$ magnitude error (Vega mags)\\
bestjsrc	& Origin of best available $J$ magnitude (tm = 2MASS, ukidss = UKIDSS, vvv = VVV, denis = DENIS, mystix = MYStIX, arias06 = \citealt{2006MNRAS.366..739A}; see Section~\ref{sec:bestphot})\\
besthmag	& Best available $H$ magnitude (Vega mags)\\
besthmerr	& Best available $H$ magnitude error (Vega mags)\\
besthsrc	& Origin of best available $H$ magnitude (codes as for bestjsrc)\\
bestkmag	& Best available $K_s$ magnitude (Vega mags)\\
bestkmerr	& Best available $K_s$ magnitude error (Vega mags)\\
bestksrc	& Origin of best available $K_s$ magnitude (codes as for bestjsrc)\\
twomassnum & Identifier from 2MASS (if available)\\
jflim	& Set to $<$ if true brightness is fainter than the given magnitude at $J$\\
jmag	& 2MASS $J$ magnitude measurement (Vega mags); where jsrc = mystix, this is the value adopted by MYStIX, which may be from 2MASS or UKIRT\\
jmerr	& 2MASS $J$ magnitude error  (Vega mags); blank for upper limits\\
jsrc	& Origin of this measurement -- 2m means 2MASS photometry matched by position by the authors;
mystix means the counterpart and photometry were adopted from the MYStIX catalogs (from 2MASS or UKIRT).\\
hflim	& Set to $<$ if true brightness is fainter than the given magnitude at $H$\\
hmag	& 2MASS $H$ magnitude measurement  (Vega mags); where hsrc = mystix, this is the value adopted by MYStIX, which may be from 2MASS or UKIRT\\
hmerr	& 2MASS $H$ magnitude error  (Vega mags); blank for upper limits\\
hsrc    & Origin of this measurement -- 2m means 2MASS photometry matched by position by the authors;
mystix means the counterpart and photometry were adopted from the MYStIX catalogs (from 2MASS or UKIRT).\\
kflim	&Set to $<$ if true brightness is fainter than the given magnitude at $K_s$\\
kmag	& 2MASS $K_s$ magnitude measurement  (Vega mags); where ksrc = mystix, this is the value adopted by MYStIX, which may be from 2MASS or UKIRT\\
kmerr	& 2MASS $K_s$ magnitude error  (Vega mags); blank for upper limits\\
ksrc    & Origin of this measurement -- 2m means 2MASS photometry matched by position by the authors;
mystix means the counterpart and photometry were adopted from the MYStIX catalogs (from 2MASS or UKIRT).\\\\
vvvnum    & Identifier from VVV\\
vvvzmag	  & VVV $Z$ magnitude (Vega mag)\\
vvvzmerr  & VVV $Z$ magnitude error (Vega mag); 0.3 mag where assigned (see text)\\
vvvymag	  & VVV $Y$ magnitude (Vega mag)\\
vvvymerr  & VVV $Y$ magnitude error (Vega mag); 0.3 mag where assigned (see text)\\
vvvjmag   & VVV $J$ magnitude (Vega mag)\\
vvvjmerr  & VVV $J$ magnitude error (Vega mag)\\
vvvhmag	  & VVV $H$ magnitude  (Vega mag)\\
vvvhmerr  & VVV $H$ magnitude error (Vega mag)\\
vvvkmag	  & VVV $K_s$ magnitude  (Vega mag)\\
vvvkmerr  & VVV $K_s$ magnitude error (Vega mag)\\
denisnum & Identifier from DENIS\\
idmag  & DENIS $i$ magnitude (Vega mag)\\
idmerr & DENIS $i$ magnitude error (Vega mag)\\
jdmag  & DENIS $J$ magnitude  (Vega mag)\\
jdmerr & DENIS $J$ magnitude error (Vega mag)\\
kdmag  & DENIS $K_s$ magnitude  (Vega mag)\\
kdmerr & DENIS $K_s$ magnitude error (Vega mag)\\
ukidssnum & Identifier from UKIDSS\\
ukidssjmag  & UKIDSS $J$ magnitude  (Vega mag)\\
ukidssjmerr & UKIDSS $J$ magnitude error (Vega mag); 0.3 mag where assigned (see text)\\
ukidsshmag  & UKIDSS $H$ magnitude (Vega mag)\\
ukidsshmerr & UKIDSS $H$ magnitude error (Vega mag); 0.3 mag where assigned (see text)\\
ukidsskmag  & UKIDSS $K$ magnitude  (Vega mag)\\
ukidsskmerr & UKIDSS $K$ magnitude error (Vega mag); 0.3 mag where assigned (see text)\\
\cutinhead{MIR Photometry \& Associated Information}
bestirac1mag & Best available IRAC-1 magnitude (Vega mag)\\
bestirac1merr & Best available IRAC-1 magnitude error (Vega mag)\\
bestirac2mag & Best available IRAC-2 magnitude (Vega mag)\\
bestirac2merr & Best available IRAC-2 magnitude error (Vega mag)\\
bestirac3mag & Best available IRAC-3 magnitude (Vega mag)\\
bestirac3merr & Best available IRAC-3 magnitude error (Vega mag)\\
bestirac4mag & Best available IRAC-4 magnitude (Vega mag)\\
bestirac4merr & Best available IRAC-4 magnitude error (Vega mag)\\
iracsrc & Origin of any of the best available IRAC fluxes 
(mystix=MYStIX papers, gl=GLIMPSE, seip=SEIP, kumar= Kumar \etal., own=this paper) \\
bestmips1flim & Set to $<$ if true brightness is fainter than the given magnitude at MIPS-1 \\
bestmips1mag & Best available MIPS-1 magnitude (Vega mag)\\
bestmips1merr & Best available MIPS-1 magnitude error (Vega mag)\\
mips1src & Origin of any the best available MIPS-1 flux
(seip=SEIP, own=this paper) \\
seipname         & Identifier from SEIP\\
seipirac1mag     & SEIP IRAC-1 magnitude (Vega mag) \\
seipirac1merr    & SEIP IRAC-1 magnitude error (Vega mag)\\
seipirac2mag     & SEIP IRAC-2 magnitude (Vega mag) \\
seipirac2merr    & SEIP IRAC-2 magnitude error (Vega mag)\\
seipirac3mag     & SEIP IRAC-3 magnitude (Vega mag) \\
seipirac3merr    & SEIP IRAC-3 magnitude error (Vega mag)\\
seipirac4mag     & SEIP IRAC-4 magnitude (Vega mag) \\
seipirac4merr    & SEIP IRAC-4 magnitude error (Vega mag)\\
seipmips1flim    & Set to $<$ if true brightness is fainter than the given magnitude at MIPS-1 in SEIP\\
seipmips1mag     & SEIP MIPS-1 magnitude (Vega mag)\\
seipmips1merr    & SEIP MIPS-1 magnitude error (Vega mag)\\
glname           & Identifier from GLIMPSE\\
glirac1mag       & GLIMPSE IRAC-1 magnitude (Vega mag) \\
glirac1merr      & GLIMPSE IRAC-1 magnitude error (Vega mag)\\
glirac2mag       & GLIMPSE IRAC-2 magnitude (Vega mag) \\
glirac2merr      & GLIMPSE IRAC-2 magnitude error (Vega mag)\\
glirac3mag       & GLIMPSE IRAC-3 magnitude (Vega mag) \\
glirac3merr      & GLIMPSE IRAC-3 magnitude error (Vega mag)\\
glirac4mag       & GLIMPSE IRAC-4 magnitude (Vega mag) \\
glirac4merr      & GLIMPSE IRAC-4 magnitude error (Vega mag)\\
kumarnum         & Identifier from \citet{2010MNRAS.407.1170K}\\
kumarirac1mag    & IRAC-1 magnitude (Vega mag)\\
kumarirac2mag    & IRAC-2 magnitude (Vega mag)\\
kumarirac3mag    & IRAC-3 magnitude (Vega mag)\\
kumarirac4mag    & IRAC-4 magnitude (Vega mag)\\
kumarclass       & YSO class from \citet{2010MNRAS.407.1170K}\\
kumarysoflag     & Set to 1 if \citet{2010MNRAS.407.1170K} identified it as a YSO\\
wisenum     & Identifier from AllWISE\\
wise1flim	& Set to $<$ if true brightness is fainter than the given magnitude at WISE-1\\
wise1mag  & AllWISE WISE-1 magnitude (Vega mag) \\
wise1merr & AllWISE WISE-1 magnitude error (Vega mag) \\
wise2flim	& Set to $<$ if true brightness is fainter than the given magnitude at WISE-2\\
wise2mag  & AllWISE WISE-2 magnitude (Vega mag) \\
wise2merr & AllWISE WISE-2 magnitude error (Vega mag) \\
wise3flim	& Set to $<$ if true brightness is fainter than the given magnitude at WISE-3\\
wise3mag  & AllWISE WISE-3 magnitude (Vega mag) \\
wise3merr & AllWISE WISE-3 magnitude error (Vega mag) \\
wise4flim	& Set to $<$ if true brightness is fainter than the given magnitude at WISE-4\\
wise4mag  & AllWISE WISE-4 magnitude (Vega mag) \\
wise4merr & AllWISE WISE-4 magnitude error (Vega mag) \\
catwisenum     & Identifier from CatWISE\\
catwise1flim	& Set to $<$ if true brightness is fainter than the given magnitude at WISE-1\\
catwise1mag  & CatWISE WISE-1 magnitude (Vega mag) \\
catwise1merr & CatWISE WISE-1 magnitude error (Vega mag) \\
catwise2flim	& Set to $<$ if true brightness is fainter than the given magnitude at WISE-2\\
catwise2mag  & CatWISE WISE-2 magnitude (Vega mag) \\
catwise2merr & CatWISE WISE-2 magnitude error (Vega mag) \\
unwisename     & Identifier from unWISE\\
unwise1flim	& Set to $<$ if true brightness is fainter than the given magnitude at WISE-1\\
unwise1mag  & unWISE WISE-1 magnitude (Vega mag) \\
unwise1merr & unWISE WISE-1 magnitude error (Vega mag) \\
unwise2flim	& Set to $<$ if true brightness is fainter than the given magnitude at WISE-2\\
unwise2mag  & unWISE WISE-2 magnitude (Vega mag) \\
unwise2merr & unWISE WISE-2 magnitude error (Vega mag) \\
\cutinhead{MYStIX and Related Catalogs}
mystixjmag       & MYStIX $J$ magnitude (Vega mag)\\
mystixjmerr      & MYStIX $J$ magnitude error (Vega mag)\\
mystixhmag       & MYStIX $H$ magnitude (Vega mag)\\
mystixhmerr      & MYStIX $H$ magnitude error (Vega mag)\\
mystixkmag       & MYStIX $K_s$ magnitude (Vega mag)\\
mystixkmerr      & MYStIX $K_s$ magnitude error (Vega mag)\\
mystixirac1mag   & MYStIX IRAC-1 magnitude (Vega mag) \\
mystixirac1merr  & MYStIX IRAC-1 magnitude error (Vega mag)\\     
mystixirac2mag   & MYStIX IRAC-2 magnitude (Vega mag)\\     
mystixirac2merr  & MYStIX IRAC-2 magnitude error (Vega mag)\\     
mystixirac3mag   & MYStIX IRAC-3 magnitude (Vega mag)\\
mystixirac3merr  & MYStIX IRAC-3 magnitude error (Vega mag)\\     
mystixirac4mag   & MYStIX IRAC-4 magnitude (Vega mag)\\     
mystixirac4merr  & MYStIX IRAC-4 magnitude error (Vega mag)\\     
povich\_mirname   & Identifier from \citet{2013ApJS..209...31P} based on MIR catalog \\
povich\_nirname   & Identifier from \citet{2013ApJS..209...31P} based on NIR catalog \\     
povich\_xrayname  & Identifier from \citet{2013ApJS..209...31P} based on X-ray catalog \\     
broos\_iauname    & Identifier from \citet{2013ApJS..209...32B} that is IAU-compliant\\
broos\_xrayname   & Identifier from \citet{2013ApJS..209...32B} based on X-ray catalog\\
broos\_mirname    & Identifier from \citet{2013ApJS..209...32B} based on MIR catalog\\
broos\_nirname    & Identifier from \citet{2013ApJS..209...32B} based on NIR catalog\\
broos\_ob3name    & Identifier from \citet{2013ApJS..209...32B} based on OB catalog\\
romine\_name      & Identifier from \citet{2016ApJ...833..193R}\\
mystixmemb       & Set to 1 if any of the preceding MYStIX papers indicates it is a member.\\
ramirez20num     & Identifier from \citet{2020AA...633A.155R}\\
spicynum         & Identifier from \citet{2021ApJS..254...33K} \\
spicyclass       & Class from \citet{2021ApJS..254...33K}\\
\cutinhead{Sparser MIR and FIR Catalogs}
akariircname & Identifier from AKARI IRC \\
akari9flux & AKARI 9 \mum\ flux in $\mu$Jy\\
akari9ferr & Error in AKARI 9 \mum\ flux in $\mu$Jy\\
akari18flux & AKARI 18 \mum\ flux in $\mu$Jy\\
akari18ferr & Error in AKARI 18 \mum\ flux in $\mu$Jy\\
akarifisname & Identifier from AKARI FIS \\
akari65flux & AKARI 65 \mum\ flux in $\mu$Jy\\
akari65ferr & Error in AKARI 65 \mum\ flux in $\mu$Jy\\
akari90flux & AKARI 90 \mum\ flux in $\mu$Jy\\
akari90ferr & Error in AKARI 90 \mum\ flux in $\mu$Jy\\
akari140flux & AKARI 140 \mum\ flux in $\mu$Jy\\
akari140ferr & Error in AKARI 140 \mum\ flux in $\mu$Jy\\
akari160flux & AKARI 160 \mum\ flux in $\mu$Jy\\
akari160ferr & Error in AKARI 160 \mum\ flux in $\mu$Jy\\
msxname & Identifier from MSX \\
msxaflux & MSX A band flux in $\mu$Jy\\
msxaferr & Error in MSX A band flux in $\mu$Jy, computed from the percentage uncertainty in the MSX catalog\\
msxb1flux & MSX B1 band flux in $\mu$Jy\\
msxb1ferr & Error in MSX B1 band flux in $\mu$Jy, computed from the percentage uncertainty in the MSX catalog\\
msxb2flux & MSX B2 band flux in $\mu$Jy\\
msxb2ferr & Error in MSX B2 band flux in $\mu$Jy, computed from the percentage uncertainty in the MSX catalog\\
msxcflux & MSX C band flux in $\mu$Jy\\
msxcferr & Error in MSX C band flux in $\mu$Jy, computed from the percentage uncertainty in the MSX catalog\\
msxdflux & MSX D band flux in $\mu$Jy\\
msxdferr & Error in MSX D band flux in $\mu$Jy, computed from the percentage uncertainty in the MSX catalog\\
msxeflux & MSX E band flux in $\mu$Jy\\
msxeferr & Error in MSX E band flux in $\mu$Jy, computed from the percentage uncertainty in the MSX catalog\\
irasname & Identifier from IRAS \\
iras1flux & IRAS-1 flux in $\mu$Jy\\
iras1ferr & Error in IRAS-1 band flux in $\mu$Jy, taken to be 20\% of the flux\\
iras2flux & IRAS-2 flux in $\mu$Jy\\
iras2ferr & Error in IRAS-2 band flux in $\mu$Jy, taken to be 20\% of the flux\\
iras3flux & IRAS-3 flux in $\mu$Jy\\
iras3ferr & Error in IRAS-3 band flux in $\mu$Jy, taken to be 20\% of the flux\\
iras4flux & IRAS-4 flux in $\mu$Jy\\
iras4ferr & Error in IRAS-4 band flux in $\mu$Jy, taken to be 20\% of the flux\\
\cutinhead{Remaining Catalogs}
vandenancker\_ysoflag & Set to 1 if \citet{1997AAS..123...63V} indicates it is a YSO.\\
vanaltenajones & Identifier from \citet{1972AA....20..425V} as tabulated by \citet{1997AAS..123...63V}\\
sag & Identifier from \citet{1978MNRAS.184..467S} as tabulated by \citet{1997AAS..123...63V}\\
walker & Identifier from \citet{1957ApJ...125..636W} as tabulated by \citet{1997AAS..123...63V}\\
othervan & Other identifier from \citet{1997AAS..123...63V}\\
zhao06num & Identifier from \citet{2006ChJAA...6..435Z}\\
zhao06mem & Set to 1 if \citet{2006ChJAA...6..435Z} indicates it is a YSO.\\
chen07num           & Identifier from \citet{2007AJ....134.1368C}\\ 
chenmembprob        & Membership probability from \citet{2007AJ....134.1368C}\\ 
chenmembflag        & Set to 1 for those stars from \citet{2007AJ....134.1368C} we took as members\\ 
kalarinum & Identifier from \citet{2015MNRAS.453.1026K} \\
rauwnum             & Identifier from \citet{2002AA...395..499R}\\ 
castronum           & Identifier from \citet{2013PhDT........36C}\\ 
pris19name          & Identifier from \citet{2019AA...623A.159P}\\ 
prisclass           & Class from \citet{2019AA...623A.159P}\\ 
damiani19name       & Identifier from \citet{2019AA...623A..25D}\\ 
damiani19bestmemb   & Set to 1 for best members from \citet{2019AA...623A..25D}\\   
damiani19memb       & Set to 1 for members from \citet{2019AA...623A..25D}\\ 
damiani04num        & Identifier from \citet{2004ApJ...608..781D}\\ 
prisinzano05num     & Identifier from \citet{2005AA...430..941P}\\   
prisinzano07num     & Identifier from \citet{2007AA...462..123P}\\   
pris07disk          & Set to 1 for disks from \citet{2007AA...462..123P}\\ 
pris07memb          & Set to 1 for members from \citet{2007AA...462..123P}\\ 
arias06num          & Identifier from \citet{2006MNRAS.366..739A}\\ 
arias06jmag         & $J$ magnitude (Vega mag) from \citet{2006MNRAS.366..739A}\\ 
arias06jmerr        & $J$ magnitude error (Vega mag) from \citet{2006MNRAS.366..739A}\\ 
arias06hmag         & $H$ magnitude (Vega mag) from \citet{2006MNRAS.366..739A}\\ 
arias06hmerr        & $H$ magnitude error (Vega mag) from \citet{2006MNRAS.366..739A}\\ 
arias06kmag         & $K_s$ magnitude (Vega mag) from \citet{2006MNRAS.366..739A}\\ 
arias06kmerr        & $K_s$ magnitude error (Vega mag) from \citet{2006MNRAS.366..739A}\\ 
arias07num          & Identifier from \citet{2007MNRAS.374.1253A}\\ 
arias07spty         & Spectral type from \citet{2007MNRAS.374.1253A}\\ 
arias07yso          & Set to 1 if \citet{2007MNRAS.374.1253A} indicated it was a YSO\\ 
wrightmem           & Set to 1 for those stars identified as members by \citet{2019MNRAS.486.2477W}\\ 
jacksonmem          & Membership probability $P_{3D}$ from \citet{2022MNRAS.509.1664J}; $-1$ means the star is in their sample but was excluded from their membership analysis (see Section~\ref{sec:jackson})\\ 
hendersonnum        & Identifier from \citet{2012ApJ...747...51H} \\ 
hendersonperiod     & Rotation period from \citet{2012ApJ...747...51H}\\   
ordenesnum          & Identifier from \citet{2022MNRAS.517.6191O}\\ 
ordeneskmag         & $Ks$ magnitude from \citet{2022MNRAS.517.6191O}\\ 
ordeneskmerr        & $Ks$ magnitude error from \citet{2022MNRAS.517.6191O}\\ 
ordenesclass        & Class from \citet{2022MNRAS.517.6191O}\\ 
ordenesperiod       & Rotation period from \citet{2022MNRAS.517.6191O}\\ 
\enddata
\tablenotetext{a}{Also delivered to IRSA.}
\end{deluxetable}

\end{document}